\documentclass[fleqn,usenatbib]{mnras}
\usepackage{newtxtext,newtxmath}
\usepackage[T1]{fontenc}
\usepackage{ae,aecompl}

\usepackage{graphicx}	
\usepackage{amsmath}	
\usepackage{amssymb}	
\usepackage{color}
\usepackage{soul}

\newcommand\textlcsc[1]{\textsc{\MakeLowercase{#1}}}
\newcommand{\quotes}[1]{``#1''}
\defcitealias{Ucci2017}{U17}
\defcitealias{Ucci2017b}{U18}

\title[Dwarf galaxies ISM]{The interstellar medium of dwarf galaxies: new insights from Machine Learning analysis of emission line spectra}

\author[G. Ucci et al.]{
G. Ucci$^{1}$\thanks{\href{mailto:graziano.ucci@sns.it}{graziano.ucci@sns.it}},
A. Ferrara$^{1,2}$,
S. Gallerani$^{1}$,
A. Pallottini$^{1,3,4,5}$,
G. Cresci$^{6}$,\newauthor
C. Kehrig$^{7}$,
L. K. Hunt$^{6}$,
J. M. Vilchez$^{7}$,
L. Vanzi$^{8}$
\\
$^{1}$Scuola Normale Superiore, Piazza dei Cavalieri 7, 56126, Pisa, Italy\\
$^{2}$Kavli IPMU, The University of Tokyo, 5-1-5 Kashiwanoha, Kashiwa 277-8583, Japan\\
$^{3}$Centro Fermi, Museo Storico della Fisica e Centro Studi e Ricerche \quotes{Enrico Fermi}, Piazza del Viminale 1, Roma, 00184, Italy\\
$^{4}$Cavendish Laboratory, University of Cambridge, 19 J. J. Thomson Ave., Cambridge CB3 0HE, United Kingdom\\
$^{5}$Kavli Institute for Cosmology, University of Cambridge, Madingley Road, Cambridge CB3 0HA, UK\\
$^{6}$INAF - Osservatorio Astrofisco di Arcetri, largo E. Fermi 5, 50127 Firenze, Italy\\
$^{7}$Instituto de Astrof\'isica de Andaluc\'ia, CSIC, Apartado de correos 3004, 18080 Granada, Spain\\
$^{8}$Department of Electrical Engineering and Center of Astro Engineering, Pontificia Universidad Catolica de Chile,\\ Av. Vicua Mackenna 4860 Santiago, Chile
}

\date{Accepted XXX. Received YYY; in original form ZZZ}

\pubyear{2018}

\begin{document}
\label{firstpage}
\pagerange{\pageref{firstpage}--\pageref{lastpage}}
\maketitle

\begin{abstract}
Dwarf galaxies are ideal laboratories to study the physics of the interstellar medium (ISM). Emission lines have been widely used to this aim. Retrieving the full information encoded in the spectra is therefore essential. This can be efficiently and reliably done using Machine Learning (\textlcsc{ML}) algorithms. Here, we apply the \textlcsc{ML} code \textlcsc{GAME} to MUSE (Multi Unit Spectroscopic Explorer) and PMAS (Potsdam Multi Aperture Spectrophotometer) Integral Field Unit (IFU) observations of two nearby Blue Compact Galaxies (BCGs): Henize 2-10 and IZw18. We derive spatially resolved maps of several key ISM physical properties. We find that both galaxies show a remarkably uniform metallicity distribution. Henize 2-10 is a star forming dominated galaxy, with a Star Formation Rate ($SFR$) of about 1.2 M$_{\odot}$ yr$^{-1}$. Henize 2-10 features dense and dusty ($A_V$ up to 5-7 mag) star forming central sites. We find IZw18 to be very metal-poor ($Z$= 1/20 $Z_{\odot}$). IZw18 has a strong interstellar radiation field, with a large ionization parameter. We also use models of PopIII stars spectral energy distribution as a possible ionizing source for the HeII $\lambda$4686 emission detected in the IZw18 NW component. We find that PopIII stars could provide a significant contribution to the line intensity. The upper limit to the PopIII star formation is 52\% of the total IZw18 $SFR$.
\end{abstract}


\section{Introduction}
Emission lines in the spectra of galaxies contain a huge amount of information on the physical properties of their Interstellar Medium (ISM). The ISM of local Blue Compact Galaxies (BCGs), a sub-class of Dwarf Galaxies, represents a particularly interesting case of study. In fact, since BCGs are low metallicity, compact, star-forming systems, they are thought to represent local analogues of early galaxies \citep{Lequeux1979,Garland2015} that will become soon observable in greater detail with forthcoming instruments (e.g., JWST). Given their small distances, local BCGs can be studied at much higher spatial resolution with respect to high redshift galaxies. Thus, ISM studies of local BCGs can be used as benchmarks for understanding the structure, formation and evolution of high redshift galaxies. The concept of BCG was firstly introduced by \citet{Zwicky1965} to refer to \quotes{quasi-stellar galaxies}, namely galaxies with pure emission line spectra, initially barely distinguishable from stars \citep{Arp1965}. Since the spectra of BCGs resemble galactic HII regions \citep{Haro1956,Zwicky1966,Markarian1967}, these galaxies are also called HII galaxies \citep{Markarian1967,Sargent1970,Melnick1985,Hazard1986,Terlevich1991}. This is the case of Henize 2-10 \citep{Corbin1993}, a Wolf-Rayet starburst galaxy with a central radio source, a metallicity of 12+log(O/H) = 8.55 $\pm$ 0.02 \citep{Esteban2014}, intense star formation rate \citep[SFR $\sim$ 1.9 M$_{\odot}$ yr$^{-1}$, see][]{Reines2011}, and the possible presence of an active massive black hole \citep[][but see also \citealt{Cresci2017}]{Reines2016}. 

BCGs have different physical properties compared to other varieties of star-forming galaxies (Arp1965,Zwicky1966). In fact, their star formation occurs in short bursts, separated by long ($\sim$ 10$^7$ years) quiescent periods \citep{Sargent1970,Searle1972,Searle1973,Kunth2000} resulting in young stellar populations \citep[with commonly accepted indications of older underlying population,][]{Tolstoy2009}, and a small amount of dust. Very low metallicity values have been indeed measured in several BCGs: IZw18 \citep[12+log(O/H) = 7.17 $\pm$ 0.04,][]{Searle1972,Dufour1988,Skillman1993}; the starburst galaxy SBS 0335-052W \citep{Izotov1990,Izotov1997,Izotov2005,Izotov2009} with 12+log(O/H) = 7.13 $\pm$ 0.02 \citep{Izotov2009}; the irregular galaxy DDO 68 \citep[12+log(O/H) = 7.21 $\pm$ 0.03,][]{Pustilnik2005,Berg2012}; the low-luminosity dwarf Leo P \citep[12+log(O/H) = 7.17 $\pm$ 0.04,][]{McQuinn2015,Skillman2013}; AGC 198691, the most metal poor galaxy known in the local Universe \citep[12+log(O/H) = 7.02 $\pm$ 0.03,][]{Hirschauer2016}.

Most of the results reported above have been obtained by adopting diagnostics based on emission-line ratios. An alternative, promising approach relies on Machine Learning (\textlcsc{ML}) techniques \citep[][hereafter \citetalias{Ucci2017} and \citetalias{Ucci2017b}]{Ucci2017,Ucci2017b}. In this work, we study the ISM physical properties of Henize 2-10 and IZw18 by processing their Integral Field Unit (IFU) observations with the \textlcsc{ML} code \textlcsc{GAME} \citepalias[GAlaxy Machine learning for Emission lines][]{Ucci2017,Ucci2017b}. The advantage of this code relies on the fact that it is possible to infer simultaneously several ISM physical properties by using all the information in the spectra. The purpose of this work is to apply \textlcsc{GAME} in the case of two well-studied examples of local BCGs in view of future applications, especially at high redshifts.

The paper is organized as follows: in Sec. \ref{sec:game}, we summarize the main features of the \textlcsc{GAME} code; in Sec. \ref{sec:henize} and \ref{sec:izw18}, we present the results for Henize 2-10 and IZw18, respectively; in Sec. \ref{sec:conclusions} we discuss and summarize the main results of this study.

\section{Inferring ISM Physical Properties}\label{sec:game}
\textlcsc{GAME} is a \textlcsc{ML} code that infers the ISM physical properties by analyzing the emission line intensities in a galaxy spectrum \citepalias[for the complete description see][]{Ucci2017,Ucci2017b}. The code is based on a Supervised \textlcsc{ML} algorithm, and it is trained with a library of 100,000 synthetic spectra spanning many different ISM phases, including HII (ionized) regions, PDRs and neutral regions \citepalias{Ucci2017,Ucci2017b}. The library of synthetic spectra is generated by using the photoionization code \textlcsc{CLOUDY v13.03} \citep{Cloudy} by varying the total (i.e., HI + HII + H$_2$) gas density ($n$), total column density ($N_H$), ionization parameter\footnote{\label{udef}In this work we adopted the following definition for the ionization parameter: $U = Q(H)/( 4\pi R_{S}^{2} n c)$, where $Q(H)$ is the ionizing photon flux, $c$ is the speed of light and $R_{S}$ is the Str\"{o}mgren radius.} ($U$), and metallicity\footnote{In the library we assumed fixed solar abundance ratios for all the elements \citepalias{Ucci2017b}.} ($Z$); the library contains PopII and PopIII stellar populations \citepalias[see][]{Ucci2017b}. The emission line library is then processed in order to account for noise in the observations \citepalias{Ucci2017b}.
Given a set of input lines in a spectrum, the \textlcsc{ML} performs a training on the library, then evaluates the line intensities to give a determination of the physical properties; each physical property is determined by the AdaBoost \textlcsc{ML} algorithm separately and independently \citepalias{Ucci2017}. The errors on the input emission line intensities and the uncertainties on the physical properties determinations have been taken into account and they have been included in the analysis \citepalias{Ucci2017b}.

It is important to notice that \textlcsc{GAME} does not use a pre-selected subset of emission line ratios (i.e., [NII]/H$\alpha$, R$_{23}$), but rather reconstructs the ISM properties of galaxies exploiting all the lines available in the input, including the faint ones. As showed in \citetalias{Ucci2017b}, the accuracy of the determination of the physical properties improves with an increasing number of emission lines.

The physical properties directly inferred by \textlcsc{GAME} in this work are: gas density ($n$), column density ($N_H$), ionization parameter ($U$), metallicity ($Z$), Far-Ultraviolet (FUV, 6 - 13.6 eV) flux ($G$)\footnote{For a given library spectrum $G$, $n$, and $U$ are directly related, but they are determined independently by the \textlcsc{ML} algorithm, as explained in the text.}, visual extinction ($A_V$). Starting from these properties it is possible to derive information also on the star formation surface density ($\Sigma_{SFR}$), and the gas mass surface density ($\Sigma_{gas}$), as detailed below.
We can use $N_H$  maps to obtain an estimate of the total gas mass contained in each spaxel of the map:

\begin{equation}
M =  \mu m_H N_H A_{spax}\,,
\label{eq:mass}
\end{equation}
where $\mu$ is the mean molecular weight (in the following we assume $\mu = 1.4$), $m_H$ is the hydrogen atom mass, and $A_{spax}$ is the spaxel area. The gas mass surface density is then

\begin{equation}
\Sigma_{gas} = \mu m_H N_H\,.
\label{eq:sigma_mass}
\end{equation}

Using these quantities, we can also infer the star formation rate surface density ($\Sigma_{SFR}$), by assuming a Schmidt-Kennicutt relation \citep{Schmidt1959,Kennicutt1998_schmidt,Krumholz2012}:

\begin{equation}
\Sigma_{SFR} = \eta \frac{\Sigma_{gas}}{t_{sf}}
\label{eq:sfr:pre}
\end{equation}
where $\eta$ is the star formation efficiency and $t_{sf}$ is the star formation time scale. As shown by \citet{Krumholz2012}, the relation is well fitted in a variety of environments by using $t_{sf} = t_{ff}$, where $t_{ff}$ is the gas free-fall time, and $\eta = 0.015$. Thus we can write eq. \ref{eq:sfr:pre} in terms of quantities directly derived from \textlcsc{GAME} (i.e., $n$ and $N_H$):

\begin{equation}
\Sigma_{SFR} = 0.015\, m_H^{3/2} N_H \sqrt{\frac{32 G n}{3 \pi}}.
\label{eq:sfr}
\end{equation}

Alternatively it is also possible to derive the SFR from the H$\alpha$ line intensity after correcting for the dust extinction, using specific calibrators \citep{Kennicutt1998_schmidt,Calzetti2007,Kennicutt2012}. In Sec. \ref{sec:sfr_he} we will show a comparison between the SFR estimated via eq. \ref{eq:sfr} and the calibration from \citet{Calzetti2008} with different dust extinction laws.
We now apply \textlcsc{GAME} to the study of two BCGs, Henize 2-10 and IZw18.

\section{Henize 2-10}\label{sec:henize}
\begin{figure}
	\centering
	\includegraphics[width=1.0\linewidth]{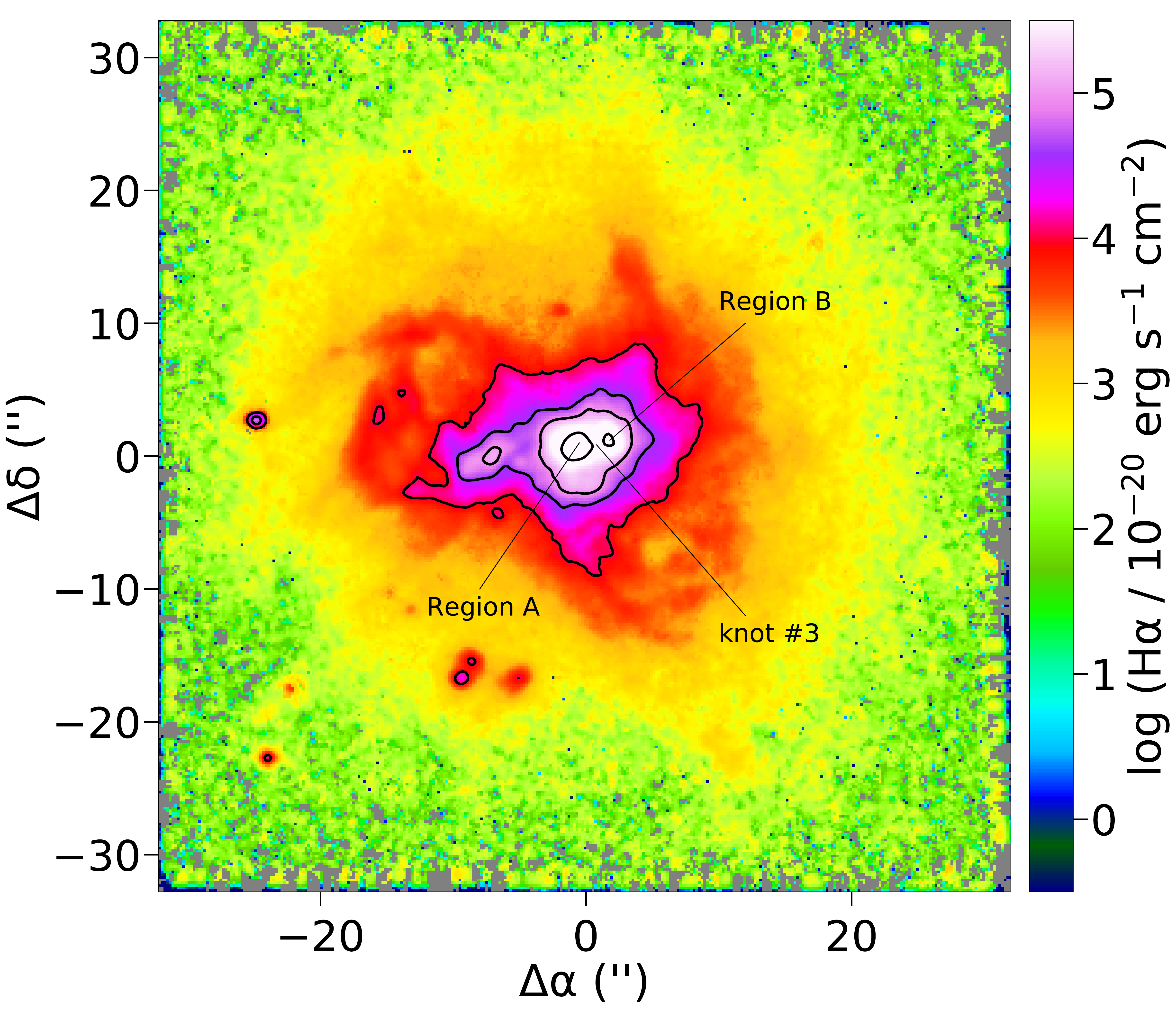}
	\caption{H$\alpha$ image of Henize 2-10 \citep{Cresci2017}. Black lines show the contours of the H$\alpha$ emission (80, 10, 4, and 1 in units of 10$^{-16}$ ergs s$^{-1}$ cm$^{-2}$). The H$\alpha$ contour levels are the same as in \citet{Cresci2017} in order to be able to directly compare their figures. Shown are the two regions A and B where the H$\alpha$ flux peaks. We also show the radio source designated by \citet{Johnson2003} as knot \#3. The field of view of the image is (64\arcsec)$^2$ corresponding to a physical size of (2.56 kpc)$^2$.}
	\label{fig:he_halpha}
\end{figure}

The blue compact local galaxy Henize 2-10 (hereafter He 2-10) can be considered as a prototype of an HII \citep{Allen1976} Wolf-Rayet \citep{Dodorico1983,Kawara1987,Conti1991,Vacca1992} starburst galaxy. It is located at a distance of 8.23 Mpc \citep{Tully2013} with a corresponding angular scale of 40 pc/\arcsec, and its core has an optical extent less than 1 kpc. He 2-10 has been extensively studied both in the optical and in the infrared wavelength ranges \citep{Vanzi1997,Vacca2002,Engelbracht2005}, up to the sub-millimeter \citep{Bayet2004,Johnson2017}. It has a stellar mass of (10 $\pm$ 3) $\times$ 10$^{9}$ M$_{\odot}$ \citep{Reines2011,Nguyen2014}. The estimated SFR is $\sim$ 1.9 M$_{\odot}$ yr$^{-1}$ \citep{Reines2011}, as revealed by classical indicators such as the H$\alpha$ \citep{Mendez1999} and the 24 $\mu$m flux \citep{Engelbracht2005}. In Fig. \ref{fig:he_halpha}, we show the H$\alpha$ image of He 2-10. The bulk of the H$\alpha$ emission is located in two central emitting regions (hereafter Region A and B) separated by $\sim 2\arcsec$ ($\sim$ 80 pc).

An unresolved non-thermal nuclear source has been found in the nucleus of He 2-10 \citep{Kobulnicky1999,Johnson2003}. This source (designated as knot \#3) is located between Regions A and B in Fig. \ref{fig:he_halpha}. \citet{Reines2011} linked this source to the presence of a compact X-ray emission detected with Chandra observations \citep{Kobulnicky2010}. The size of this emitting region has been constrained to be < 1 pc $\times$ 3 pc \citep{Reines2012} and recently associated with the presence of a weakly accreting super massive black hole having mass log(M/M$_{\odot}$) $\sim$ 6.3 \citep{Reines2016}. The origin of the X-ray emission in He 2-10 is still debated \citep{Cresci2017}. We further discuss this point in Appendix \ref{sec:sf_dominated} where we show that on scales larger than $\sim$ 4 pc from the center, the radiation field is dominated by stars.

\subsection{ISM physical properties in He 2-10}\label{sec:phys_he}
In this section, we present a systematic overview of the ISM physical properties in He 2-10, inferred by applying \textlcsc{GAME} \citepalias{Ucci2017,Ucci2017b} to MUSE \citep[Multi Unit Spectroscopic Explorer,][]{Bacon2010} optical integral field observations taken from \citet{Cresci2017}. We have a total of 321 $\times$ 328 = 105,288 spaxels with a spatial sampling of $0.2\arcsec$ $\times$ $0.2\arcsec$. The spectral resolution (i.e., $R = \lambda / \Delta\lambda$) goes from 1750 at 4650 \AA{} to 3750 at 9300 \AA{} \citep{Cresci2015,Cresci2017}. The emission lines used as input for \textlcsc{GAME} are reported in Table \ref{table:lines_henize} along with the fraction of spaxels with a detected and fitted line, having SNR > 3. For details on the line fitting procedure we refer the reader to \citet{Cresci2017}.

\begin{table}
	\caption{Emission lines and wavelengths used to analyze the galaxy He 2-10. The third column reports the fraction of spaxels with a detected and fitted line, having SNR > 3.}
	\centering
	\vspace{2mm}
	\begin{tabular}{c c c}
		\hline\hline
		line & wavelength [\AA] & fraction of spaxels\\
		\hline
		H$\beta$ & 4861 & 20 \%\\
        {[O III]} & 5007 & 15 \%\\
        He I & 5876 & 3 \%\\
        {[O I]} & 6300 & 2 \%\\
        H$\alpha$ & 6563 & 36 \%\\
        {[N II]} & 6584 & 18 \%\\
        He I & 6678 & 1 \%\\
        {[S II]} & 6717 & 15 \%\\
        {[S II]} & 6731 & 11 \%\\
        {[S III]} & 9069 & 2 \%\\
		\hline
		\hline
	\end{tabular}
	\label{table:lines_henize}
\end{table}

\begin{figure*}
    \raggedleft
    \includegraphics[width=0.48\textwidth]{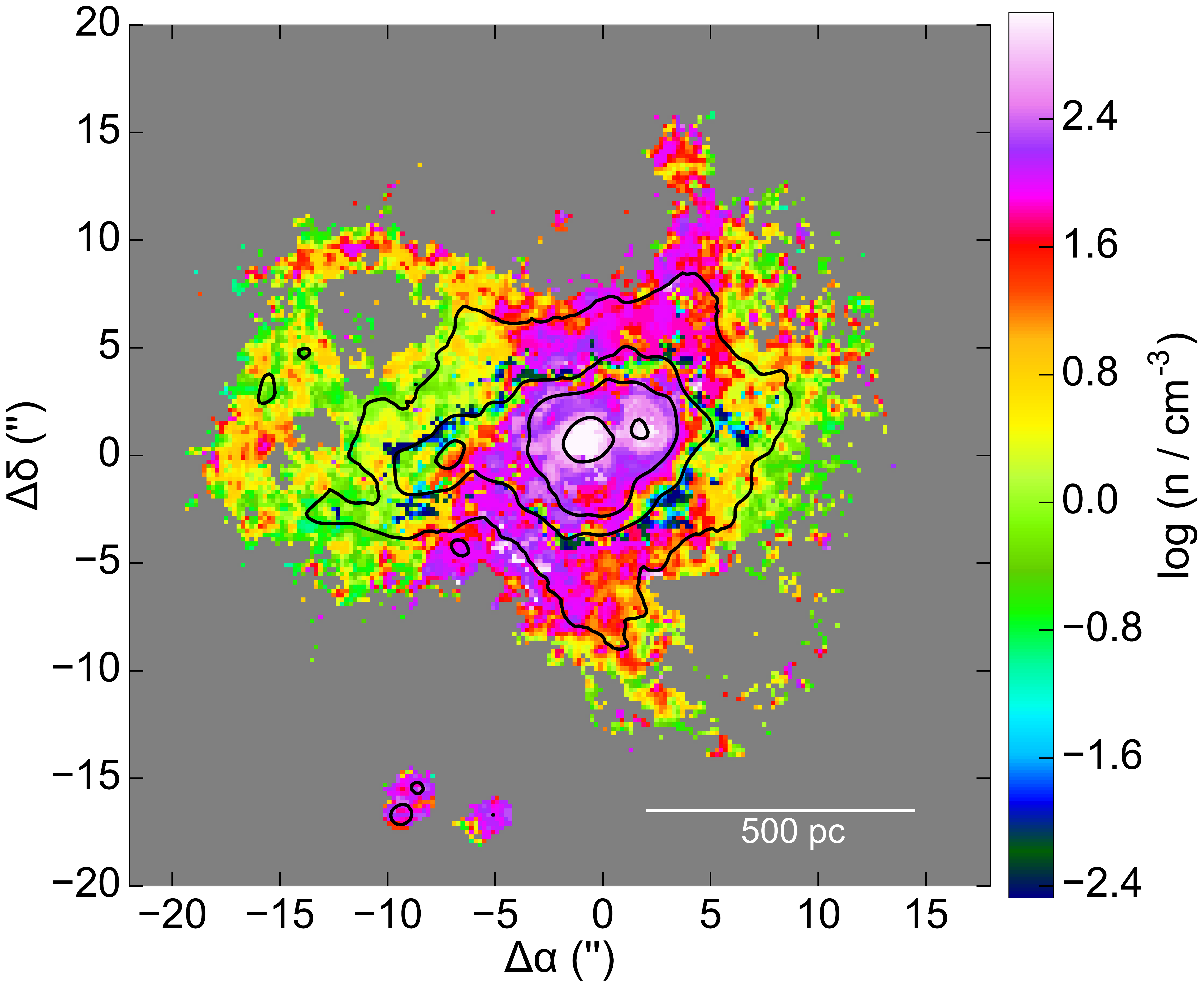}
    \raggedleft
    \includegraphics[width=0.482\textwidth]{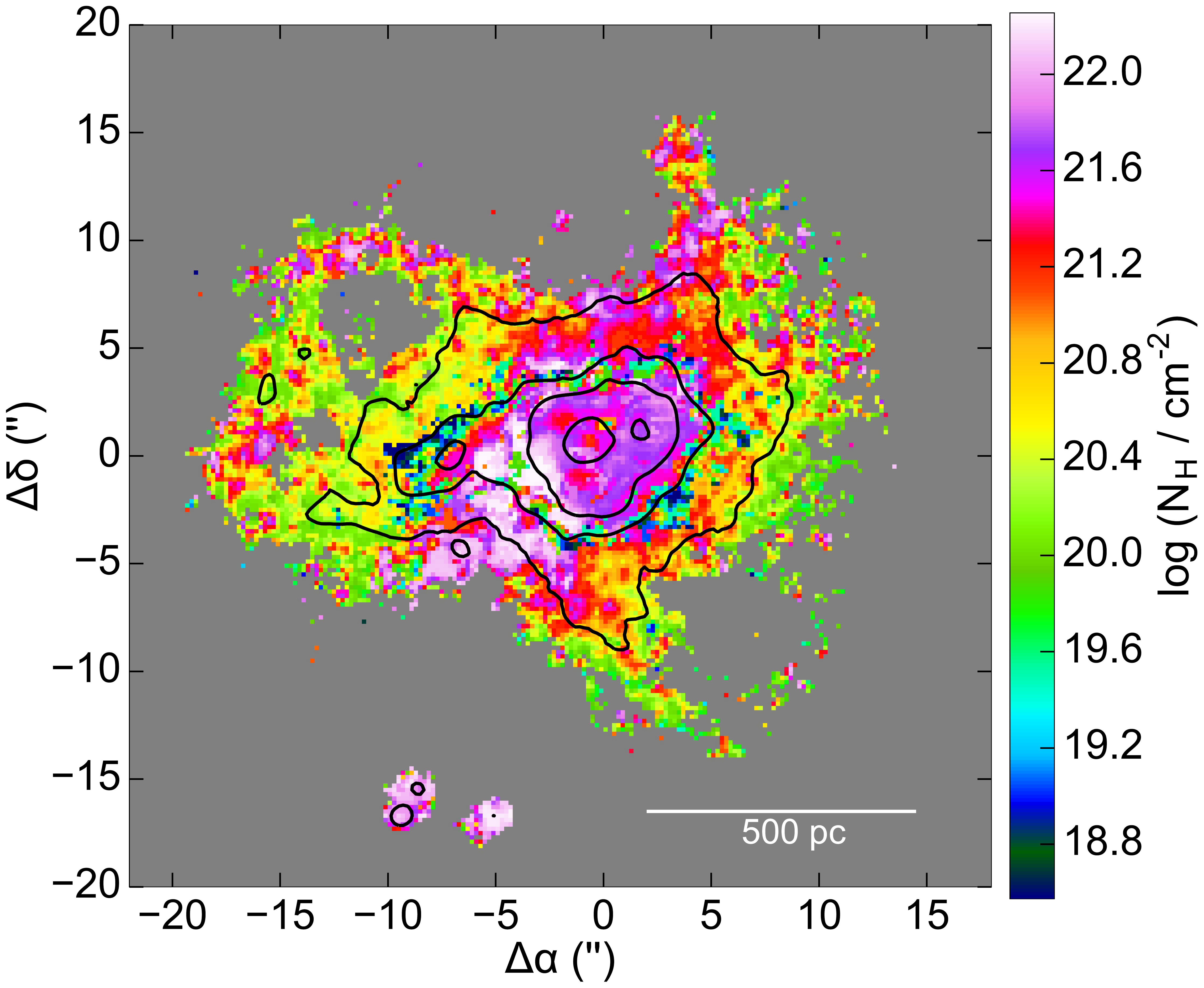}
    \raggedleft
    \includegraphics[width=0.482\textwidth]{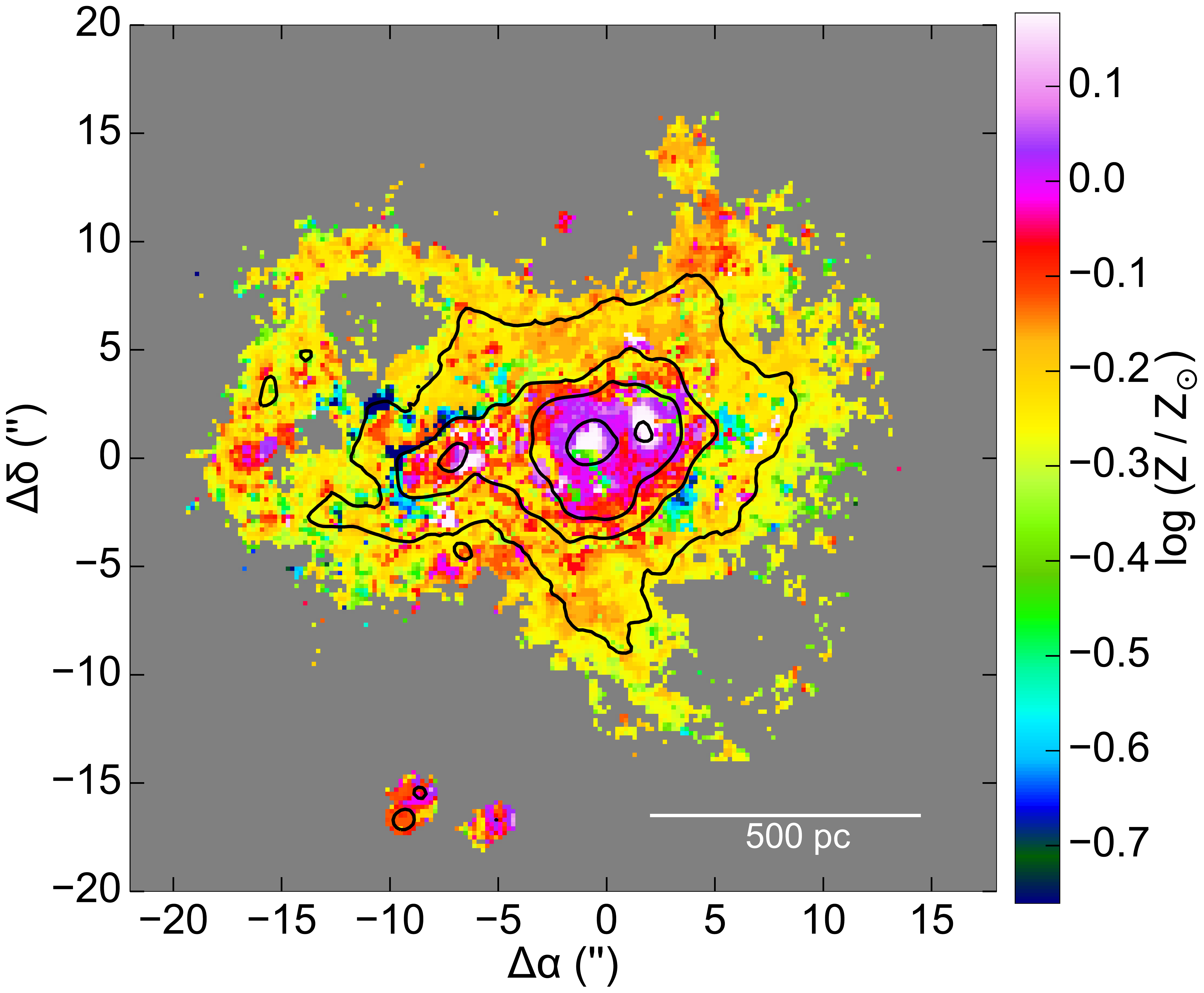}
	\raggedleft
	\includegraphics[width=0.478\textwidth]{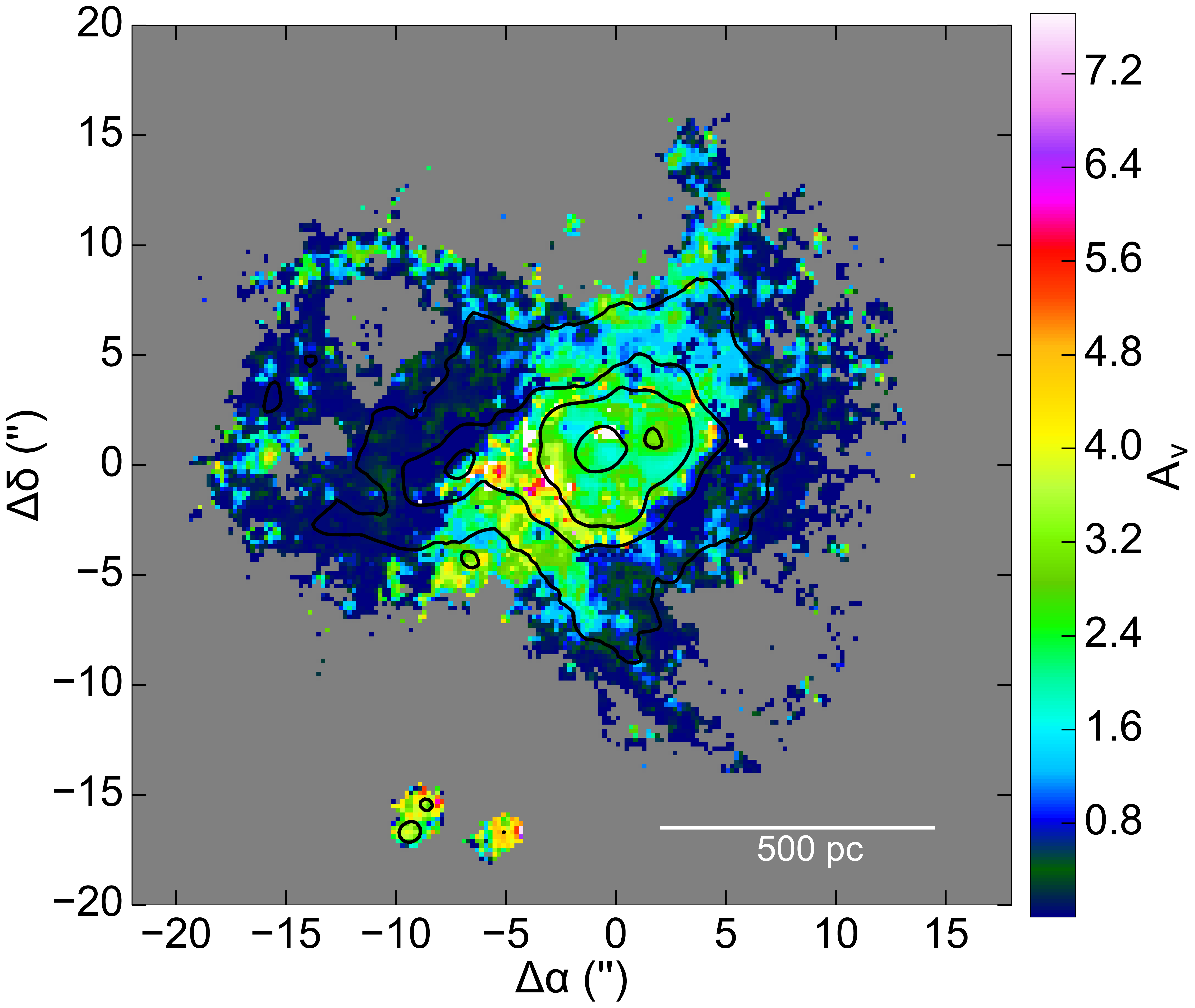}
    \raggedright
    \includegraphics[width=0.485\textwidth]{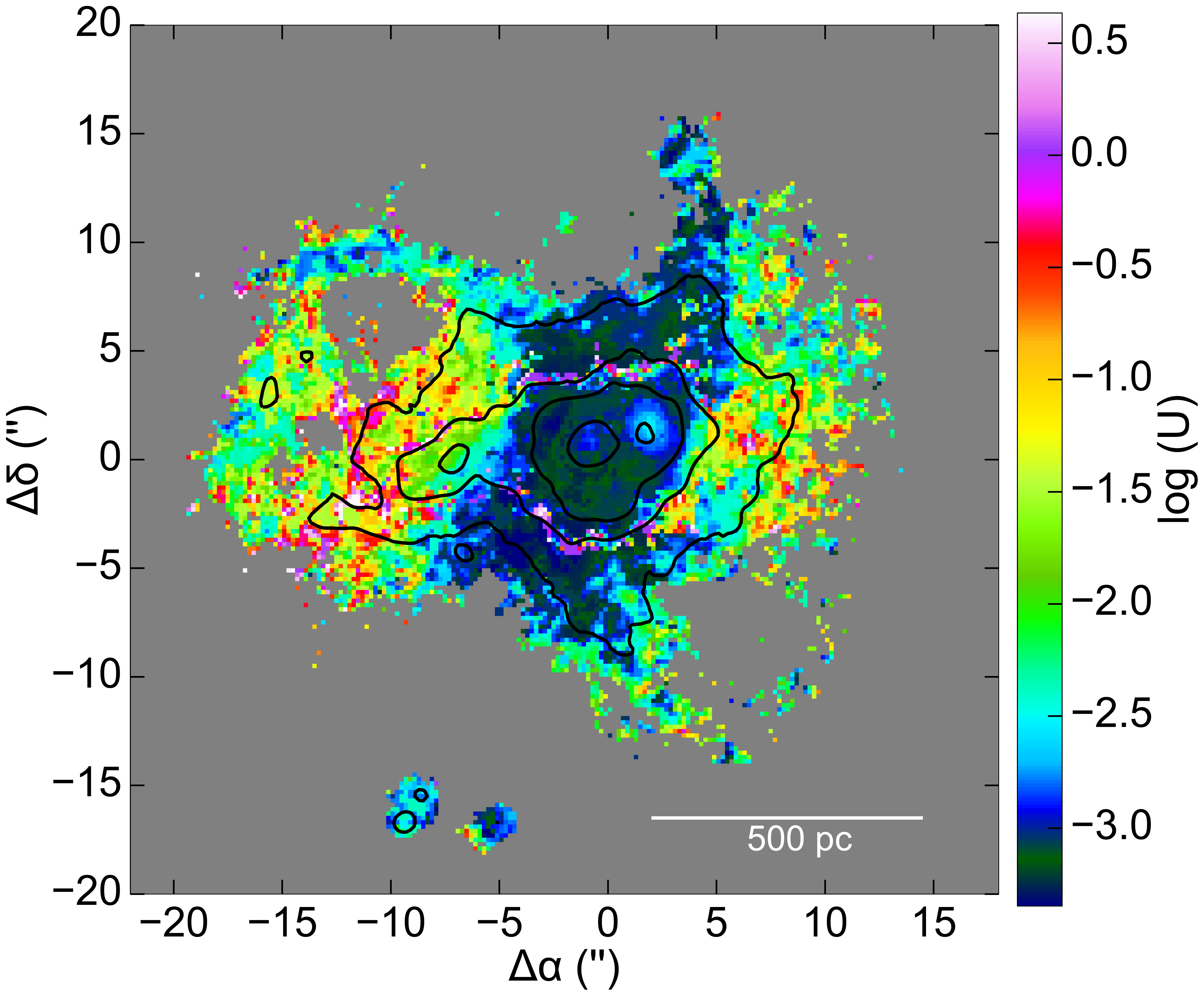}
    \includegraphics[width=0.485\textwidth]{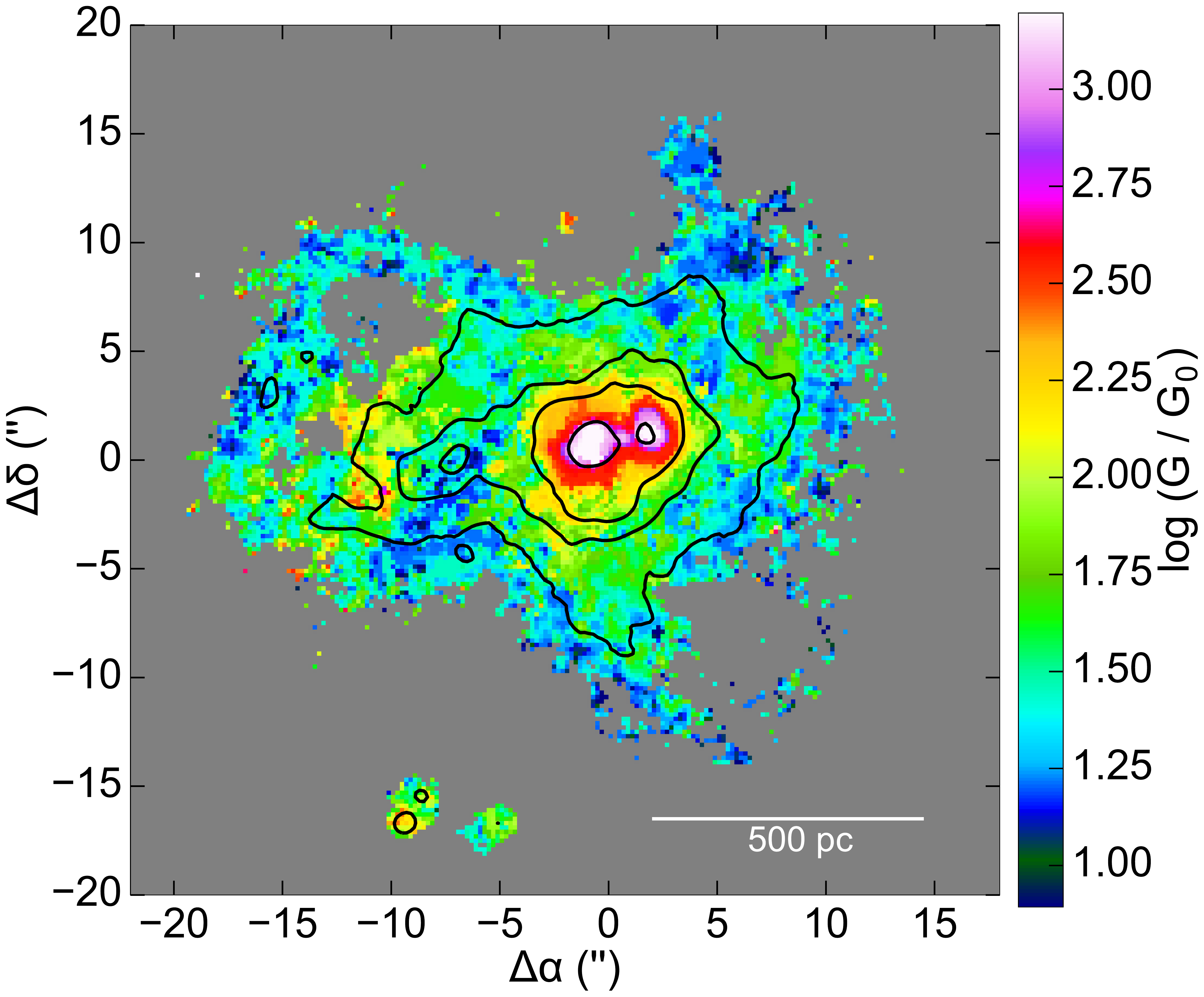}
    \caption{Maps of the ISM physical properties inferred for the local galaxy He 2-10 using the code \textlcsc{GAME} \citep{Ucci2017,Ucci2017b}: density ($n$), column density ($N_H$), metallicity ($Z$), visual extinction ($A_V$), ionization parameter ($U$) and the FUV flux in Habing units ($G/G_{0}$). As in Fig. \ref{fig:he_halpha}, black lines show the contours of the H$\alpha$ emission line (80, 10, 4, and 1 in units of 10$^{-16}$ ergs s$^{-1}$ cm$^{-2}$).}
	\label{fig:He_2-10_physical}
\end{figure*}

\begin{figure*}
	\centering
	\includegraphics[width=1.0\linewidth]{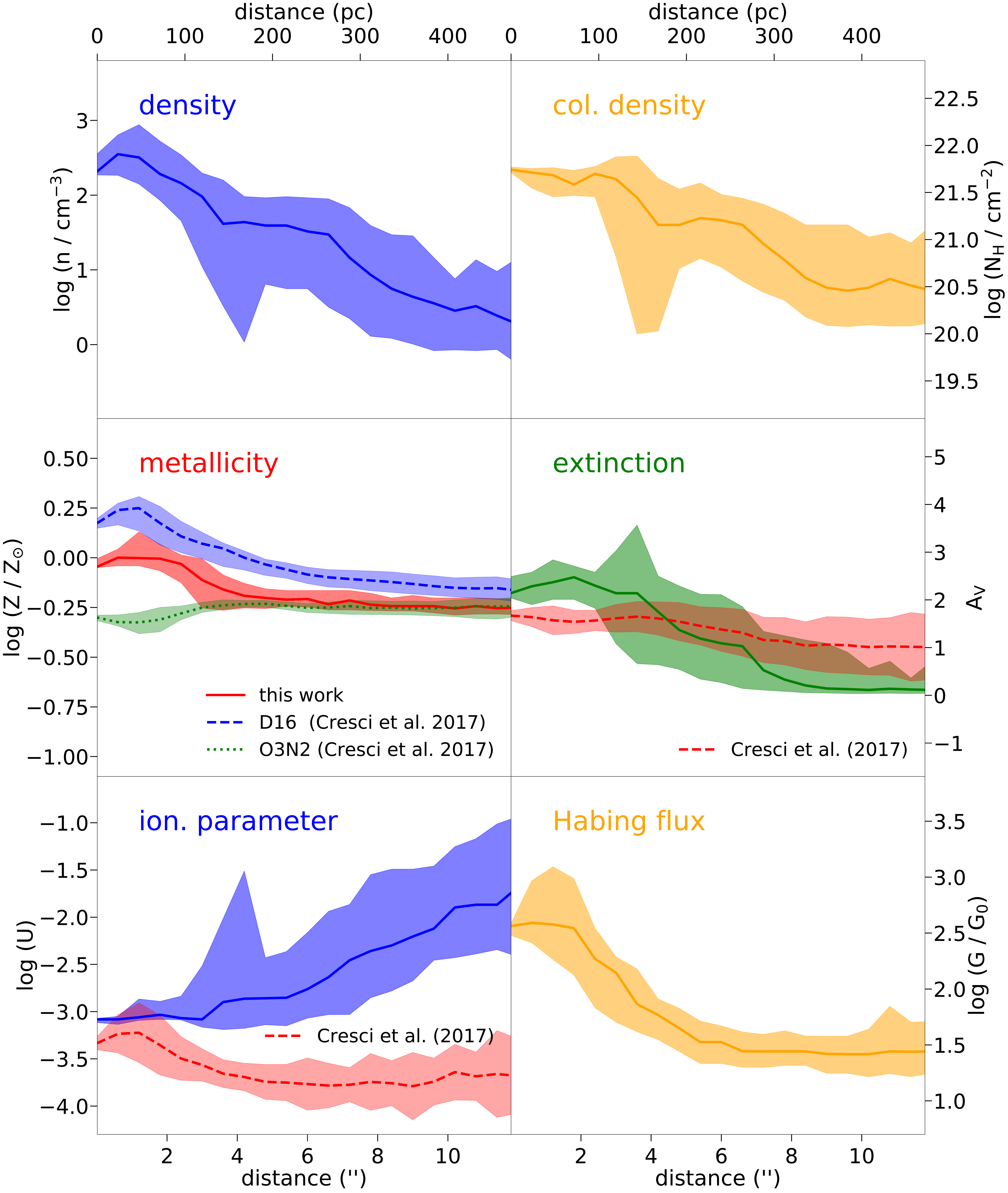}
	\caption{Median radial profiles (median of the inferred physical property values in annular bins of 0.6\arcsec) of the He 2-10 ISM physical properties obtained from the maps shown in Fig. \ref{fig:He_2-10_physical}, centered on ($\Delta\alpha$,$\Delta\delta$)=(0,0). Upper and lower shaded regions denote respectively the 2nd and 3rd quartiles. For metallicity, extinction and ionization parameter, we also report the radial profiles as inferred by \citet{Cresci2017}.}
	\label{fig:he_radials_physical}
\end{figure*}

\begin{figure*}
	\centering
    \includegraphics[width=1.0\textwidth]{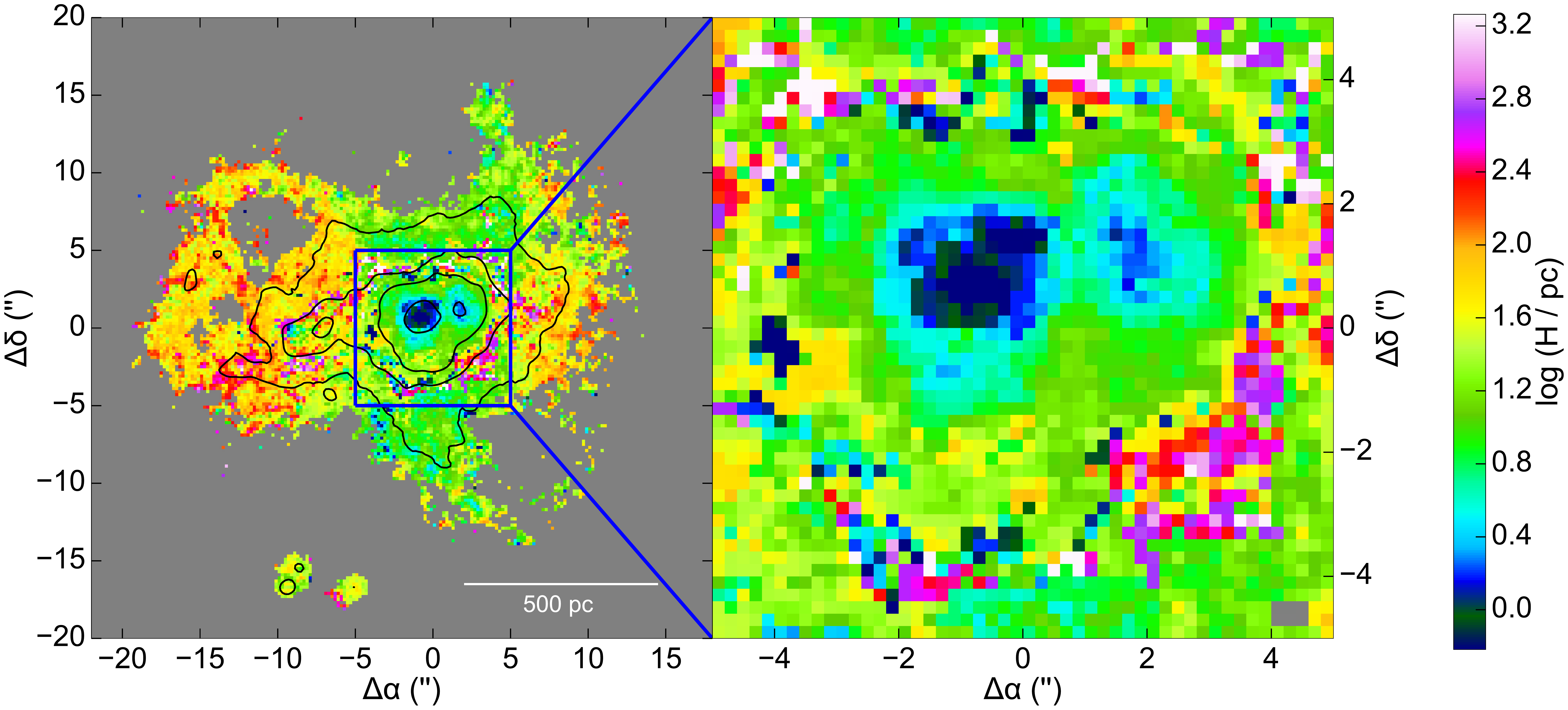}
    \caption{\textit{Left panel}: \quotes{effective} scale height defined as $H$ = $N_H$/$n$. As in Fig. \ref{fig:he_halpha}, black lines show the contours of the H$\alpha$ emission line (80, 10, 4, and 1 in units of 10$^{-16}$ ergs s$^{-1}$ cm$^{-2}$). \textit{Right panel}: zoom on the blue box.}
	\label{fig:He_2-10_radius}
\end{figure*}

\begin{figure*}
	\centering
    \includegraphics[width=0.485\textwidth]{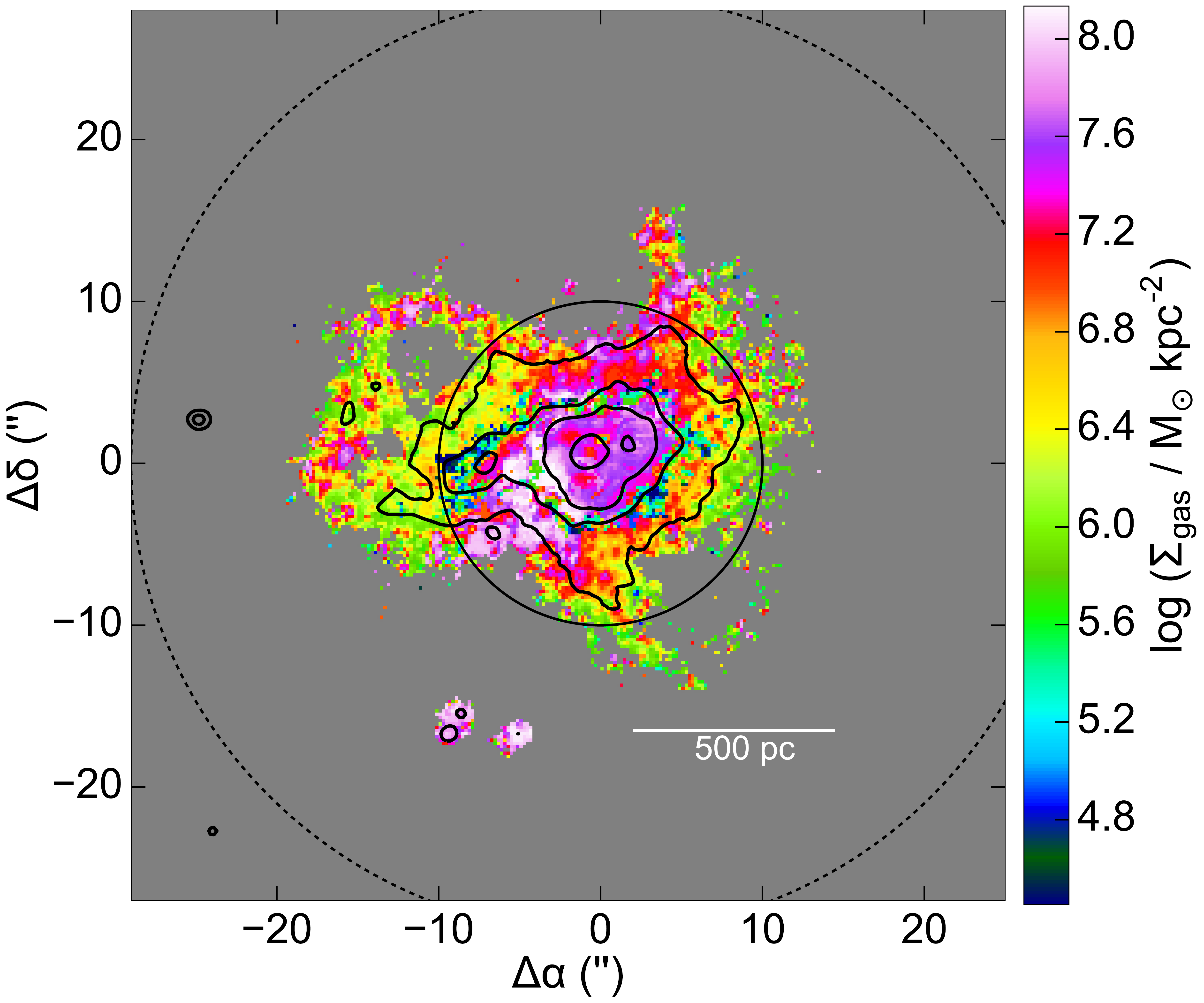}
    \includegraphics[width=0.48\textwidth]{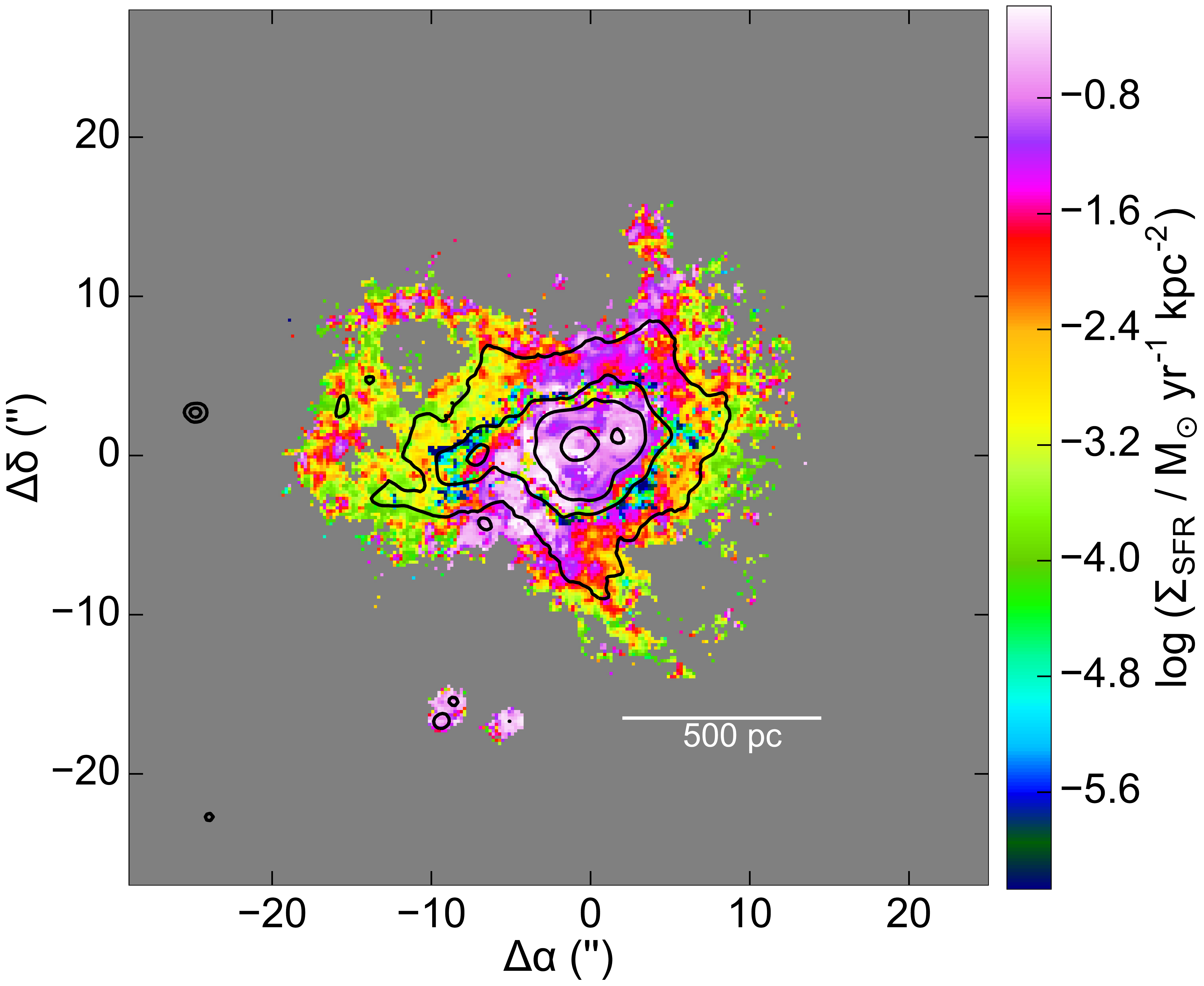}
    \caption{Gas mass surface density ($\Sigma_{gas}$, \textit{left panel}) and Star Formation surface density ($\Sigma_{SFR}$, \textit{right panel}) for the local galaxy He 2-10. As in Fig. \ref{fig:he_halpha}, black lines show the contours of the H$\alpha$ emission line (80, 10, 4, and 1 in units of 10$^{-16}$ ergs s$^{-1}$ cm$^{-2}$). Inner and outer black circles denote the areas covered by the CO observations of \citet{Vanzi2009} and \citet{Kobulnicky1995}, respectively (see text for details).}
	\label{fig:He_2-10_inferred}
\end{figure*}

\begin{figure*}
	\centering
	\includegraphics[width=0.9\linewidth]{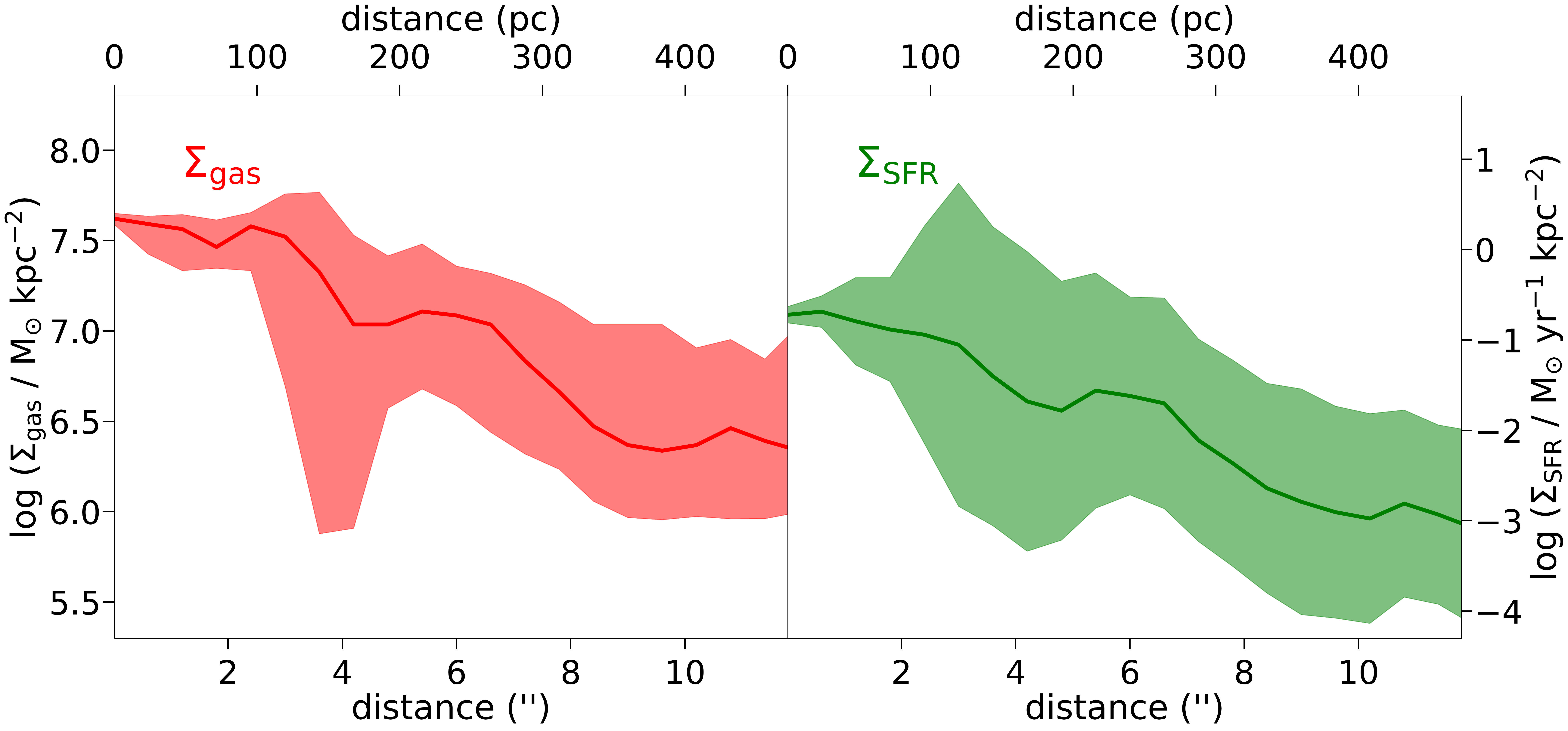}
	\caption{Median radial profiles (median of the inferred physical property values in annular bins of 0.6\arcsec) of He 2-10 $\Sigma_{gas}$ and $\Sigma_{SFR}$ (\textit{left} and \textit{right panel}, respectively) obtained from the maps shown in Fig. \ref{fig:He_2-10_inferred}, and centered on ($\Delta\alpha$,$\Delta\delta$)=(0,0). Upper and lower shaded regions denote respectively the 2nd and 3rd quartiles.}
	\label{fig:he_radials_inferred}
\end{figure*}

The resulting maps (median radial profiles\footnote{Because the physical properties distributions in the inferred maps are relatively spread, we preferred to use the median instead of the average in order to prevent an increase of the radial profiles purely driven by the presence of few high values.} centered on ($\Delta\alpha$,$\Delta\delta$)=(0,0)) of the inferred physical properties are reported in Fig. \ref{fig:He_2-10_physical} and \ref{fig:He_2-10_inferred} (Fig. \ref{fig:he_radials_physical} and \ref{fig:he_radials_inferred}). As observations outside the center of the galaxy become noisy, we selected only the lines with SNR > 3 (see Table \ref{table:lines_henize}). Then we analyzed the spaxels with at least 6 lines satisfying this condition (see discussion in Appendix D of \citetalias{Ucci2017b}). These spaxels are the ones plotted in each panel of Fig. \ref{fig:He_2-10_physical} and \ref{fig:He_2-10_inferred}. In Appendix \ref{sec:sigmas} we also report the uncertainties on the physical properties computed with GAME on all the spaxels visualized.

\subsubsection{Gas density}\label{sec:he_n}
The gas density map ($n$) in the upper left panel of Fig. \ref{fig:He_2-10_physical} shows that the highest density is reached close to Regions A and B, where log($n$/cm$^{-3}$) $\sim$ 3.3. For distances $d$ $\gtrsim$ 5\arcsec ($\gtrsim$ 200 pc) from the center, the median density radial profile in the galaxy decreases until a distance of $\sim$ 350 pc around a value of log($n$/cm$^{-3}$) $\sim 0.5$ (see Fig. \ref{fig:he_radials_physical}). The upper right panel of Fig. \ref{fig:He_2-10_physical} shows that column densities in the He 2-10 central region (i.e., within $10\arcsec$ from the center) vary from $\sim$ 10$^{20}$ cm$^{-2}$ up to $\sim$ 3 $\times$ 10$^{22}$ cm$^{-2}$, consistently with results found in literature \citep{Baas1994,Kobulnicky1995,Meier2001,Santangelo2009,Cresci2017}.

The left panel of Fig. \ref{fig:He_2-10_radius} shows the map of the the effective scale height of the disk, defined as $H$ = $N_H/n$. In the right panel, we zoom in the central region [5\arcsec $\times$ 5\arcsec, i.e. 200 $\times$ 200 pc$^2$]. where the uncertainties are smaller (see Fig. \ref{fig:He_2-10_sigma} in Appendix \ref{sec:sigmas}). In the central region, we find a relatively small height of the disk, i.e. -0.5 $\lesssim$ log($H$/pc) $\lesssim$ 1.5 (0.3 $\lesssim$ $H$/pc $\lesssim$ 30). This is the typical dimension of Giant Molecular Clouds (GMCs) that we are resolving in the center of the galaxy. Region A and B, with their high values of densities and metallicities, and given also the high H$\alpha$ and FUV flux $G$, can be identified as the sites of intense star formation.

\subsubsection{Metallicity}\label{sec:he_Z}
The metallicity map and the corresponding radial profile are reported in the central left panels of Fig. \ref{fig:He_2-10_physical} and \ref{fig:he_radials_physical}, respectively. We find that Region A and B ($d$ $\lesssim 4\arcsec$) are characterized by super-solar values (log($Z/Z_{\odot}$) $\sim$ 0.2 - 0.3), while at larger distances (4\arcsec $\lesssim$ $d$ $\lesssim$ 10\arcsec) the metallicity is slightly sub-solar (i.e., log($Z/Z_{\odot}$) $\sim$ -0.15) and remarkably uniform (within a factor of $\sim$ 2). The obtained average metallicity (12+log(O/H) $\sim$ 8.6) is the one expected from the mass-metallicity relation for a galaxy with a stellar mass of M$_{*}$ $\sim$ 10$^{10}$ M$_{\odot}$, i.e. 12+log(O/H) $\sim$ 8.7 \citep{Sanchez2017}.

\begin{figure}
	\centering
    \includegraphics[width=0.95\linewidth]{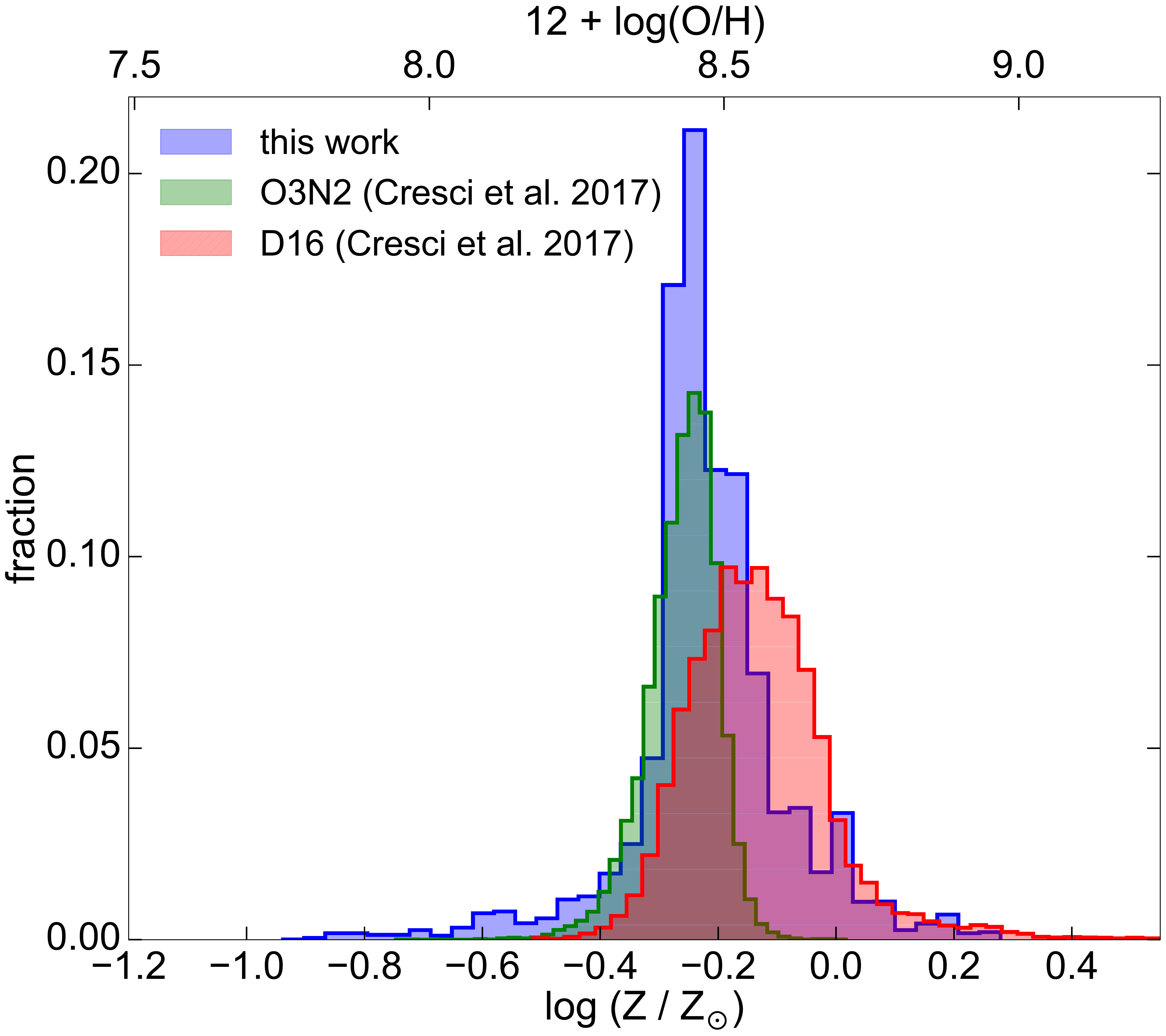}
    \caption{Distributions of the He 2-10 metallicity (central left panel of Fig. \ref{fig:He_2-10_physical}). The blue shaded histogram reports the inferred values for the metallicity obtained in this work. The green and red shaded regions show the results obtained by \citet{Cresci2017} using the $O3N2$ line ratio and the diagnostic by \citet{Dopita2016} ($D16$), respectively.}
	\label{fig:hist_metals_he}
\end{figure}

\begin{figure*}
	\centering
    \includegraphics[width=0.49\textwidth]{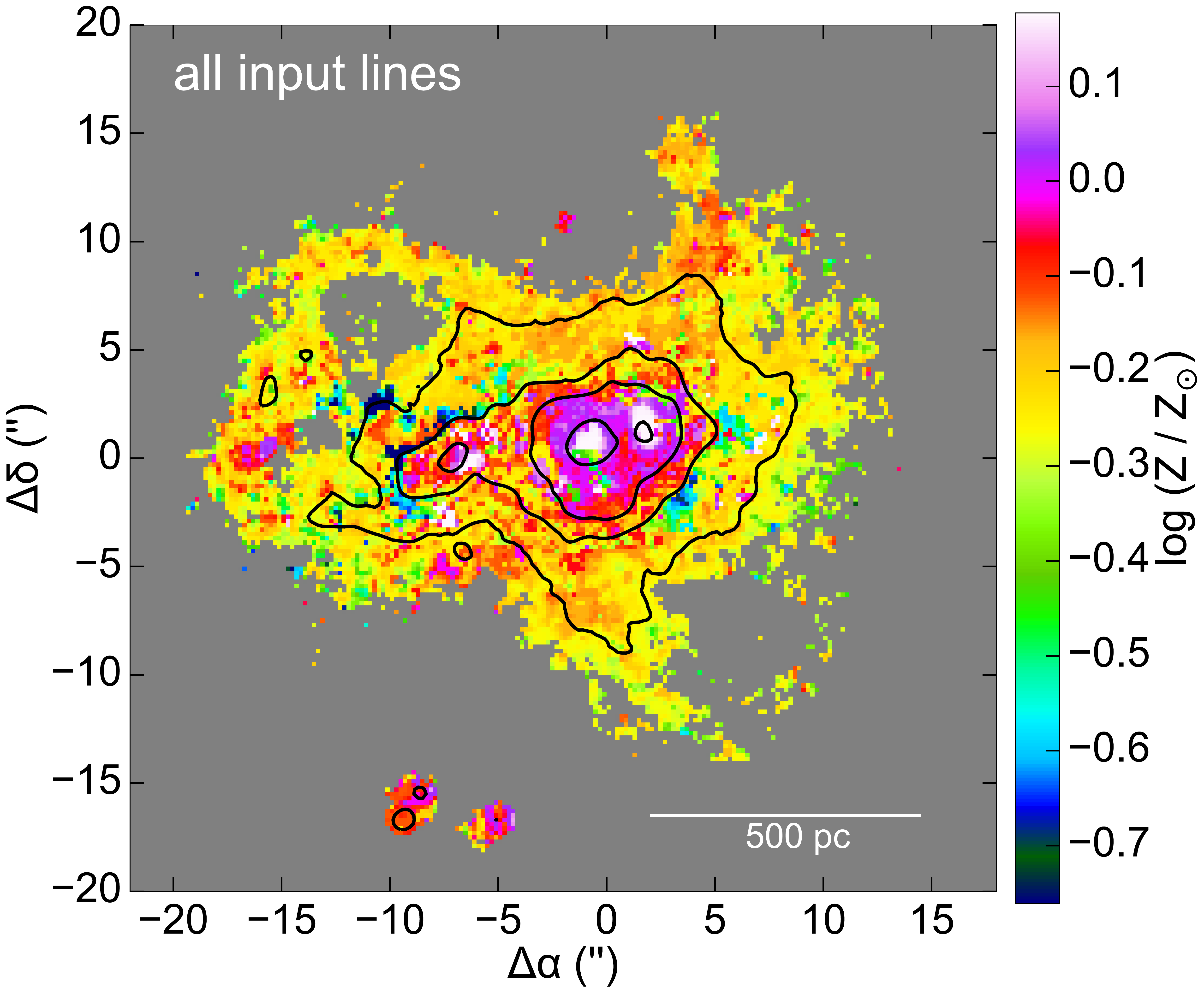}
    \includegraphics[width=0.49\textwidth]{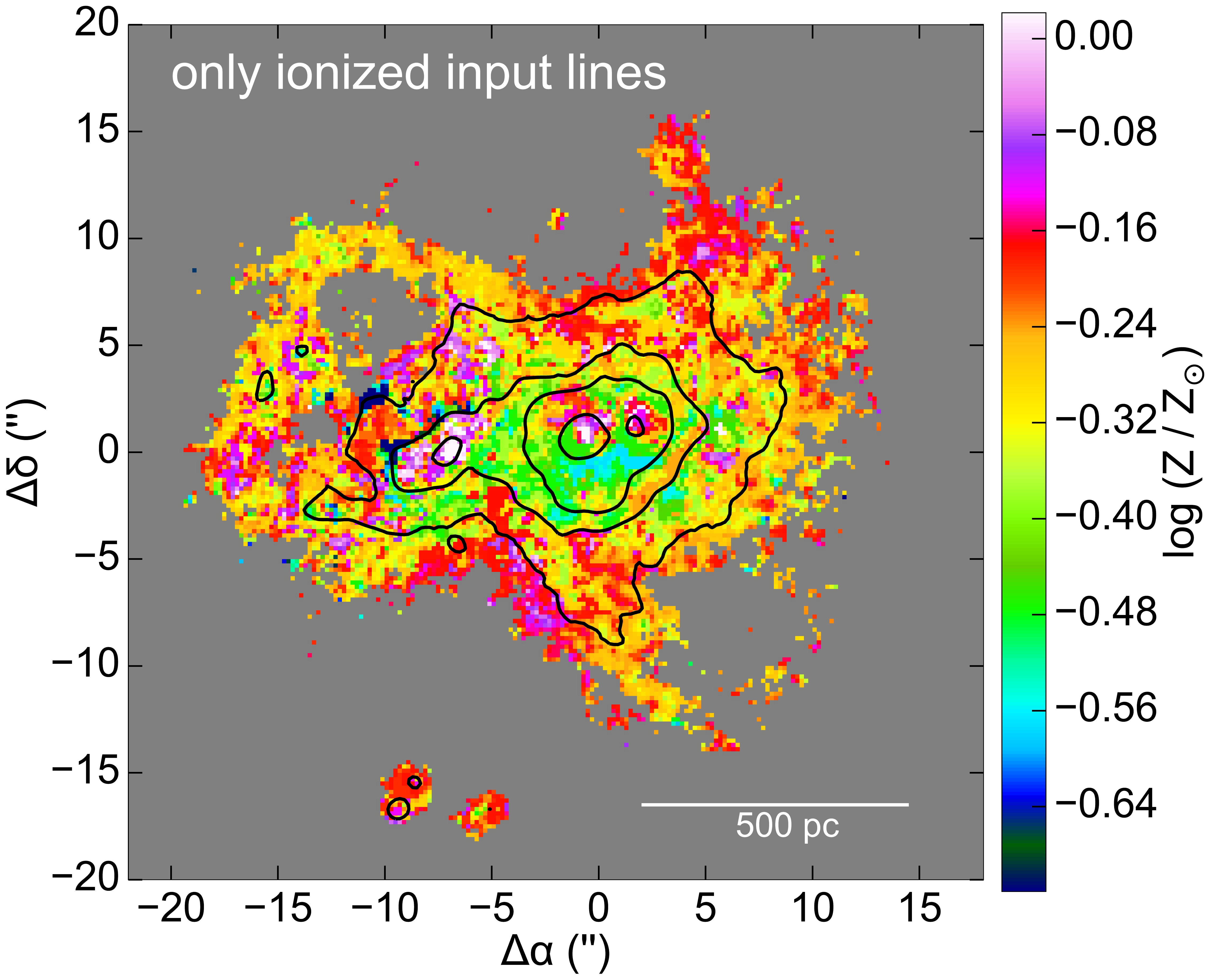}
    \caption{\textit{Left panel:} map of the ISM metallicity ($Z$) inferred for He 2-10 using all the input lines reported in Table \ref{table:lines_henize}. \textit{Right panel:} map of $Z$ inferred for He 2-10 using only the ionized input lines reported in Table \ref{table:lines_henize} (i.e., excluding the [OI] and [SII] lines). As in Fig. \ref{fig:he_halpha}, black lines show the contours of the H$\alpha$ emission line (80, 10, 4, and 1 in units of 10$^{-16}$ ergs s$^{-1}$ cm$^{-2}$).}
	\label{fig:comparison_metallicities}
\end{figure*}

\begin{figure}
	\centering
    \includegraphics[width=0.95\linewidth]{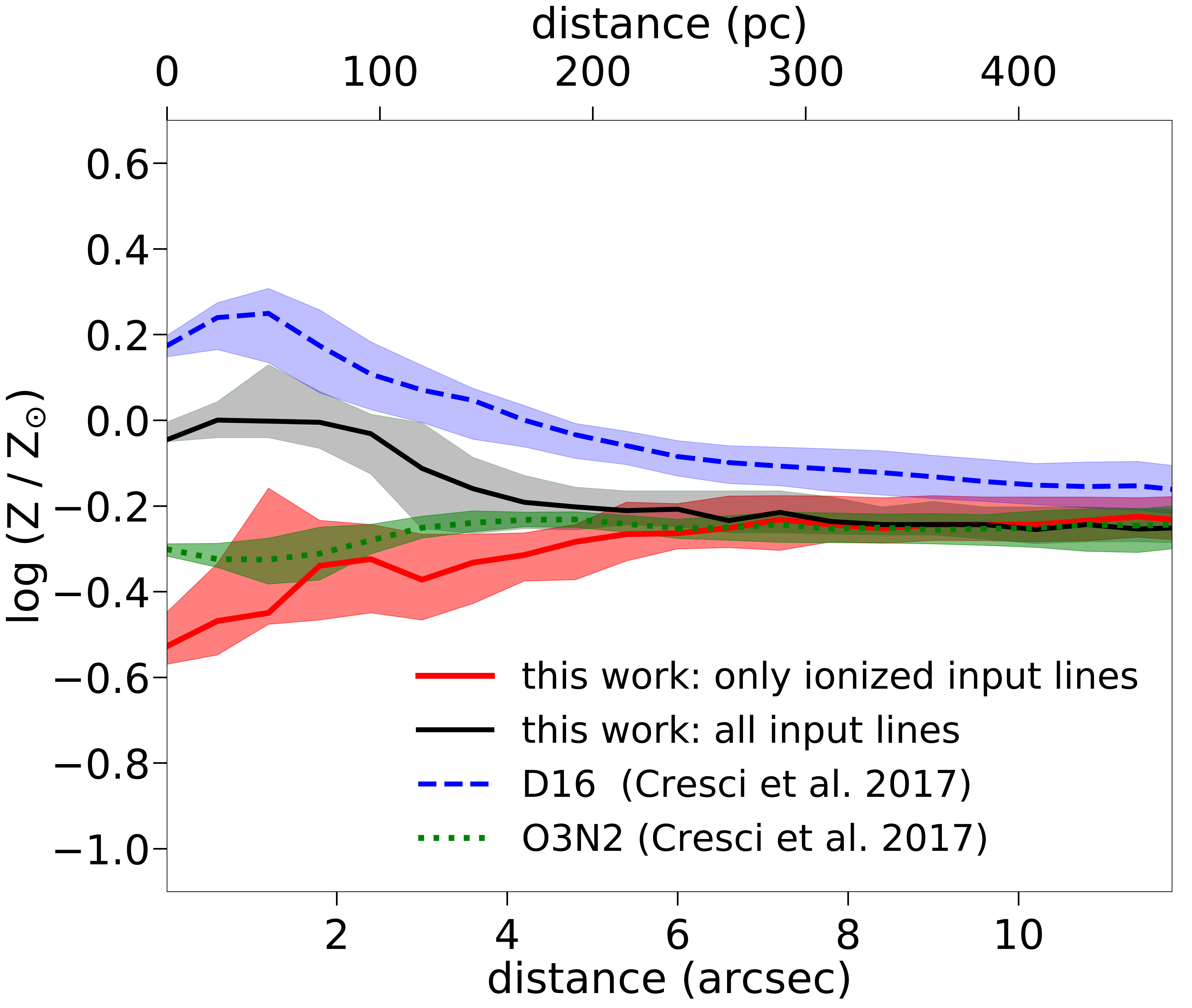}
    \caption{Median radial profiles (median of the inferred physical property values in annular bins of 0.6\arcsec) of the He 2-10 metallicity ($Z$) obtained from the maps shown in Fig. \ref{fig:comparison_metallicities}, centered on ($\Delta\alpha$,$\Delta\delta$)=(0,0). Upper and lower shaded regions denote respectively the 2nd and 3rd quartiles. We also report the radial profiles as inferred by \citet{Cresci2017}.}
	\label{fig:radial_comparison_metallicities}
\end{figure}

We also compare our results with the radial profiles of \citet{Cresci2017} who adopted the D16 \citep{Dopita2016} and the O3N2 \citep{Curti2017} calibrators (Fig. \ref{fig:he_radials_physical}). The metallicities obtained with \textlcsc{GAME} are between the results obtained through the O3N2 and D16 calibrators. In particular, in Fig. \ref{fig:hist_metals_he} we compare the distributions of the derived metallicity values inferred through the 3 different methods. The distribution obtained by \textlcsc{GAME} covers the same values inferred with the O3N2 and D16 methods, but spans a larger range (i.e., -1.0 $\lesssim$ log($Z/Z_{\odot}$) $\lesssim$ 0.3). The standard deviations of the metallicity values inferred with the three approaches are respectively: $\sigma$(\textlcsc{GAME}) $\sim$ 0.13, $\sigma$(O3N2) $\sim$ 0.06, $\sigma$(D16) $\sim$ 0.10. The difference in the results obtained between this work and \citet{Cresci2017} can be ascribed to the fact that while the O3N2 and D16 diagnostics account for ionized regions only, the \textlcsc{GAME} results are based on a library spanning several phases of the ISM \citepalias{Ucci2017,Ucci2017b}. In fact, the physical properties are inferred taking into account not only ionized gas constraints (see Table \ref{table:lines_henize}), but also lines coming from neutral (i.e., [OI]) and partially neutral regions (i.e., [SII]). In order to test this hypothesis, we have repeated the analysis removing the [OI] and [SII] lines, and we have checked whether the metallicity measurements are affected. In Fig. \ref{fig:comparison_metallicities}, we report the metallicity inferred with \textlcsc{GAME} using two different sets of input emission lines: one with all the input lines reported in Table \ref{table:lines_henize} (\textit{left panel}), and one using only purely ionized input lines (\textit{right panel}), i.e. removing the [OI] and [SII] lines. In Fig. \ref{fig:radial_comparison_metallicities} we show the median radial profiles of the maps reported in Fig. \ref{fig:comparison_metallicities}, compared with the previous estimates by \citet{Cresci2017}. It is evident that the metallicity values inferred from ionized lines only (especially in the central region, $d$ $\lesssim$ 5\arcsec), are lower than the ones obtained using the full set of input emission lines (i.e., the one that contains also neutral and partially neutral emission lines).

\subsubsection{Dust extinction}\label{sec:av_he}
In Fig. \ref{fig:He_2-10_physical}, we show the map of the visual extinction $A_V$. In the central region of the galaxy (i.e., d $\lesssim 5\arcsec$) $A_V$ ranges between 1 and 5 magnitudes while in the outer regions (i.e., d $\gtrsim 8\arcsec$), the $A_V$ median radial profile flattens to $\sim$ 0.1 mag (see Fig. \ref{fig:he_radials_physical}), although with a large dispersion.

\citet{Vacca1992} measured $E(B-V)$ = 0.54 (corresponding to $A_V$ = 1.7 mag, assuming $R_V = 3.1$) from the Balmer decrement in optical slit spectroscopy. \citet{Cresci2017} obtained from optical lines $A_V$ $\sim$ 2.3 in Region B. \citet{Cabanac2005} found $A_V$ = 1.25 ($A_V$ = 10.5) over an aperture of $2.4\arcsec$ $\times$ $15.6\arcsec$ centered on the nucleus from the H$\alpha$/Br$\gamma$ (Br$\gamma$/Br10) ratio \citep[see also][]{Kawara1989}. From the Br12/Br$\gamma$ ratio \citet{Cresci2010} obtained $A_V$ = 1.5 - 12 mag in a region within $2-3\arcsec$ from the center. The inferred extinction found in literature from IR observations is higher with respect to the $A_V$ estimated from the optical wavelength range. Since the central region of He 2-10 is a complex system characterized by the presence of absorbing material (i.e., gas and dust) mixed with emitting sources \citep{Martin-Hernandez2006}, classic diagnostics based on optical lines do not allow $A_V$ measurements into dust-embedded star forming regions, as IR data do.

Fig. \ref{fig:He_2-10_physical} shows that $A_V$ mildly correlates with $n$, as quantified in Fig. \ref{fig:correlation} with a correlation coefficient $C$ = 0.64. In the star-forming Regions A and B, we obtain $A_V \gtrsim 2-3$ mag. The visual extinction reaches a peak in the SW part centered at ($\Delta\alpha$,$\Delta\delta$) = (-4,-3), at $d \sim 4\arcsec$ from the center, where there is a region highly correlated with high column densities ($N_H \gtrsim 10^{22}$ cm$^{-2}$). It is a region with $\Sigma_{gas} > 10^8$ M$_{\odot}$ kpc$^{-2}$ and where most of the gas mass is located (e.g., Fig. \ref{fig:He_2-10_inferred}), and already associated in \citet{Cresci2017} with an accreting cloud detected in CO observations.

\subsubsection{Interstellar radiation field}
We characterize the radiation field of He 2-10 in terms of the FUV flux $G$ and the ionization parameter $U$, whose maps are shown in the bottom right and left panel of Fig \ref{fig:He_2-10_physical}, respectively. From the median radial profile, the FUV flux in Habing units\footnote{$G_0$ = 1.6 $\times$ 10$^{-3}$ erg s$^{-1}$ cm$^{-2}$ \citep{Habing1968}.} is $G/G_0\sim$ 10$^3$ in the center of the galaxy (in correspondence of the highly star forming regions A and B), while at larger distances (i.e., $d \gtrsim 4\arcsec$) it flattens around $G/G_0$ $\sim$ 10$^{1.5}$. For what concerns the ionization parameter, $U$ $\sim$ 10$^{-3}$ in the central regions ($d \lesssim 4\arcsec$), while at larger distances $U$ $\sim$ 10$^{-1.5}$. Our results are consistent with the ones reported by \citet{Cresci2017} in the inner region of the galaxy while, at $d \gtrsim 2\arcsec$, \textlcsc{GAME} infers $U$ values up to two orders of magnitude larger\footnote{In the $U$ map, we inferred values of the ionization parameter up to $\sim 10^{1.5}$, about 0.5-1.0 dex higher than the most extreme values found in the literature. However the spaxels with these extreme values (i.e., log($U$) > 1) are found to be $\lesssim$ 5\% of the total number and highly correlated with low density regions. For further discussion see also Sec. \ref{sec:radiation_izw18}.}. In what follows, we analyze the physical origin of our findings.

\begin{figure}
	\centering
    \includegraphics[width=1.0\linewidth]{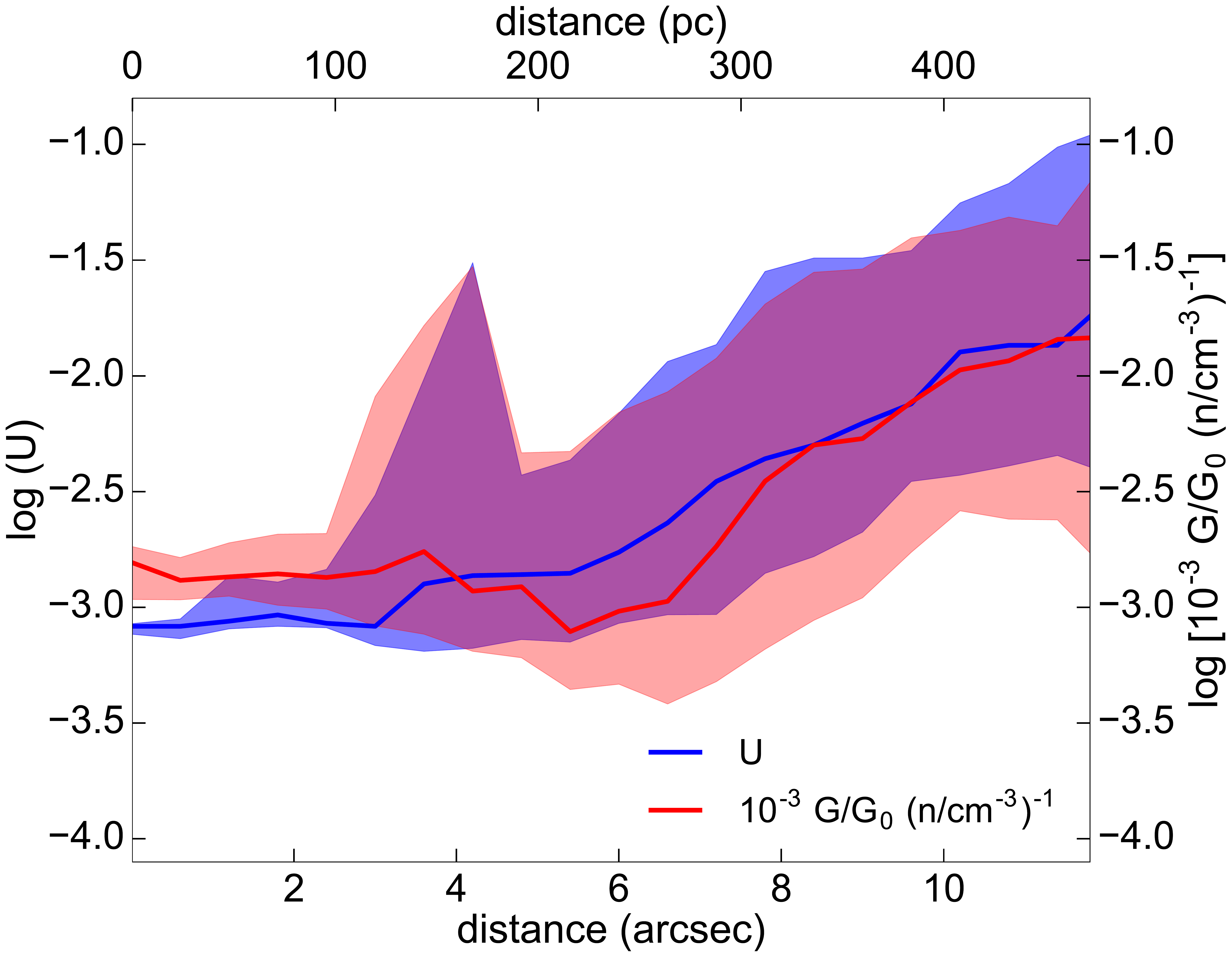}
    \caption{Median radial profile of the He 2-10 ratio between the scaled FUV flux and the density obtained from the maps shown in Fig. \ref{fig:He_2-10_physical} (see text for details). Upper and lower shaded regions denote respectively the 2nd and 3rd quartiles. For comparison purposes, we also show the radial profile for the ionization parameter $U$ that follows the same trend.}
	\label{fig:radial_govern}
\end{figure}

The $U$ profile depends on $G$ and $n$ according to the following relation\footnote{The ratio between the number of FUV photons ($\propto G$) and the number of ionizing photons ($\propto U n$), depends on the shape of the stellar spectra that constitutes the \textlcsc{GAME} library that, in its turn, depends on the stellar metallicity. Given the almost flat metallicity radial profile, this ratio is not expected to vary along the disk of the galaxy.}: $G/G_0 \sim 10^3 U n$. In Fig. \ref{fig:radial_govern}, we compare the radial median profile of $U$ (blue line/shaded region) with the one of $10^{-3} G/G_0~n^{-1}$ (red line/shaded region), where $U$, $G$, and $n$ are the values inferred by \textlcsc{GAME} independently. This figure shows that the expected physical trend is correctly recovered by the \textlcsc{ML} algorithm. At small distances ($d < 4\arcsec$), large values of the FUV flux $G/G_0\sim 10^3$ are compensated by much larger values of the gas density ($n \sim 10^3$ cm$^{-3}$), thus providing a small value of $U$ ($\sim 10^{-3}$). Vice versa, at larger distances, the strong decrease of the gas density (log($n$/cm$^{-3}$) $\sim$ 0, because of the strong anti-correlation with $U$, see Fig. \ref{fig:correlation}) and the flattening of the FUV flux at the value $G/G_0\sim 10^{1.5}$, combine into a large U value ($\gtrsim 10^{-2}$).

\begin{figure*}
	\centering
    \includegraphics[width=0.49\textwidth]{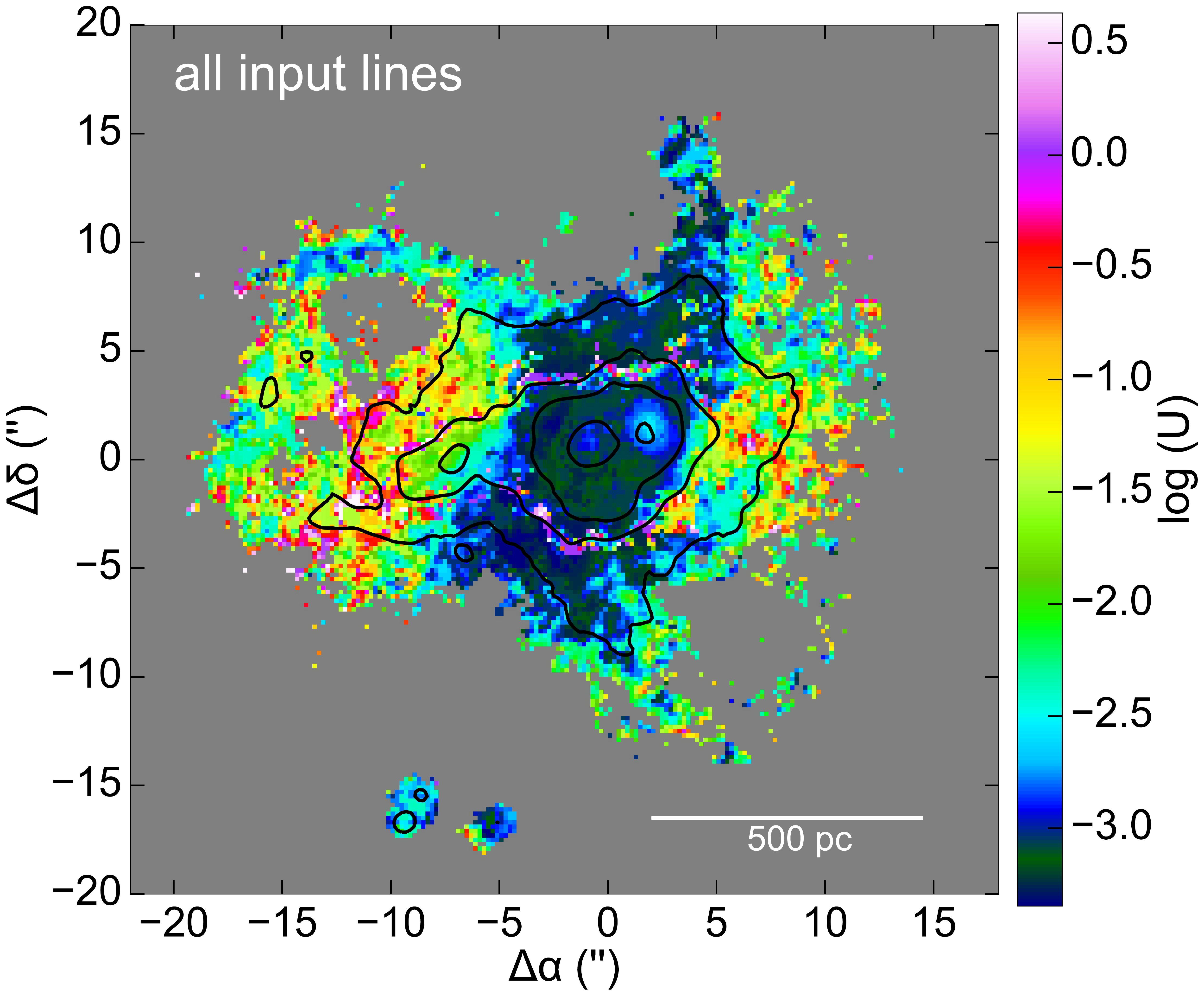}
    \includegraphics[width=0.49\textwidth]{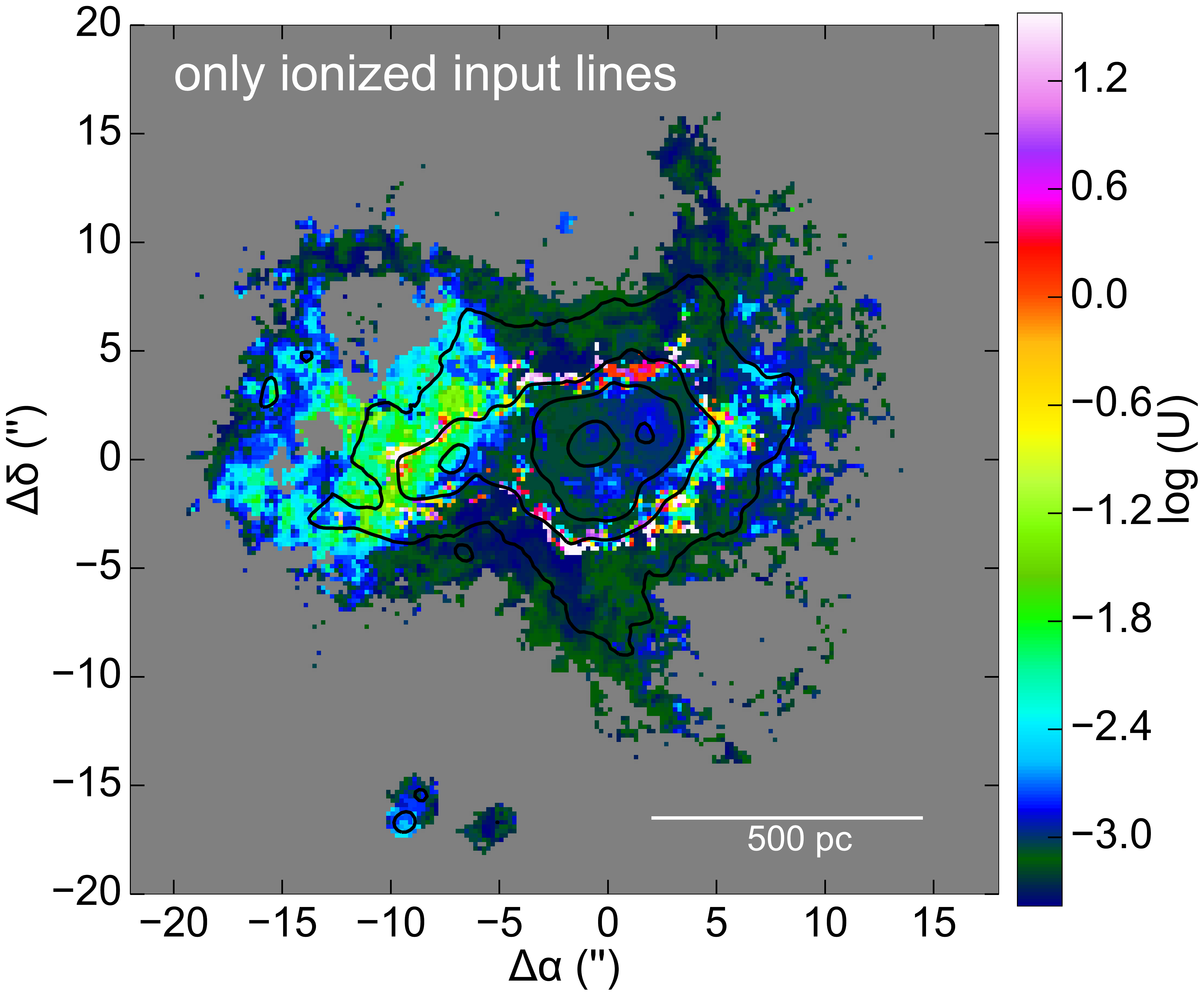}
    \caption{\textit{Left panel:} map of the ISM ionization parameter ($U$) inferred for He 2-10 using all the input lines reported in Table \ref{table:lines_henize}. \textit{Right panel:} map of $U$ inferred for He 2-10 using only the ionized input lines reported in Table \ref{table:lines_henize} (i.e., excluding the [OI] and [SII] lines). As in Fig. \ref{fig:he_halpha}, black lines show the contours of the H$\alpha$ emission line (80, 10, 4, and 1 in units of 10$^{-16}$ ergs s$^{-1}$ cm$^{-2}$).}
	\label{fig:comparison_ionization}
\end{figure*}

\begin{figure}
	\centering
    \includegraphics[width=0.95\linewidth]{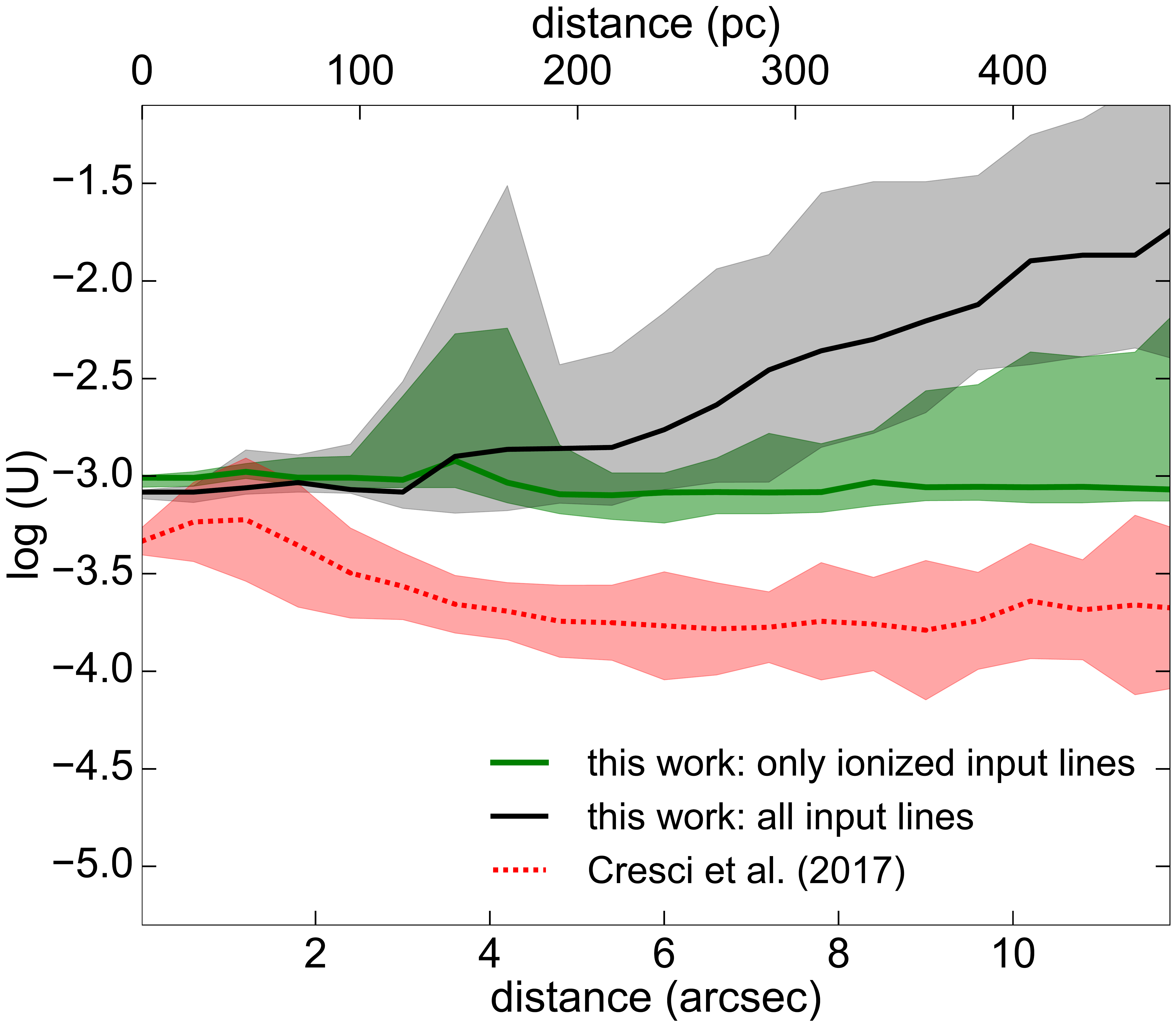}
    \caption{Median radial profiles (median of the inferred physical property values in annular bins of 0.6\arcsec) of the He 2-10 ionization parameter ($U$) obtained from the maps shown in Fig. \ref{fig:comparison_ionization}, centered on ($\Delta\alpha$,$\Delta\delta$)=(0,0). Upper and lower shaded regions denote respectively the 2nd and 3rd quartiles. We also report the radial profiles as inferred by \citet{Cresci2017}.}
	\label{fig:radial_comparison_ionization}
\end{figure}

\citet{Cresci2017} used the line ratio [SIII]/[SII] to derive $U$, using the calibrations of \citet{Kewley2002}. Our results are consistent with their findings only within few arcsec ($\sim 2"$) from the center of the galaxy, while our $U$ values are larger in the outer regions. However, note that the $U$ determination has large typical uncertainties. Thus, the $U$ determination cannot be considered reliable in the outer region. The poor determination of $U$ with respect to the other variables from \textlcsc{GAME} has been already discussed in \citetalias[][]{Ucci2017}. However, the discrepancy between our findings and the results from \citet{Cresci2017} (see Fig. \ref{fig:he_radials_physical}), following what we demonstrated in Sec. \ref{sec:he_Z}, could be ascribed to the fact that we used the full set of input emission lines available with no assumptions about the underlying emitting phase (i.e., without assuming emitting HII regions only). As evident from Fig. \ref{fig:radial_comparison_ionization} (see also Fig. \ref{fig:comparison_ionization}), the difference between the radial profiles assuming only ionized lines, reduces to $\sim$ 0.5 dex (i.e, a factor of three). If we assume instead the full set of available lines, at large radii ($d \gtrsim 9$\arcsec), the difference is $\gtrsim$ 1 dex.

\subsubsection{Gas mass and SFR surface density}\label{sec:sfr_he}
We computed the gas mass surface density $\Sigma_{gas}$ (see eq. \ref{eq:sigma_mass}) by considering the spaxel area $A_{spax}$ = 64 pc$^2$.

\citet{Kobulnicky1995} estimated a molecular (atomic) hydrogen mass of 1.6$\pm$0.1 $\times$ 10$^{8}$ (1.9$\pm$0.1 $\times$ 10$^{8}$) M$_{\odot}$ from CO(1-0) and CO(2-1) detections. More recent observations based on CO(3-2) emission \citep{Vanzi2009} provide a molecular mass of 1.4 $\times$ 10$^{7}$ M$_{\odot}$. In Fig. \ref{fig:He_2-10_inferred}, the solid and dashed lines correspond to the sizes of the regions covered by the CO observations of \citet{Kobulnicky1995} (a radius of $\sim 23\arcsec$) and \citet{Vanzi2009} (a radius of $\sim 10\arcsec$), respectively. By summing up the contribution from spaxels within the outer and inner regions, we obtain a total gas mass of 1.4 $\times$ 10$^{8}$ M$_{\odot}$ and 1.9 $\times$ 10$^{7}$ M$_{\odot}$, respectively, in good agreement with previous estimates.

For what concerns the SFR, we adopt two different approaches. Firstly, by using eq. \ref{eq:sfr}, we compute the SFR surface density shown in the right panel of Fig. \ref{fig:He_2-10_inferred}. By summing up the contribution from all spaxels we obtained SFR = 1.2 M$_{\odot}$ yr$^{-1}$. Secondly, we infer the SFR from the H$\alpha$ map and we correct it for the attenuation through the Calzetti law. For a stellar population with constant SFR over the past 100 Myr, forming stars with an IMF consisting of two power laws (slope $-1.3$ in the range 0.1 - 0.5 M$_{\odot}$ and slope $-2.3$ in the range 0.5 - 120 M$_{\odot}$), \citet{Calzetti2008} reports:

\begin{equation}
\text{SFR(M}_{\odot} \text{yr}^{-1}) = 5.3 \times 10^{-42} \text{L(H}\alpha) \text{(erg/s)},
\label{eq:calzetti_sfr}
\end{equation}
where L(H$\alpha$) is the intrinsic luminosity corrected for absorption and interstellar dust attenuation\footnote{Variations of SFR $\pm 20\%$ for younger/older ages and different metallicities are also possible.}. Given the fluxes for the H$\alpha$ line and the map of the extinction $A_V$ (see Fig. \ref{fig:He_2-10_physical}), the un-corrected SFR is $\sim$ 0.05 M$_{\odot}$ yr$^{-1}$. We then corrected the H$\alpha$ fluxes using the \citet{Cardelli1989} extinction curve (assuming $R_V = 3.1$). By using eq. \ref{eq:calzetti_sfr} with corrected H$\alpha$ luminosities, and summing up the contribution from the spaxels, we obtained a SFR of $\sim$ 1.44 M$_{\odot}$ yr$^{-1}$. This estimate is consistent with the one by \citet{Reines2011} that, by using the H$\alpha$ \citep{Mendez1999} and 24 $\mu$m \citep{Engelbracht2005} fluxes, corrected for the Calzetti \citep{Calzetti2007} attenuation, found 1.9 M$_{\odot}$ yr$^{-1}$. By using the SMC \citep{Prevot1984,Bouchet1985,Gordon2003} extinction curve, we can provide another estimate of the SFR. With the same approach described before, by using our extinction map to correct the H$\alpha$ flux, we obtain: SFR$_{SMC}$ $\sim$ 1.26 M$_{\odot}$ yr$^{-1}$. This result underlines the strong dependence of the inferred SFR on the assumed extinction curve.

\subsection{Physical correlations}\label{sec:maps_correlation}

\begin{figure}
	\centering
    \includegraphics[width=0.95\linewidth]{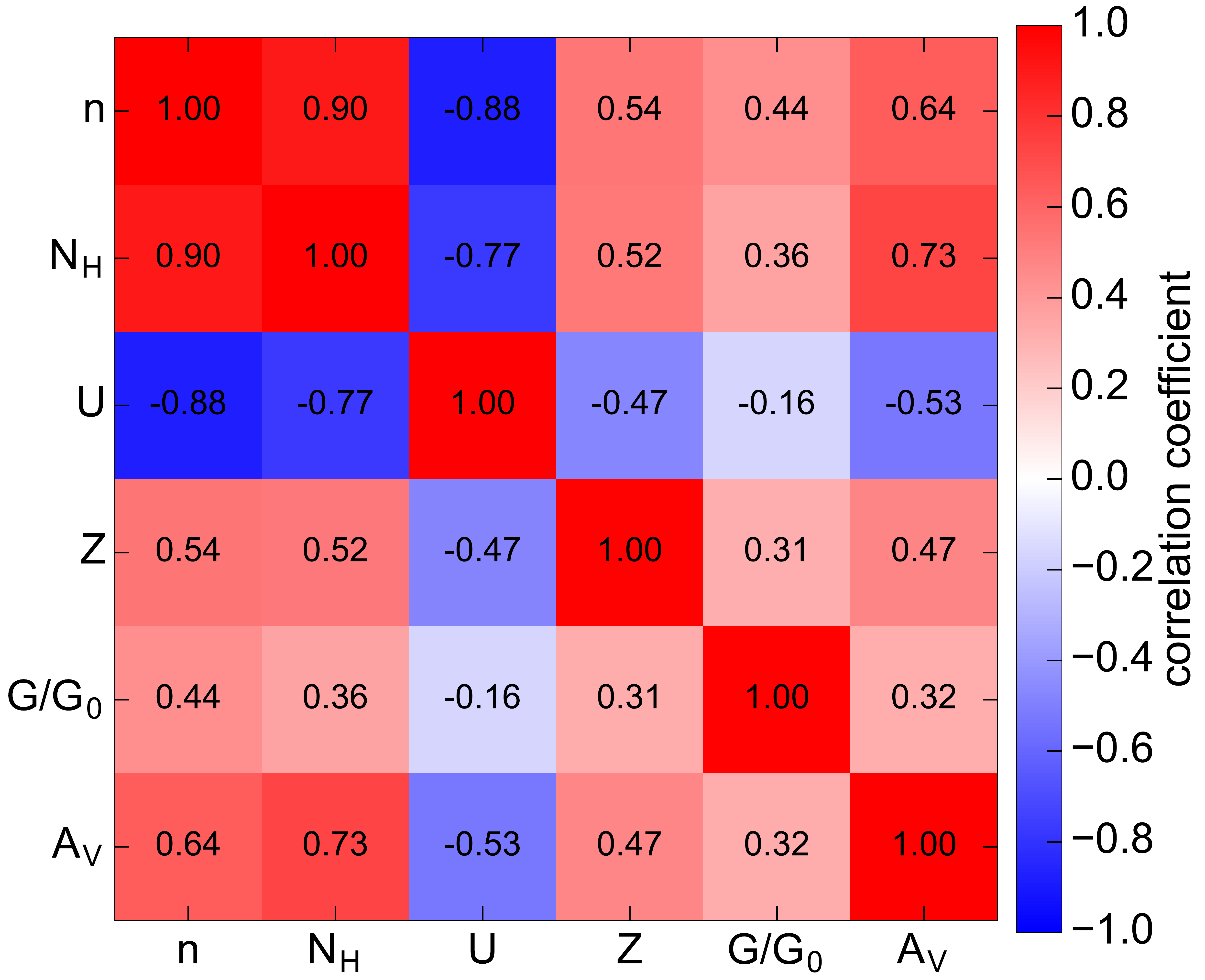}
    \caption{Pearson Correlation coefficient $C$ between the value inferred for the physical properties reported in Fig. \ref{fig:He_2-10_physical} for He 2-10. There is strong correlation (anti-correlation) for 0.5 < $C$ < 1.0 (-1.0 < $CC$ < -0.5).}
	\label{fig:correlation}
\end{figure}

We report in Fig. \ref{fig:correlation} the Pearson correlation coefficients $C$ between the various physical properties inferred by \textlcsc{GAME}. Given that the spatial resolution of emission line fluxes is limited by the seeing \citep[$\sim$ 0.68\arcsec,][]{Cresci2017}, in order to have correlation coefficients determinations from independent spaxels, $C$ are computed from 4$\times$4 re-binned maps for the physical properties.

Although determined independently, we find a clear correlation ($C$ $\sim$ 0.90) between the density ($n$) and the column density ($N_H$). Moreover, the ionization parameter ($U$) anti-correlates both with the density ($C \sim$ -0.88) and the column density ($C \sim$ -0.77), due to the dependence of $U$ on the inverse of $n$. Another expected correlation, such as the one between $A_V$ and $N_H$ is recovered, and we also find $C \sim 0.90$ between $A_V$ and the product $N_H Z$ (see Appendix \ref{sec:av_nh}).

\section{IZw18}\label{sec:izw18}

\begin{figure}
	\centering
	\includegraphics[width=1.0\linewidth]{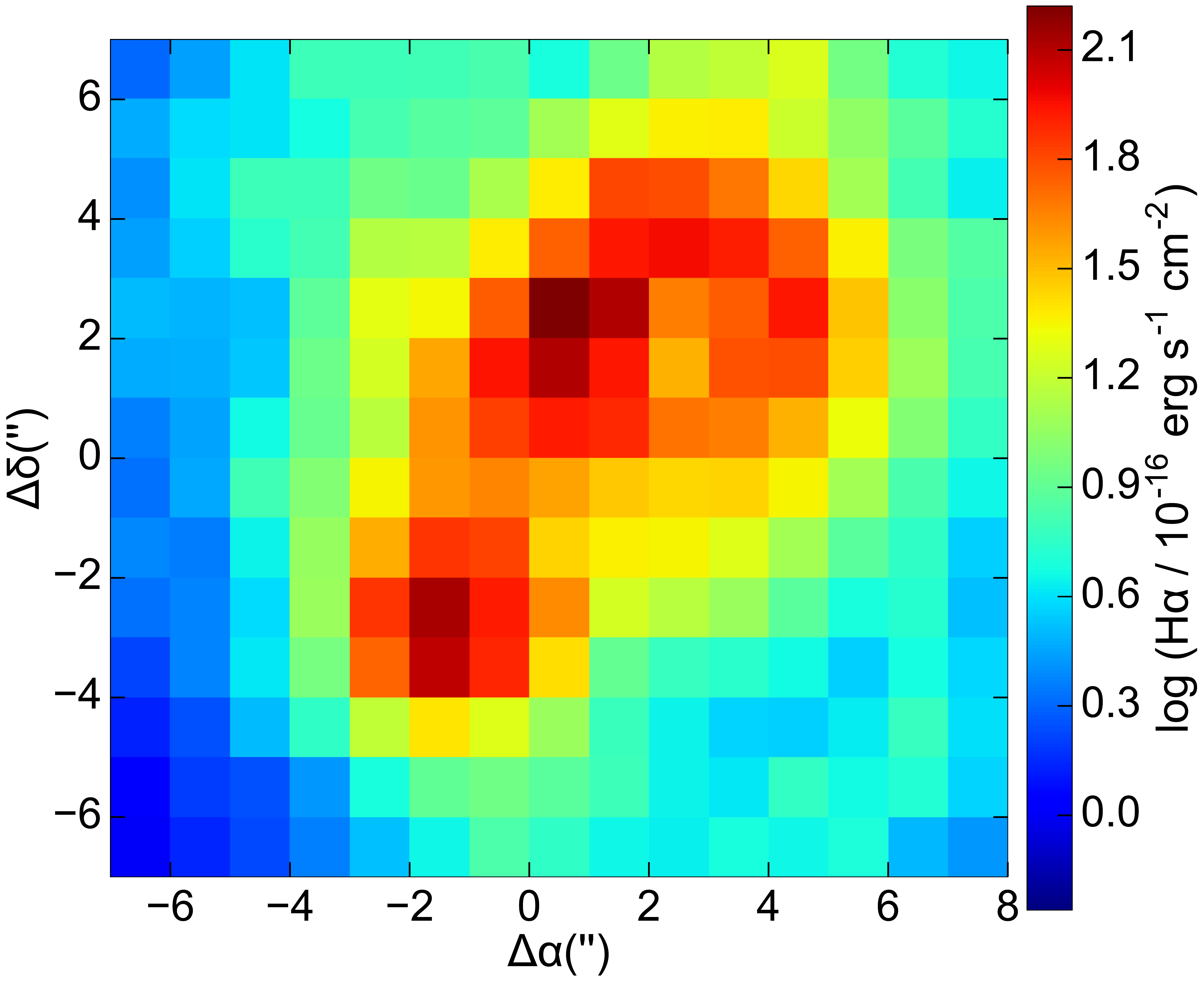}
	\caption{H$\alpha$ image of the galaxy IZw18 \citep{Kehrig2016} with a scale of 88 pc/\arcsec.}
	\label{fig:izw18_image}
\end{figure}

Since its discovery \citep{Zwicky1966}, the HII galaxy I Zwicky 18 (hereafter IZw18, also known as Mrk 116), is one of the most intriguing blue dwarf galaxies in the local Universe. \citet{Zwicky1966} described it as a system made by two compact galaxies, recognized now as two compact star forming regions associated with HII regions usually indicated as the NW and SE components \citep{Davidson1989}. It is located at a distance of 18.2 Mpc \citep{Aloisi2007} with a corresponding angular scale of 88 pc/\arcsec. It is considered an analogue of high-redshift galaxies \citep[and references therein]{Lebouteiller2013} and after the first studies \citep{Sargent1970,Searle1972,Searle1973}, it continues to keep attraction thanks to its extremely low metallicity: Z/Z$_{\odot} \sim$ 0.026 \citep{Kunth1986,Pagel1992,Vilchez1998,Izotov1999,Izotov1999b,Lebouteiller2013,Kehrig2016}. Since the metallicity is homogeneously distributed \citep{Kehrig2016}, IZw18 should host an underlying old population of intermediate/low-mass stars distributed over the whole area of the galaxy \citep{Vilchez1998}, although older/fainter ones are more easily detectable only at the system's periphery \citep{Contreras2011}.

\begin{figure*}
    \raggedleft
    \includegraphics[width=0.48\textwidth]{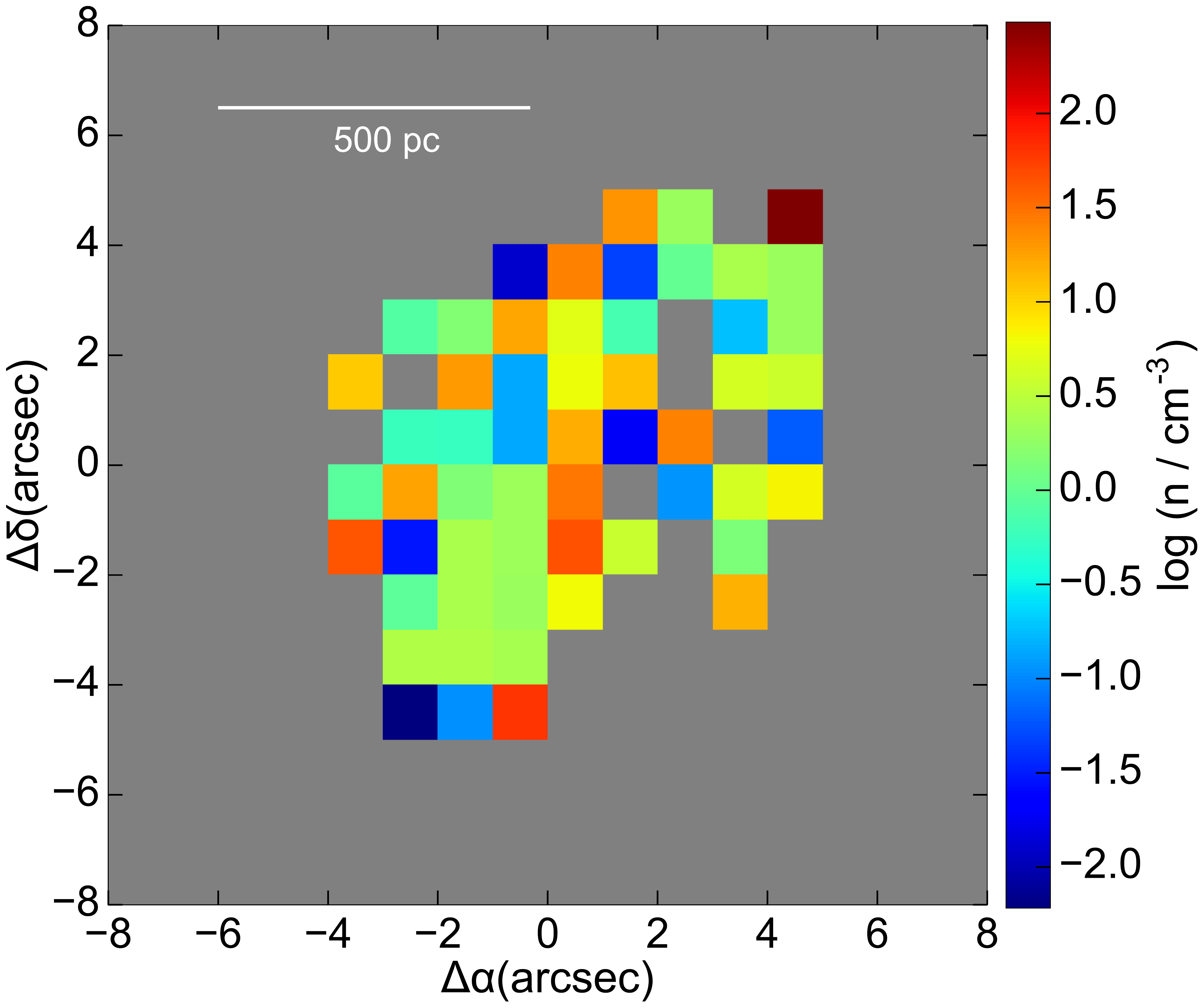}
    \raggedleft
    \includegraphics[width=0.485\textwidth]{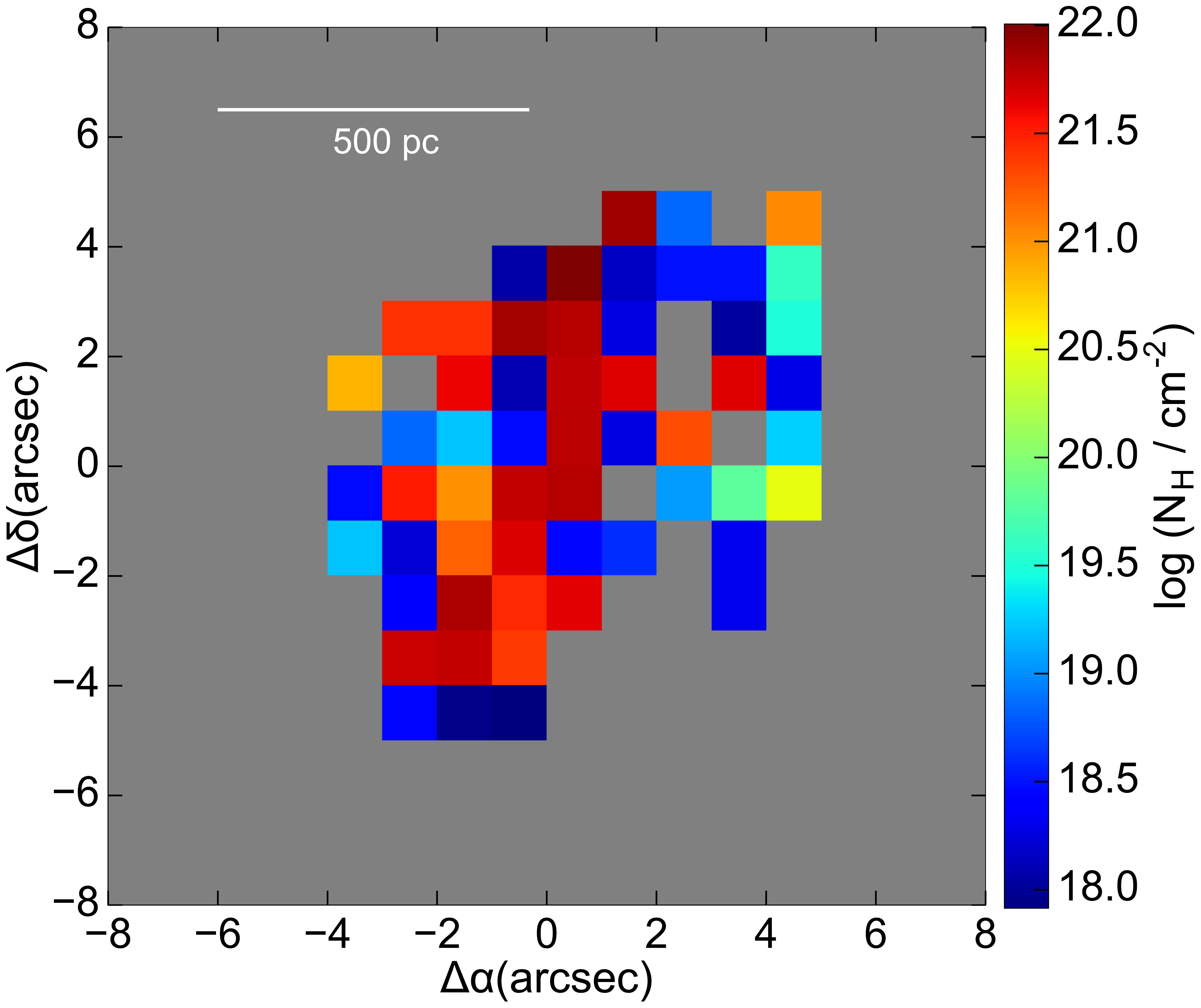}
    \raggedleft
    \includegraphics[width=0.489\textwidth]{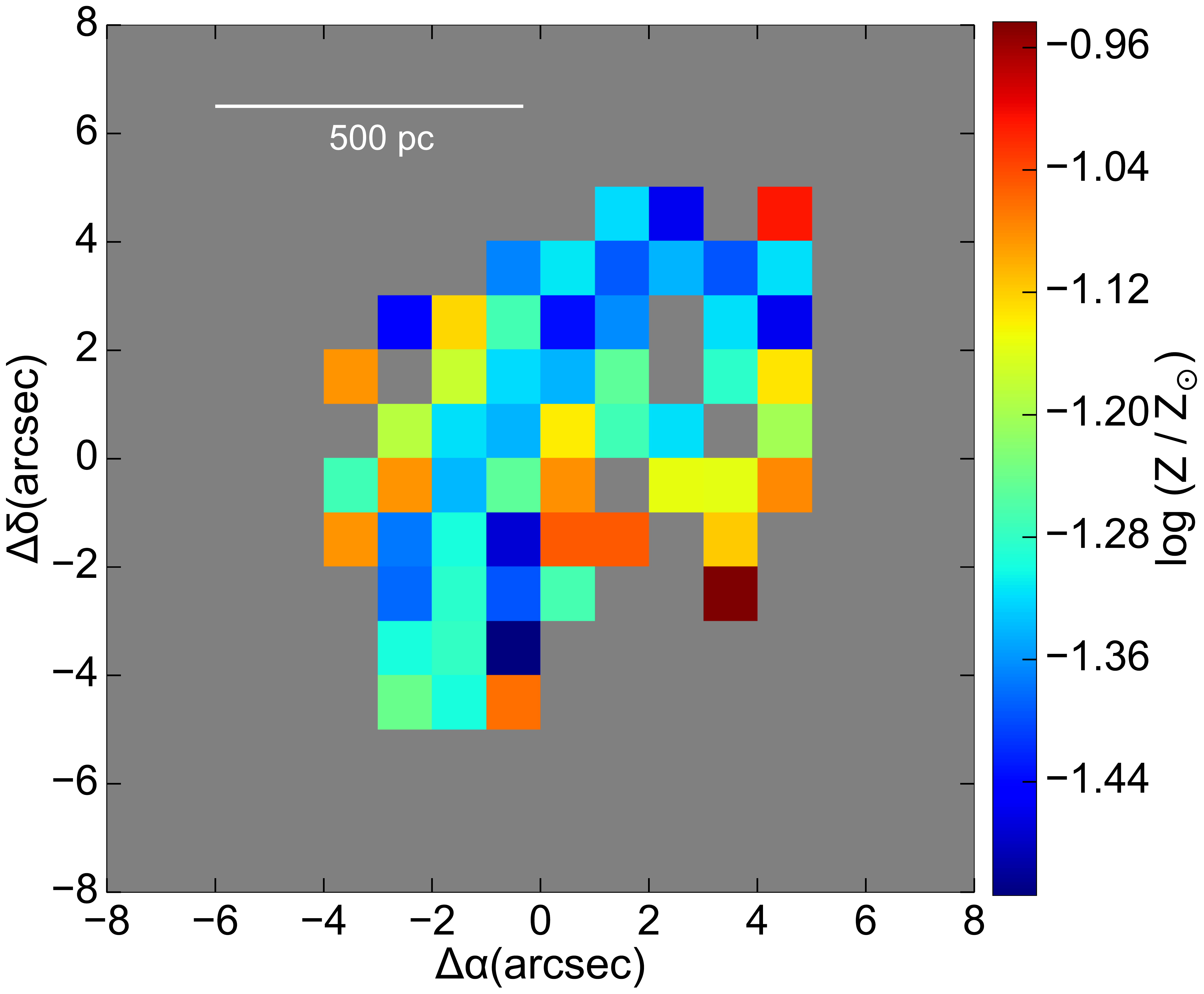}
    \raggedleft
	\includegraphics[width=0.485\textwidth]{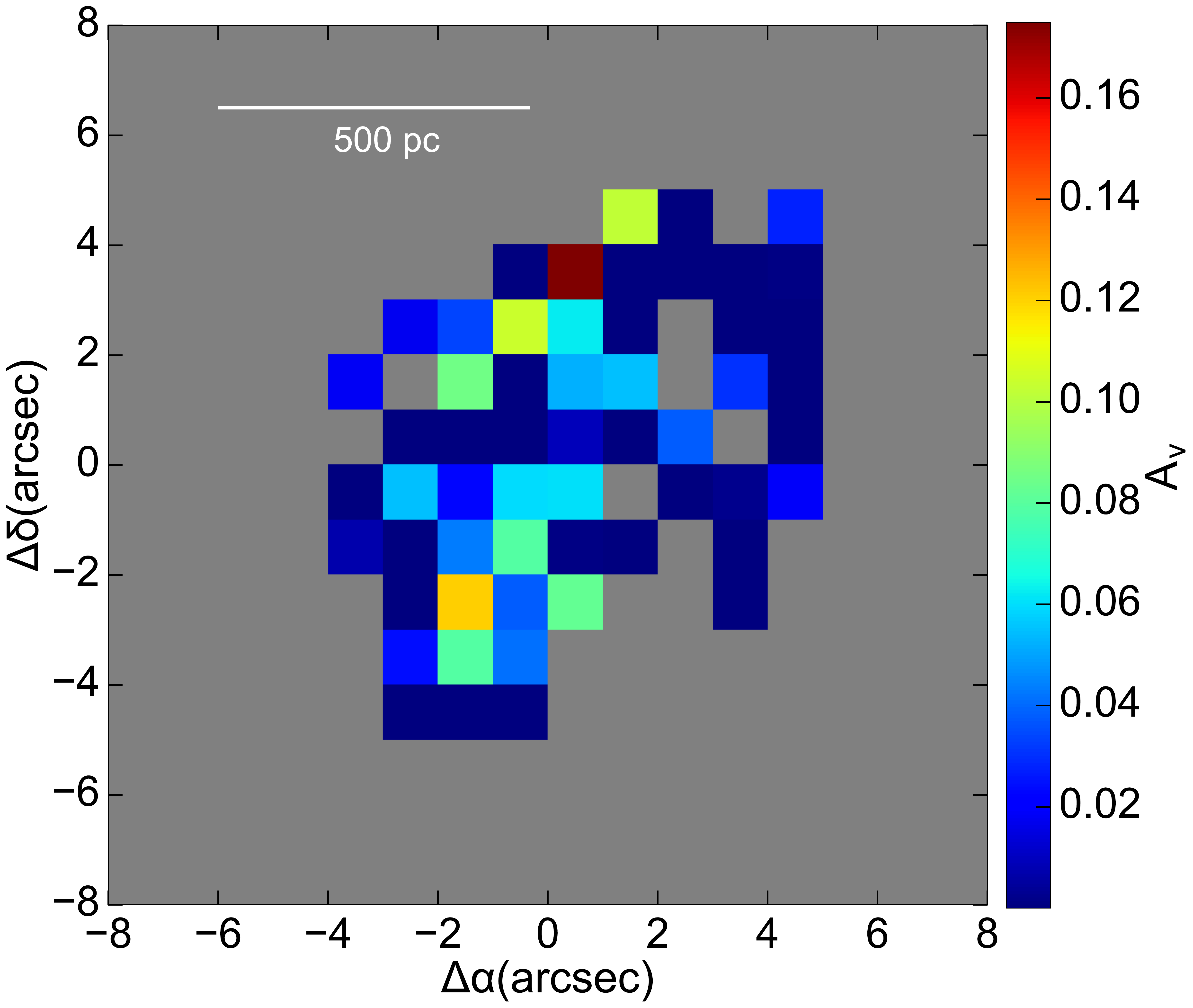}
    \raggedright
    \includegraphics[width=0.485\textwidth]{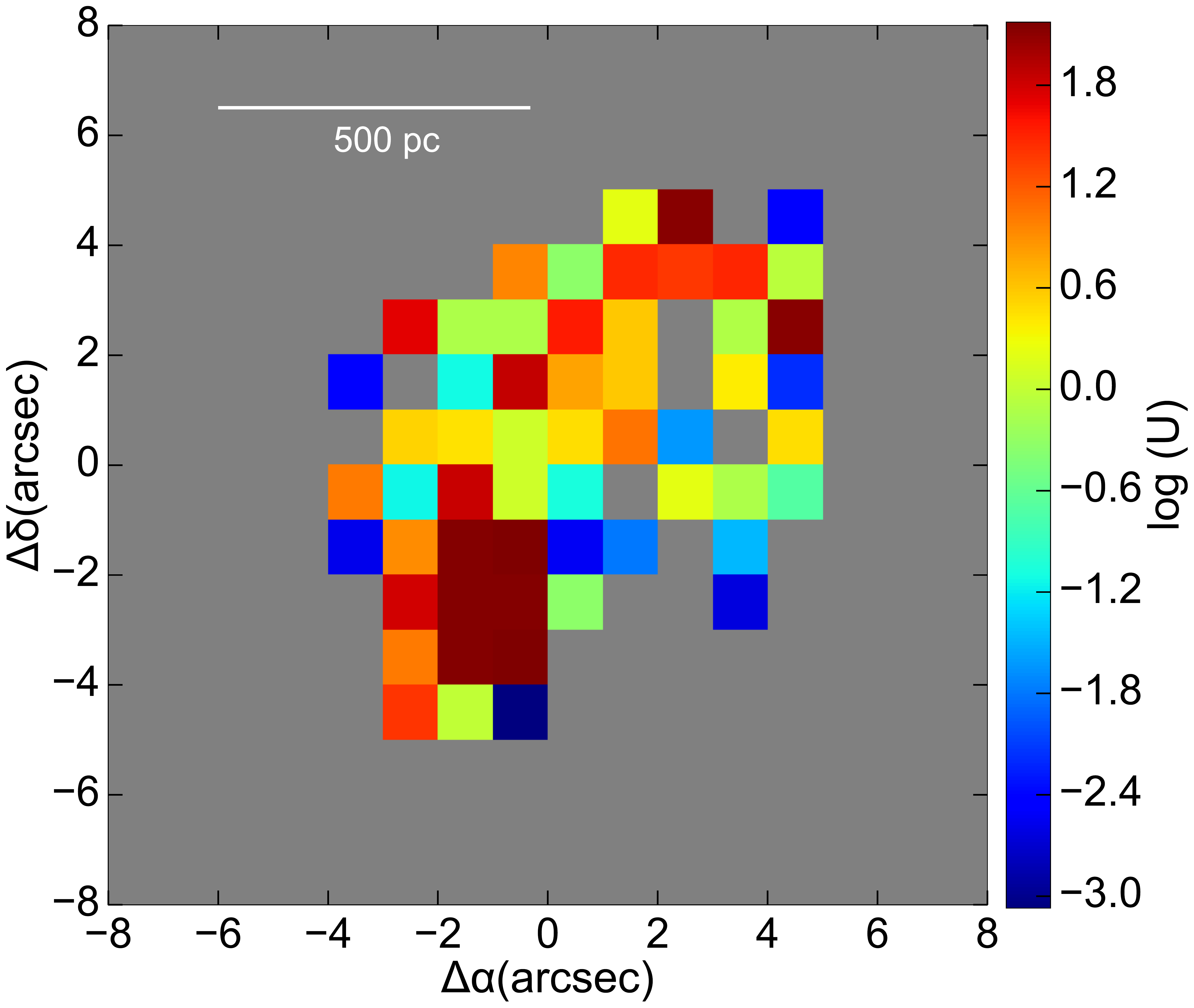}
    \includegraphics[width=0.47\textwidth]{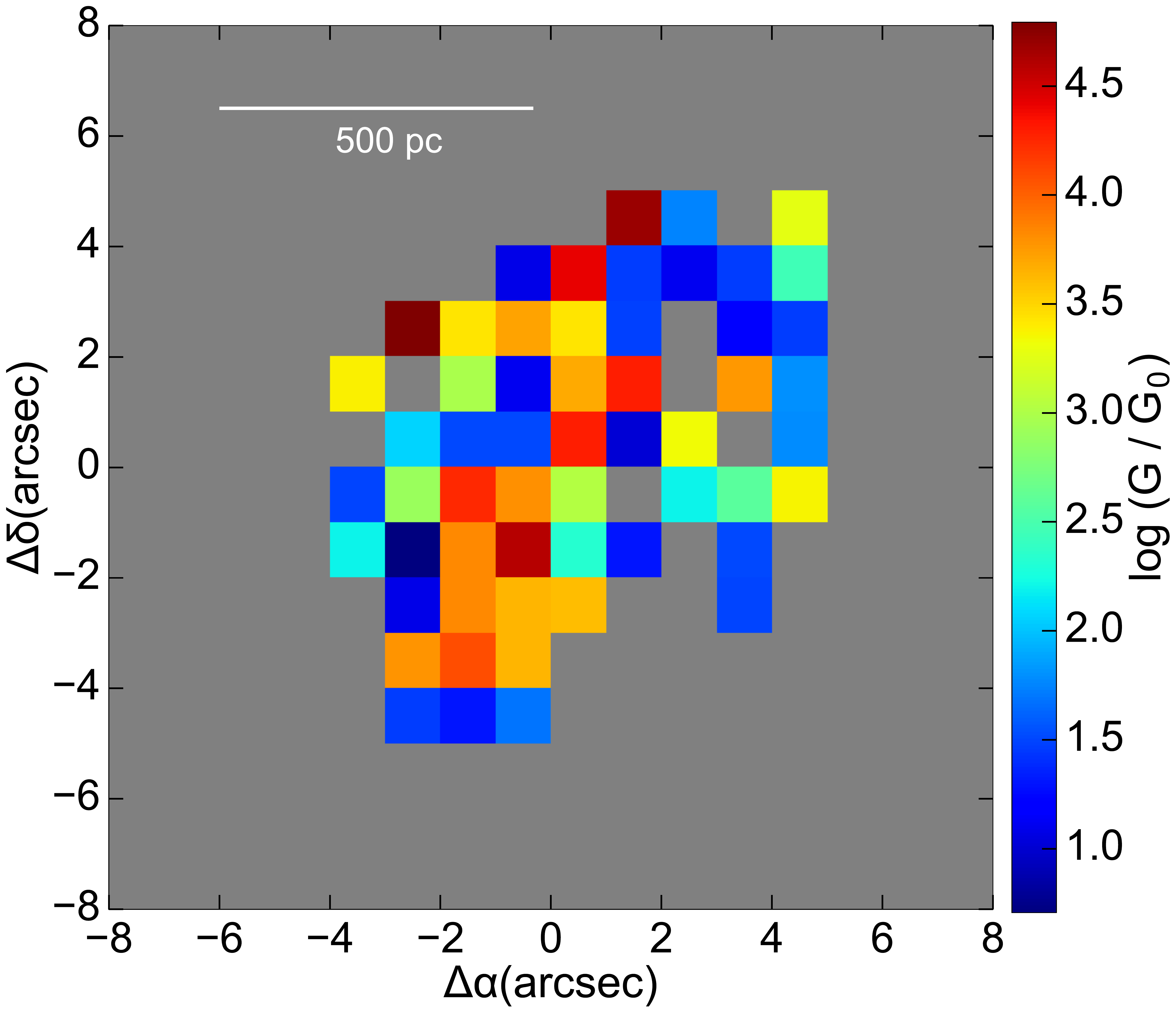}
    \caption{Maps of the ISM physical properties inferred for the galaxy IZw18 using the code \textlcsc{GAME} \citep{Ucci2017,Ucci2017b}: density ($n$), column density ($N_H$), metallicity ($Z$), visual extinction ($A_V$), ionization parameter ($U$) and the FUV flux in Habing units ($G/G_{0}$).}
	\label{fig:izw18_physical}
\end{figure*}

\begin{figure*}
	\centering
    \includegraphics[width=0.475\textwidth]{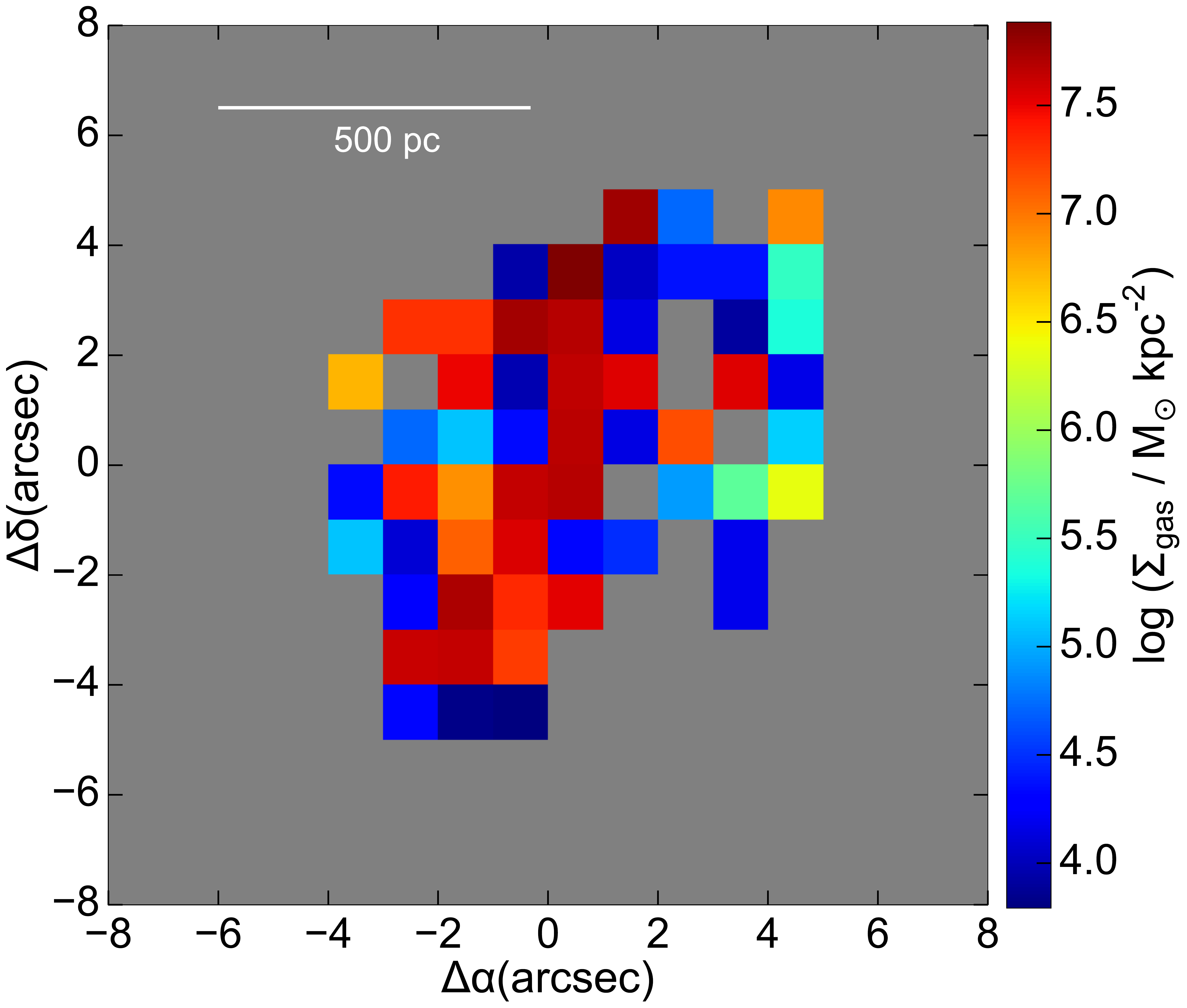}
    \includegraphics[width=0.49\textwidth]{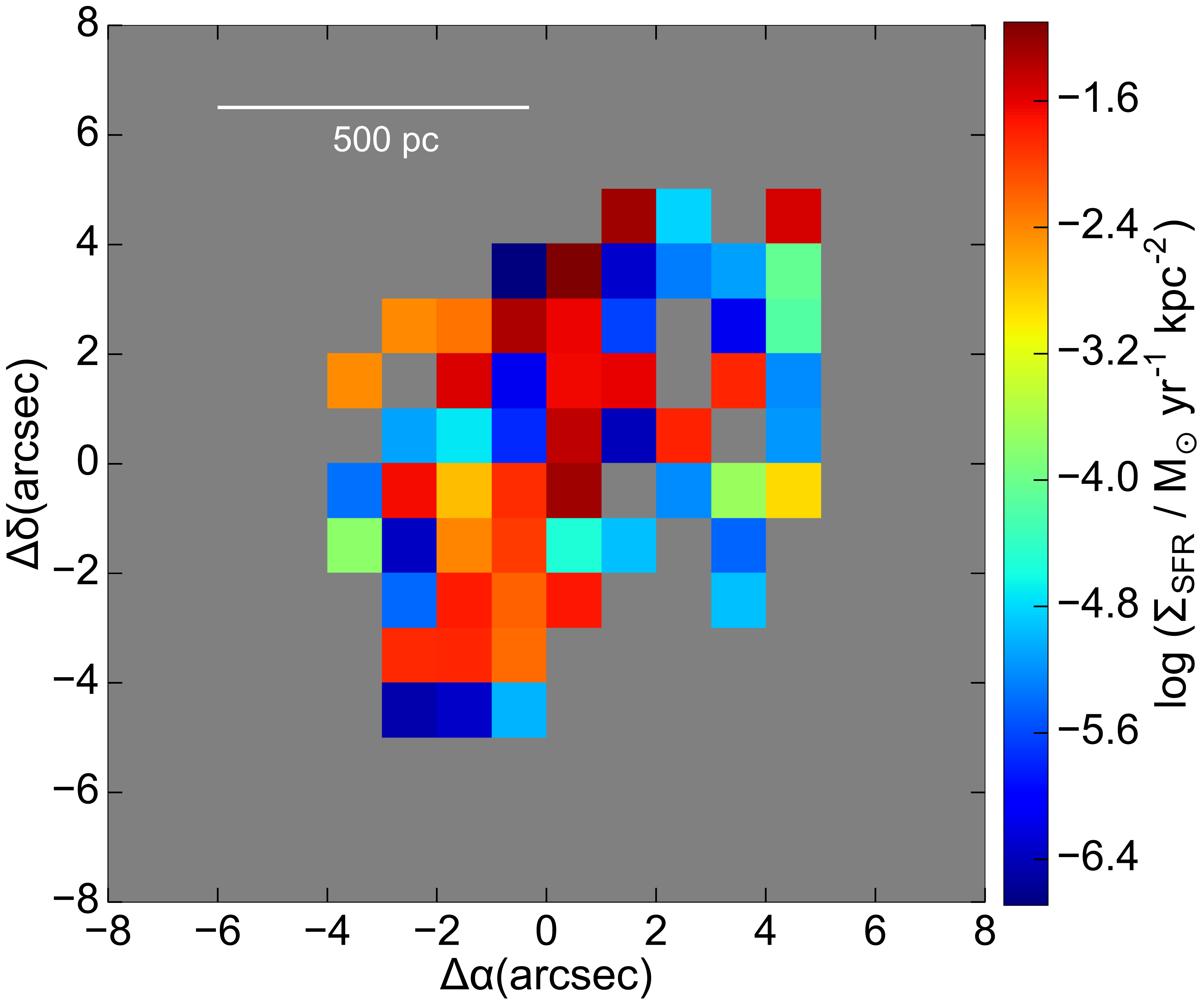}
    \caption{Gas mass surface density ($\Sigma_{gas}$, \textit{left panel}) and Star Formation surface density ($\Sigma_{SFR}$, \textit{right panel}) for the galaxy IZw18.}
	\label{fig:izw18_inferred}
\end{figure*}

\begin{table}
	\caption{Emission lines and wavelengths used to analyze the galaxy IZw18. The third column reports the fraction of spaxels with a detected and fitted line (no SNR cuts applied).}
	\centering
	\vspace{2mm}
	\begin{tabular}{c c c}
		\hline\hline
		line & wavelength [\AA] & fraction of spaxels\\
		\hline
		{[O III]} & 3727 & 29 \%\\
        {[O III]} & 4363 & 17 \%\\
        H$\beta$ & 4861 & 95 \%\\
        {[O III]} & 5007 & 98 \%\\
        {[O I]} & 6300 & 6 \%\\
        H$\alpha$ & 6563 & 99 \%\\
        {[N II]} & 6584 & 13 \%\\
        {[S II]} & 6717 & 25 \%\\
        {[S II]} & 6731 & 22 \%\\
        \hline
		\hline
	\end{tabular}
	\label{table:lines_izw18}
\end{table}

\subsection{ISM physical properties in IZw18}\label{sec:phys_izw18}
In this section, we present the ISM physical properties in IZw18 inferred by applying \textlcsc{GAME} to IFU data taken from \citet{Kehrig2016}. IZw18 data were collected by the Potsdam Multi-Aperture Spectrophotometer \citep[PMAS;][]{Roth2005,Roth2010} on the 3.5 m telescope at the Calar Alto Observatory. For more details about the observations and the data reduction we refer the reader to \citet{Kehrig2013,Kehrig2015,Kehrig2016}. For IZw18 we have a total of 16 $\times$ 16 = 256 spectra. By using the input emission lines reported in Table \ref{table:lines_izw18}, we obtained the maps of the physical properties shown in Fig. \ref{fig:izw18_physical} and \ref{fig:izw18_inferred}. $A_V$, $\Sigma_{SFR}$ and $\Sigma_{gas}$ are computed as in Sec. \ref{sec:av_he} and \ref{sec:sfr_he}, respectively. In the case of IZw18, given the lower number of spaxels available with respect to Henize 2-10, we did not apply cuts on SNR in our analysis.

The average uncertainties on the physical properties reported in Fig. \ref{fig:izw18_physical} are $\sigma(x)/x \sim 0.4-0.6$, where $x$ is the physical property expressed in logarithm. As in the case of He 2-10, we find that the metallicity is the inferred physical property with the lowest uncertainties ($\sigma[log(Z)] \lesssim 0.1$, corresponding to a relative uncertainty $\sigma(Z)/Z \lesssim 0.3$).

\subsubsection{Gas density}
In almost all spaxels we infer density values $\lesssim$ 10$^2$ cm$^{-3}$ \citep[see also][]{Kehrig2016}. The average density is $n$ $\sim$ 10$^{0.7}$ cm$^{-3}$. For the column density, we infer an average value of $N_H$ $\sim$ 10$^{21}$ cm$^{−2}$, in agreement with \citet{vanZee1998} that report a gas column density exceeding 10$^{21}$ cm$^{-2}$ mainly located in the SE and NW active star forming regions of IZw18.

\subsubsection{Metallicity}
\begin{figure}
	\centering
	\includegraphics[width=0.95\linewidth]{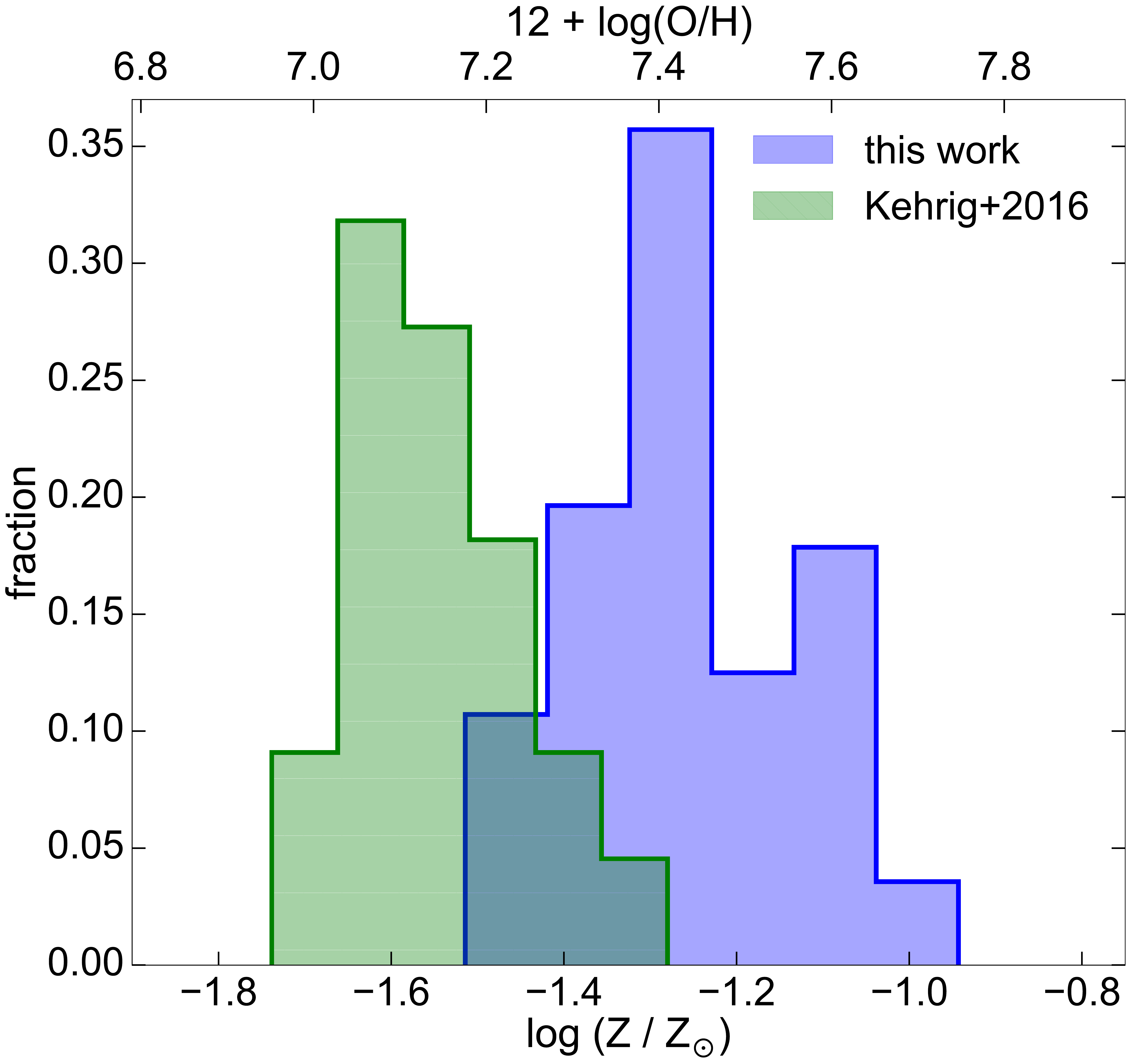}
	\caption{Distribution of metallicity for the spaxels reported in the central left panel of Fig. \ref{fig:izw18_physical}. The blue and green shaded histograms report the results obtained in this work and in \citet{Kehrig2016}, respectively.}
	\label{fig:hist_metals_izw18}
\end{figure}

In Fig. \ref{fig:izw18_physical} and \ref{fig:hist_metals_izw18} we show the map and the metallicity distribution, respectively. No significant metallicity gradient is present: the maximum variation among individual spaxels is in fact within a factor of $\sim$ 3. Although the number of available spaxels is too small for an accurate analysis, this suggests a uniform distribution over spatial scales of hundreds of parsecs \citep{Kehrig2016}. The error-weighted mean metallicity of all spaxels inferred with \textlcsc{GAME} is log($Z/Z_{\odot}$) = -1.32 $\pm$ 0.01, corresponding to 12+log(O/H) = 7.37 $\pm$ 0.01 ($\sim$ 1/20 solar)\footnote{Solar gas-phase metallicity is taken to be 12+log(O/H) = 8.69 \citep{Allende2001,Asplund2004}, i.e. we assumed the following relation: log($Z/Z_{\odot}$) = 12+log(O/H) - 8.69.}. This result is a factor of two larger than the estimate of \citet{Kehrig2016} who found, directly with the electron temperature, 12+log(O/H) = 7.11 $\pm$ 0.01 ($\sim$ 1/40 solar) (see Fig. \ref{fig:hist_metals_izw18}). This mismatch could be due to the discrepancies between different methods \citep[up to $\sim$ 0.15 dex, see][]{Kewley2008} or to the well known difference up to 0.4 $\div$ 0.5 dex between direct methods and techniques based on photoionization models \citep{Stasinska2002,Kennicutt2003,Garnett2004,Kewley2008}.

\subsubsection{Dust extinction}
The map of the visual extinction $A_V$ is reported in Fig. \ref{fig:izw18_physical}. We obtain very low values for the visual extinction ($A_V$ $\lesssim$ 0.20) with $A_V$ $\sim$ 0.01 mag for the majority of the spaxels.

\subsubsection{Interstellar radiation field}\label{sec:radiation_izw18}
Regions with higher ionization parameter are preferentially located in the SE knot where we obtained values as high as $U$ $\sim$ 10$^2$. In the rest of the galaxy we found log($U$) $\sim$ 0.5 - 1. The most extreme values for $U$ in local analogs of high redshift galaxies can approach unity \citep[e.g.,][]{Bian2016,Bian2017,Bian2018}. In fact, assuming the relation L(H$\alpha$/erg s$^{-1}$) = $1.37 \times 10^{-12}$ $Q(H)$ \citep{Osterbrock2006}, where $Q(H)$ is the ionizing photon rate in units of s$^{-1}$, and using the usual definition for $U$ (see footnote \ref{udef}), we obtain in the SW region log($U$) $\sim$ -0.5 - 0.3 (assuming $R_S$ to be 88 pc). However, the \textlcsc{GAME} output physical property most affected by large uncertainties is the ionization parameter, due to the poorer accuracy in its determination with optical lines only as input \citepalias[see also][]{Ucci2017,Ucci2017b}. A better constraint of $U$ requires a larger number of emission lines \citepalias[e.g., Fig. D5 in][]{Ucci2017b}, spanning additional wavelength ranges (i.e., UV+optical+IR). Therefore, the extremely high values of $U$, together with the high $G$ we found could be considered as an additional qualitative indication of the high underlying ionizing flux in this galaxy.

The spatial distribution of $G$ is more uniform across the galaxy, although it is possible to associate higher fluxes in the SE region where log($G/G_0$) $\gtrsim$ 3.5. The prominent ionizing flux in the SE knot in our maps, denotes the presence of a clumpy HII component, consisting of numerous knots containing blue stars \citep{Dufour1996}. Such high ionizing fluxes may be originated from the hard SED found in the NW knot, coming from the presence of PopIII stars. (see Sec. \ref{sec:popiii}).

\subsubsection{Gas mass and SFR surface density}\label{sec:izw18_sfr}
Using the same approach as in Sec. \ref{sec:sfr_he}, we obtain $\Sigma_{gas}$ and $\Sigma_{SFR}$ in Fig. \ref{fig:izw18_inferred}. Given the lack of reliable determinations of $N_H$ and $n$ in some spaxels (i.e., grey areas in Fig. \ref{fig:izw18_physical} and \ref{fig:izw18_inferred}), we can only infer lower limits, SFR $\gtrsim$ 0.007 M$_{\odot}$ yr$^{-1}$, and for the total gas mass M$_{gas}$ $\gtrsim$ 6.9 $\times$ 10$^{6}$ M$_{\odot}$. A previous estimate of $SFR = 0.1$ M$_{\odot}$ yr$^{-1}$ was obtained by \citet{Hunt2005}, while \citet{Aloisi1999} provides $SFR = 6 \times 10^{-2}$ M$_{\odot}$ yr$^{-1}$. \citet{Legrand2000} and \citet{Legrandetal2000}, by using a spectrophotometric and chemical model, showed that the photometric properties of IZw18 can be reproduced assuming a continuous SFR as low as 10$^{-4}$ M$_{\odot}$ yr$^{-1}$ lasting for a Hubble time, with a recent burst of star formation superimposed. \citet{Recchi2004} report a SFR of 4-6 $\times$ 10$^{-3}$ M$_{\odot}$ yr$^{-1}$. These $SFR$ values are required to generate the blue color of IZw18 \citep{vanZee1998,Fisher2014} and the high $G$ and $U$ we found. Galactic winds, produced by stellar winds and SNe could be able to create an outflow and eject part of the ISM. In fact, for such a small galaxy, a $SFR$ of 0.02 M$_{\odot}$ yr$^{-1}$ can already accelerate the gas and produce the UBV colors, and most of the ionizing luminosity in IZw18 \citep{Martin1996}. The starburst inside this galaxy could also trigger a galactic outflow, with the metals leaving the galaxy far more easily than the unprocessed gas \citep{Maclow1999}.

\subsection{A PopIII-like SED in IZw18 NW?}\label{sec:popiii}
IZw18 is characterized by an extended nebular HeII $\lambda$4686 emission in its NW component \citep{Kehrig2015}. This implies the presence of a hard ionizing field (i.e., photon energies $\geq$ 54.4 eV) in this region. This field has been ascribed in the literature \citep{Kehrig2015,Kehrig2016} to the presence of stars with a spectral energy distribution typical of PopIII models \citep{Schaerer2002,Raiter2010}. In fact, it has been shown \citep{Kehrig2015} that the HeII-ionizing energy budget cannot be explained by conventional excitation sources (i.e., WRs, shocks, X-ray binaries). Vice versa, one of the possible explanations for this emission is the presence of peculiar very hot, ionizing stars which may be (nearly) metal-free \citep{Kehrig2015,Kehrig2016}. To further investigate the origin of the HeII $\lambda$4686 emission, we adopt a modified version of \textlcsc{GAME} that also predicts the intensity of HeII $\lambda$4686 emission line.

\begin{figure*}
	\centering
    \includegraphics[width=0.335\textwidth]{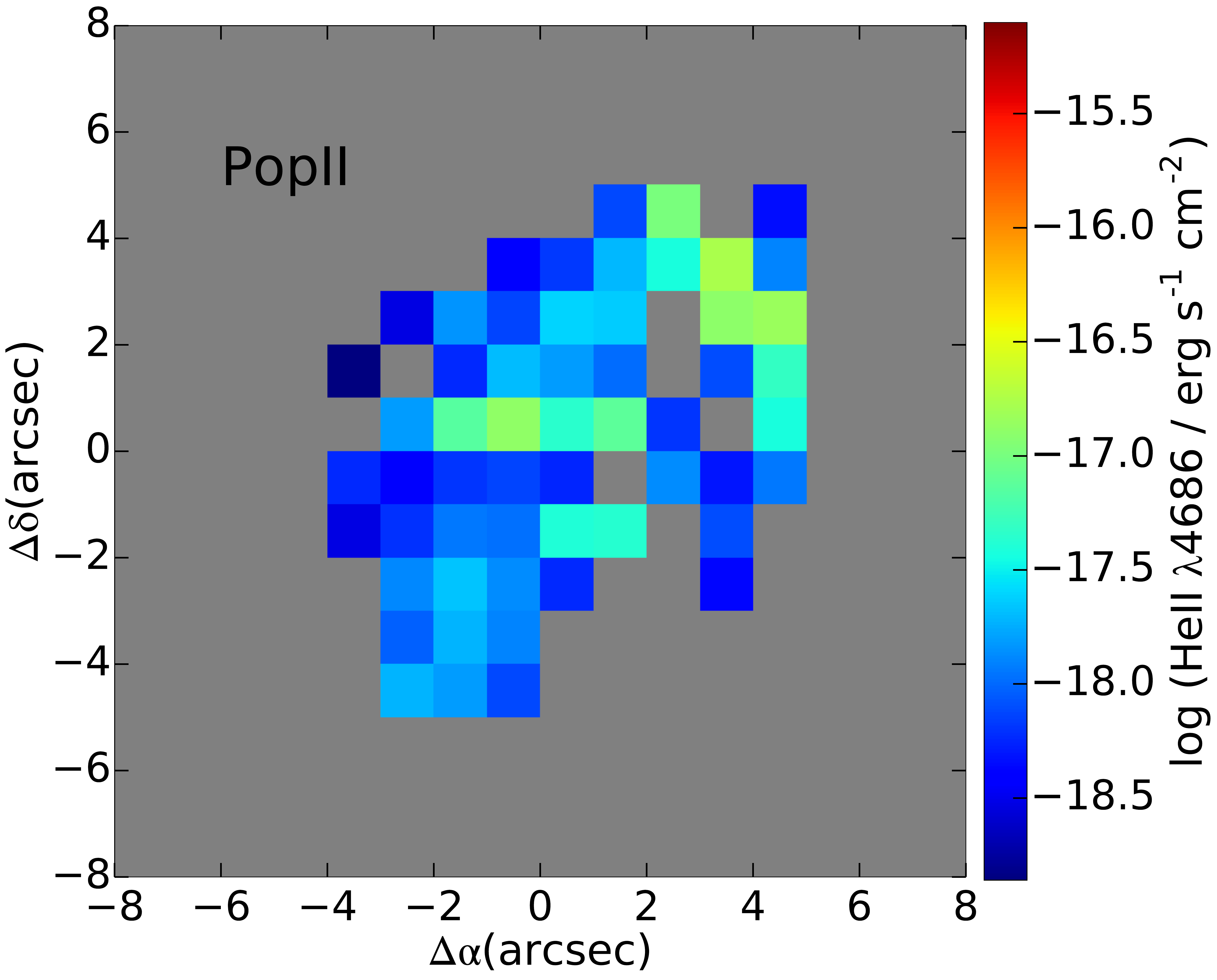}
    \includegraphics[width=0.335\textwidth]{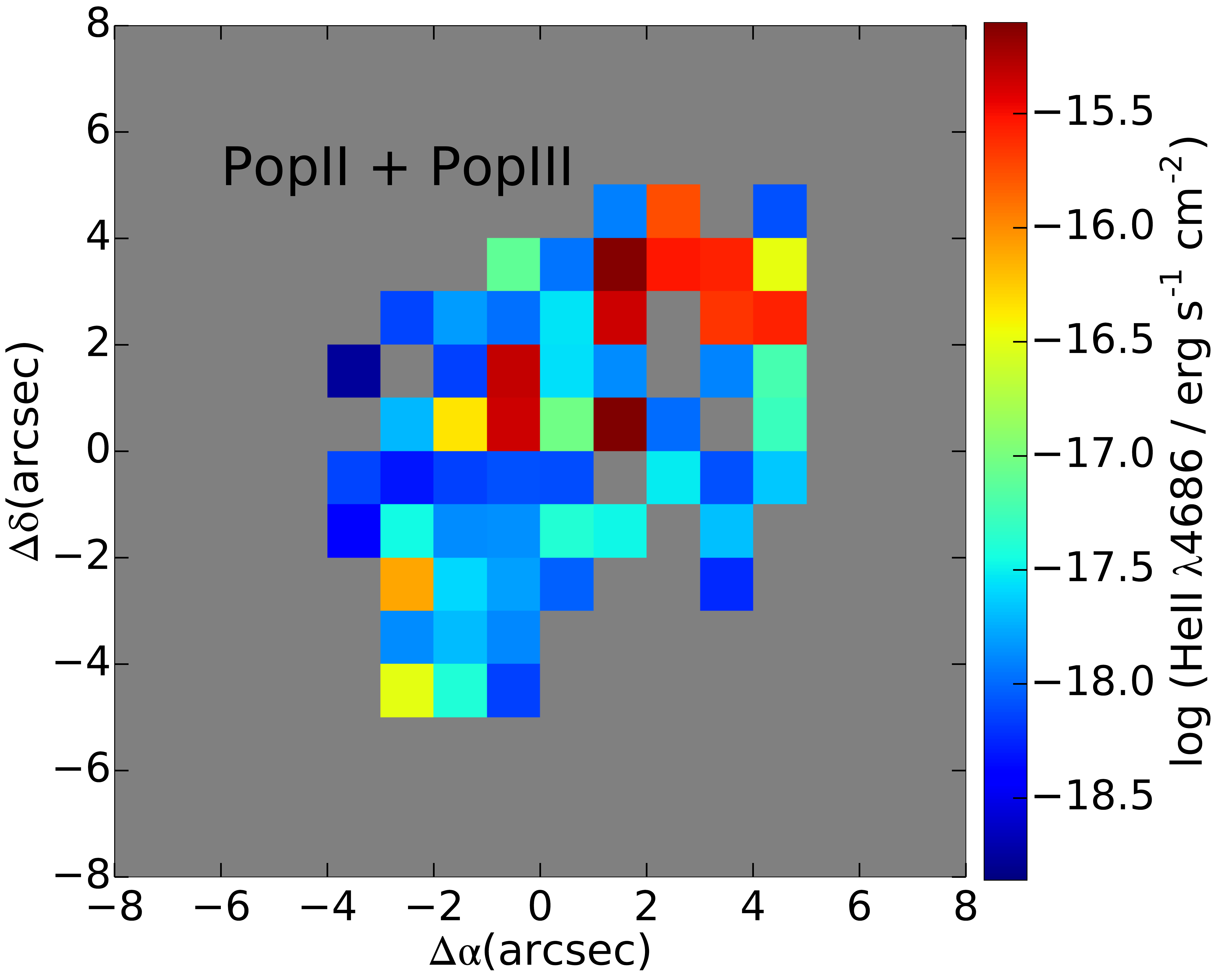}
    \includegraphics[width=0.317\textwidth]{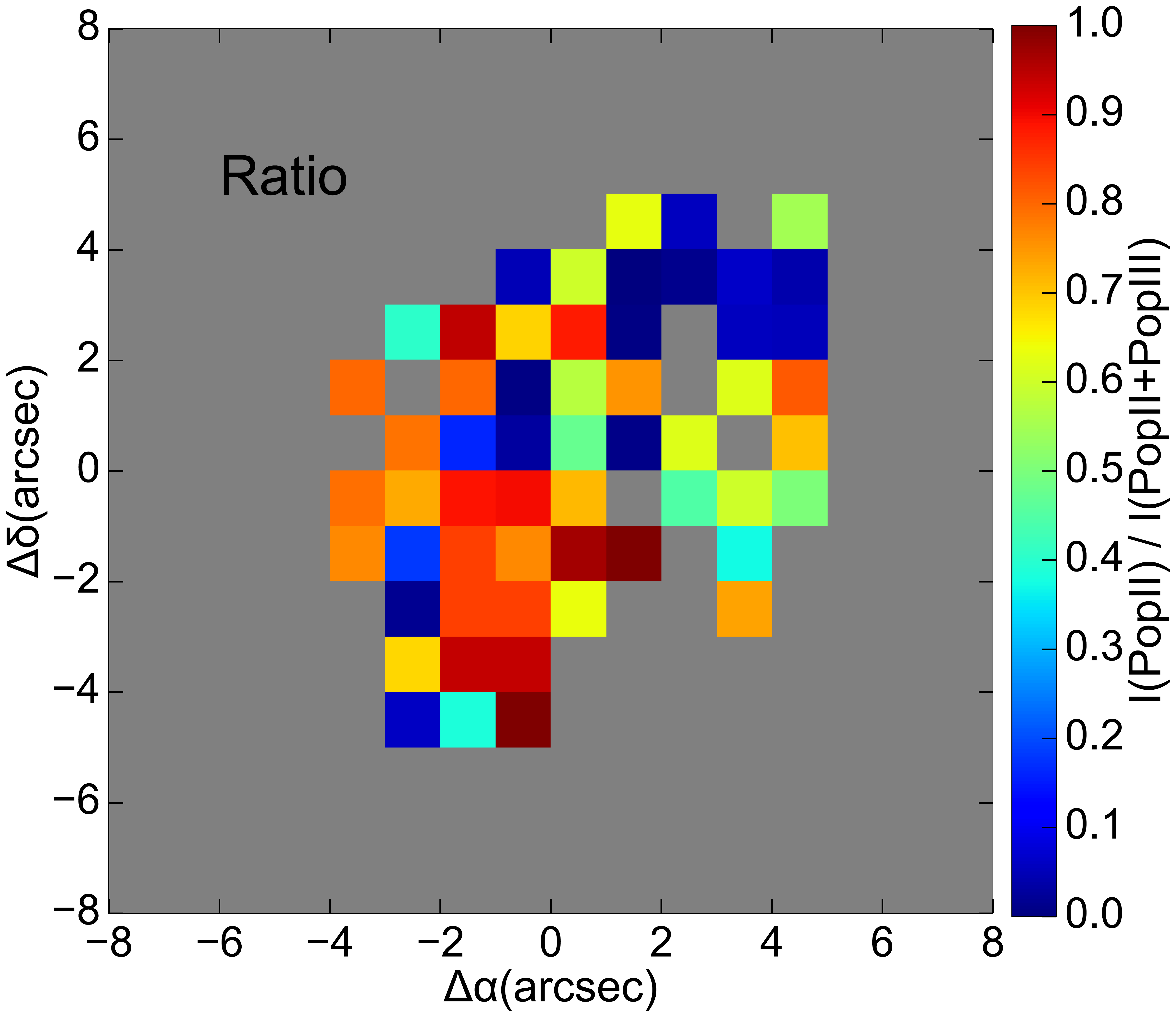}
    \caption{Predicted HeII$\lambda$4686 line emission by \textlcsc{GAME} for the spaxels of IZw18 using respectively the library containing only PopII stars (\textit{left panel}), and the full library containing also the contribution from PopIII (\textit{central panel}). \textit{Right panel}: ratio between the predicted intensities in the \textit{left} and \textit{central} panels showing a negligible contribution from PopII stars in the NW region.}
	\label{fig:izw18_heii4686}
\end{figure*}

In the left panel of Fig. \ref{fig:izw18_heii4686}, we report the predicted HeII $\lambda$4686 emission line intensities by considering a reduced library containing only PopII stars \citepalias[analogous to the ones included in][]{Ucci2017}. In the central panel is shown the HeII $\lambda$4686 line intensity, obtained by including also PopIII stars \citepalias{Ucci2017b}. With purely PopII stellar models, it is not possible to obtain emission line intensities compatible with observations \citep[e.g., Fig. 2 of][]{Kehrig2015}: the inferred values are more than one order of magnitude smaller than the observed ones. Adding in the library the contribution from PopIII stars, the HeII $\lambda$4686 intensities can reach values $\gtrsim$ 10$^{-16.0}$ erg s$^{-1}$ cm$^{-2}$, in agreement with observations in the NW region of IZw18. In Fig. \ref{fig:izw18_heii4686} we also show the ratio between the PopII and the (PopII+PopIII) contribution to the HeII $\lambda$4686 intensities. We found that there is a negligible contribution from PopII stars in the NW region, where the ratio is $\ll$ 1, denoting the possible evidence of a PopIII population. Although there is X-ray emission dominated by a single point source located in the NW region with no significant contribution of faint diffuse emission, the HeII emission is essentially due to hot stars \citep{Pequignot2008,Lebouteiller2017}. Our result suggests that PopII stellar populations can not account for the observed HeII$\lambda$4686 flux, while PopIII stars, may provide a substantial contribution to the HeII emission.

We exploit this result to provide an estimate of the PopII/PopIII relative contribution to the total SFR of IZw18. First, we select the spaxel reported in the right panel of Fig. \ref{fig:izw18_heii4686} for which the ratio is approximately 1. Given that in these spaxels no PopIII stars contribution is required to explain the HeII emission, we attributed their SFR to PopII stars only. Summing up their contribution, we obtain a value of SFR$_{PopII}$ $\sim$ 0.0031 M$_{\odot}$ yr$^{-1}$. In the remaining spaxels, where the ratio is $\gtrsim$ 1, the SFR resulting from the combination of PopII and PopIII bursts is $\sim$ 0.0039 M$_{\odot}$ yr$^{-1}$. We conclude that PopIII stars could explain at most 52\% of the total IZw18 SFR (SFR$_{PopIII}$ $\lesssim$ 0.007 M$_{\odot}$ yr$^{-1}$, see Sec. \ref{sec:izw18_sfr}).

\section{Conclusions}\label{sec:conclusions}
We presented a comprehensive investigation of the ISM physical properties in two Blue Compact Galaxies (BCGs): the starburst galaxy Henize 2-10, and the metal-poor dwarf IZw18. This study was possible taking advantage of a Machine Learning (\textlcsc{ML}) code called \textlcsc{GAME} \citepalias[GAlaxy Machine learning for Emission lines, see][]{Ucci2017,Ucci2017b}, specifically designed to infer the physical properties of multi-phase ISM. We applied \textlcsc{GAME} to MUSE and PMAS IFU observations, obtaining the resolved structure of gas density ($n$), column density ($N_H$), ionization parameter ($U$), metallicity ($Z$), FUV flux ($G$), visual extinction ($A_V$), gas mass surface density ($\Sigma_{gas}$), SFR surface density ($\Sigma_{SFR}$).

\textlcsc{GAME} proved to be extremely reliable: although each physical property is treated separately and independently \citepalias[i.e., a different \textlcsc{ML} model is constructed for each of the physical properties, see][]{Ucci2017b}, the overall morphology of the inferred maps agree with one each other and the expected correlations between physical properties such as $n$ with $N_H$ (see Fig. \ref{fig:correlation}) hold (see also Appendix \ref{sec:av_nh}).

Our main conclusions for the galaxy Henize 2-10 can be summarized as follows:

\begin{itemize}
\item It is a star-forming dominated galaxy with $SFR \gtrsim 1.2$ M$_{\odot}$ yr$^{-1}$, and a total gas mass of $M_{gas} \sim 1.9 \times 10^7$ M$_{\odot}$.\\

\item The ionization parameter in the central regions ($d \lesssim 4\arcsec$) is $U \sim 10^{-3}$, while at larger distances $U \gtrsim 10^{-2}$. The interstellar radiation field in the FUV band $G$ is also strong, with a flat radially median profile and intensities around $G/G_0 \sim$ 10$^{1.5}$, while $G$ reaches its peak in Regions A and B ($G/G_0 \sim 10^3$).\\

\item The visual extinction is very high, i.e. $\langle A_V \rangle \gtrsim 3$, with $A_V$ in some spaxels exceeding 5-7 mag. These high extinctions are especially found on a region forming a ring around the center of the galaxy.\\

\item The radially median profile for the metallicity is flat with slightly sub-solar values (i.e., log($Z/Z_{\odot}) \sim -0.2$). However, the $Z$ distribution inferred in this work covers a quite large range (i.e., -1.0 $\lesssim$ log($Z/Z_{\odot}$) $\lesssim$ 0.3) denoting the heterogeneous morphology of He 2-10 in terms of physical properties across its extension.\\

\item The gas density in Regions A and B reaches values up to $n \sim 10^{3.3}$ cm$^{-3}$. Then the radially median profile flattens around a value of $n \sim 1$ cm$^{-3}$. Typical column densities are $N_H \sim 10^{20}-10^{22}$ cm$^{-2}$.

Defining an effective scale height of the disk, as $H$ = $N_H/n$, we obtained 0.3 $\lesssim H/\text{pc} \lesssim 30$ in the central region, denoting a relatively small height of the disk. In Regions A and B we are looking at the typical dimension of Giant Molecular Clouds (GMCs), and we found high values of densities and metallicities, together with high H$\alpha$ and FUV flux $G$. These evidences suggest that these regions host dense, dusty and highly star forming sites.
\end{itemize}

The main results for IZw18 are the following:

\begin{itemize}
\item  Summing up the contribution from the analyzed spaxels, we obtained lower limits for the Star Formation Rate and total gas mass respectively of $SFR \gtrsim 0.007$ M$_{\odot}$ yr$^{-1}$ and $M_{gas} \sim 6.9 \times 10^{6}$ M$_{\odot}$. The total gas mass of IZw18 has also been estimated by \citet{Schneider2016} who found $M_{\rm tot} = M_{\rm HI} + M_{\rm H_2} + M_{\rm HII} = 9.2 \times 10^7$ M$_{\odot}$, with the different phases contributing approximately equal fractions to the total gas mass. A total atomic hydrogen mass of $\sim 10^8$ M$_{\odot}$ has been measured by \citet{Lelli2012} throughout the entire main body, with a radius of $\sim$ 8\arcsec. Both these estimates encompass a region larger than that covered by our spectra; thus our estimate is a lower limit, and consistent with the column densities of $N_H = 2 \times 10^{21}$ cm$^{-2}$ found by us and those in HI measured by \citet{Lelli2012} with a 2\arcsec beam. In fact, by integrating the \citet{Lelli2012} HI radial column density distribution up to $\sim$ 4\arcsec (the area covered by our spectral maps), the resulting mass is 8-30 $\times$ 10$^6$ M$_{\odot}$, slightly higher but consistent with our results for a lower limit.\\

\item The interstellar radiation field is very intense: in the SE knot we obtained very high values for the ionization parameter as high as $U \sim 10^2$, and FUV fluxes $G/G_0 \sim 10^{3.5}$. These high values for $U$, are an additional indication of the underlying high ionizing flux in IZw18, given the large uncertainties associated with the ionization parameter estimates \citepalias[see][]{Ucci2017b}. The prominent ionizing flux in the SE knot in our maps, can be in fact associated with the presence of a clumpy HII component, consisting of numerous knots containing blue stars \citep{Dufour1996}.\\

\item We confirmed the extreme poor metal content of IZw18: log($Z/Z_{\odot}$) = -1.32 $\pm$ 0.01 (i.e., 12+log(O/H) = 7.37 $\pm$ 0.01, corresponding to $\sim$ 1/20 solar). In this galaxy the metallicity remains also remarkably flat: $-1.5 \lesssim$ log($Z/Z_{\odot}$) $\lesssim -1.0$, with the inferred values being within a factor of three. These low metallicity correspond also to low visual extinction across the galaxy and low dust-to-gas mass ratios: $A_V \lesssim 0.20$ mag and $\mathcal{D} \sim 10^{-3.5}$, respectively.\\

\item IZw18 presents a low gas density environment, with most of the spaxels having log($n$/cm$^{-3}$) $\lesssim$ 1. We obtained column densities of $N_H \sim 10^{21}$ cm$^{-2}$.\\

\item We verified if the extended nebular HeII $\lambda$4686 emission in the NW component of IZw18 found by \citet{Kehrig2015,Kehrig2016} could be ascribed to PopIII stars. We implemented a modified version of \textlcsc{GAME} that is able to predict, in addition to the ISM physical properties, also the value for the intensity of a given emission line (i.e., HeII $\lambda$4686). Then, we compared the predicted HeII lines with observations finding that if PopIII stars are not taken into account, the inferred intensities for the HeII line are too low compared with observed values. Therefore, PopIII stars may provide a substantial contribution to the HeII emission.
\end{itemize}

\section*{Acknowledgements}
We thank the anonymous referee for her/his careful reading of our manuscript, and for useful comments and
suggestions. We thank A. Bolatto for useful discussions and suggestions. AF acknowledges support from the ERC Advanced Grant INTERSTELLAR H2020/740120. This research was partially supported by the Munich Institute for Astro- and Particle Physics (MIAPP) of the DFG cluster of excellence "Origin and Structure of the Universe".

\bibliographystyle{mnras}
\bibliography{bibliography_local_galaxies}

\appendix

\section{Photon flux in He 2-10 is star-forming dominated}\label{sec:sf_dominated}
In order to investigate the nature of the dominant radiation source, it is possible to use the Baldwin, Phillips \& Terlevich (BPT) diagram \citep{Baldwin1981}, used to distinguish between star-forming regions and active galactic nuclei. The analysis in this section closely follows the one presented in \citet{Cresci2017} (i.e., their Fig. 5), but here we also show the applicability of \textlcsc{GAME} to He 2-10 and with slightly different arguments how the stellar component dominates the radiation field in the galaxy. The BPT diagram makes use of two pairs of lines close in wavelength in order to minimize the effects of dust and continuum subtraction uncertainties. The line ratios involved in the BPT diagram are usually the [N II]/H$\alpha$ ratio, primarily sensitive\footnote{The BPT lines are also sensitive to the ionization parameter, N and O abundances, and density- vs ionization-bounded HII regions.} to the shape of the ionizing spectrum just above 1 Rydberg, and the [O III]/H$\beta$ ratio that probes the 1-3 Rydberg range.

\begin{figure*}
	\centering
    \includegraphics[width=0.95\textwidth]{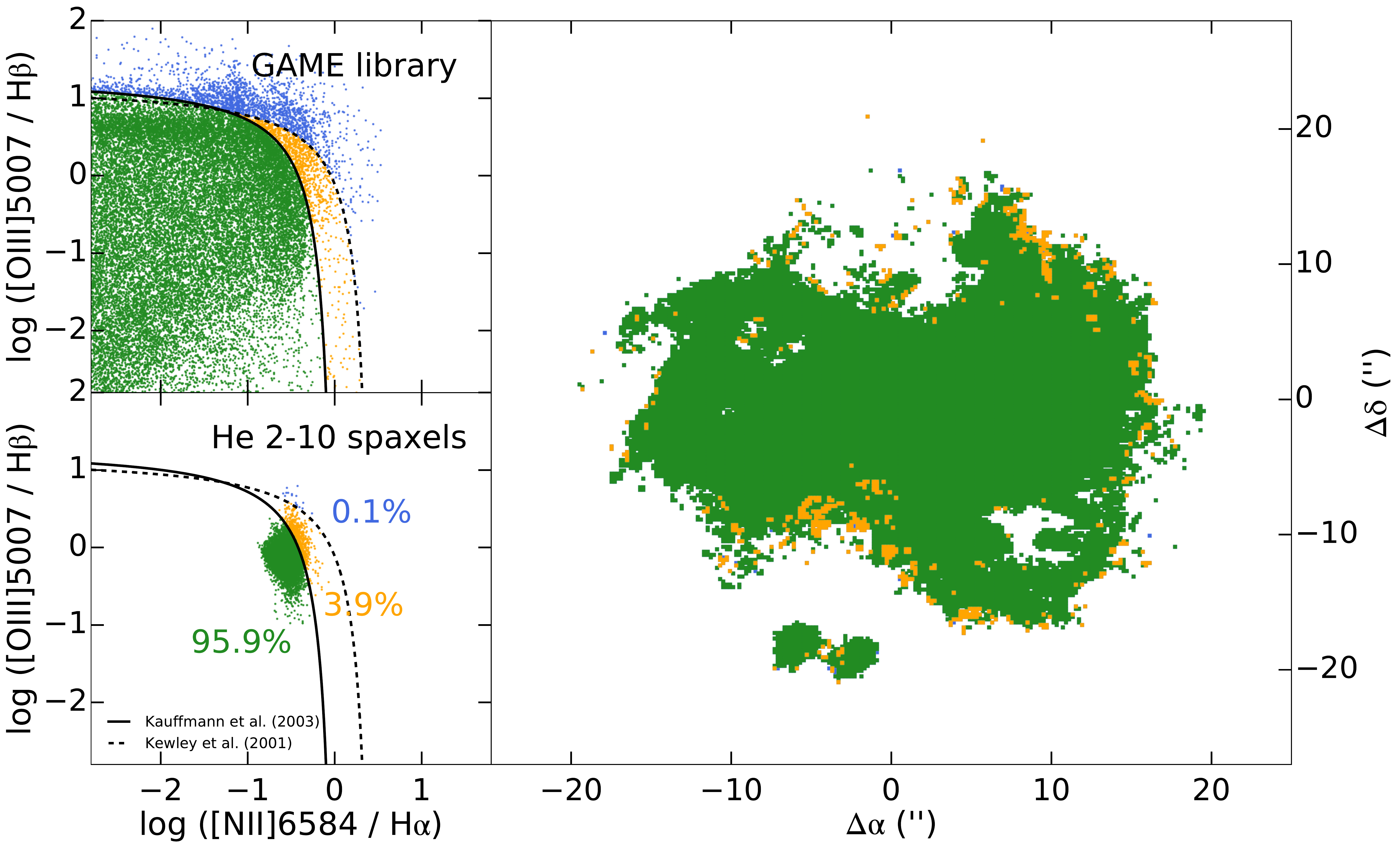}
    \caption{\textit{Upper left panel}: BPT diagram for the models contained within the library of synthetic models of \textlcsc{GAME}. As expected, because the assumed input spectra are stellar, nearly all the models falls within the Star-Forming region. \textit{Lower left panel}: BPT diagram for the observed spaxels of He 2-10 where we also added the percentages of the different region in the observed line ratios. \textit{Right panel}: Map for He 2-10 color-coded based on the BPT diagram. In green are reported the spaxels having line ratios below the line reported in \citet{Kauffmann2003} denoting regions dominated by star formation.}
	\label{fig:bpt_diagram}
\end{figure*}

The BPT diagram for the models within the \textlcsc{GAME} library \citepalias{Ucci2017b} and the spaxels in He 2-10 are respectively shown in the upper and lower left panels in Fig. \ref{fig:bpt_diagram}. Here, we report three zones \citep{Kewley2001,Kauffmann2003,Kewley2006}: (a) regions dominated by star formation, (b) intermediate regions, (c) AGNs. Star forming regions are below the \citet{Kauffmann2003} line (black solid line in Fig. \ref{fig:bpt_diagram}):

\begin{equation}
\text{log([OIII]/H}\beta) = \frac{0.61}{\text{log([NII]/H}\alpha) - 0.05} + 1.3.
\end{equation}

Intermediate regions fall between this and the \citet{Kewley2001} line (dashed line in Fig. \ref{fig:bpt_diagram}):

\begin{equation}
\text{log([OIII]/H}\beta) = \frac{0.61}{\text{log([NII]/H}\alpha) - 0.47} + 1.19.
\end{equation}

Regions dominated by AGN contribution are above the two lines. In the right panel of Fig. \ref{fig:bpt_diagram}, we marked these three zones with different colours. The map results to be dominated by green dots (i.e., more than 95\% of spaxels), meaning that almost all the line emitting gas is ionized by young stars. This result confirms the assumption that the models contained within the library used by \textlcsc{GAME} \citepalias[i.e., photoionization models coming only from stars without X-ray sources or AGNs,][]{Ucci2017b} can be applied to this galaxy.

In order to further validate this result we started from the radiation energy density of the stars contained within the galaxy, assumed to be a uniformly emitting sphere of radius $r$:

\begin{equation}
u = \frac{L_{*}}{4\pi r^2 c},
\label{eq:energy_density}
\end{equation}
where $L_{*} = \epsilon M_{*}$ is the luminosity produced by the stars inside the galaxy, $M_{*}$ is the total stellar mass contained in the galaxy, and $\epsilon$ the conversion between the stellar mass itself and the luminosity ($L_{\odot} M_{\odot}^{-1}$). By considering the previous estimates for the stellar mass contained in He 2-10 $M_{*} = 10^9$ M$_{\odot}$ \citep{Reines2011,Nguyen2014} and a conservative value based on the mass-luminosity relation for a sub-solar mass of $\epsilon = 0.1$ $L_{\odot} M_{\odot}^{-1}$ \citep{Kuiper1938,Salaris2005}, we obtained $L_{*} = 10^8$ L$_{\odot}$. We further computed the distance $d$ at which the radiation energy density of the stars is equal to that produced by the X-ray source by simply imposing:

\begin{equation}
\frac{L_X}{4 \pi d^2 c} = \frac{L_*}{4 \pi R^2 c},
\end{equation}
where $L_X$ is the luminosity of the X-ray source and $R$ is the radius of the uniformly emitting sphere, that we assumed to be $R = 6\arcsec = 240$ pc, based on the effective radius of the inner component of the galaxy provided by \citet{Nguyen2014}. Given the recent estimate of $L_X \approx 10^{38}$ erg s$^{-1}$ of \citet{Reines2016}, we obtain that d $\sim$ 4 pc. Of course more realistic values (i.e., greater values) for $\epsilon$ due to young massive stars and the application of bolometric corrections \citep{Marconi2004} for $L_X$ up to $\sim$ 10 \citep[e.g.,][]{Vasudevan2009} mean that the distance $d$ could remain the same or become even smaller (if one assumes $\epsilon$ $\sim$ 1). The region where the energy density of the X source is relevant is very small compared to the effective radius of the galaxy. The galaxy radiation field is therefore completely dominated by the stellar component. This confirm both the results by \citet{Cresci2017} and guaranteeing the applicability of \textlcsc{GAME} to this source.

\section{Maps of uncertainties for He 2-10}\label{sec:sigmas}
\textlcsc{GAME} determines 10,000 different values of each physical property \citepalias{Ucci2017b}. The associated uncertainties are given by the standard deviations of these 10,000 realizations.

\begin{figure*}
    \raggedleft
    \includegraphics[width=0.48\textwidth]{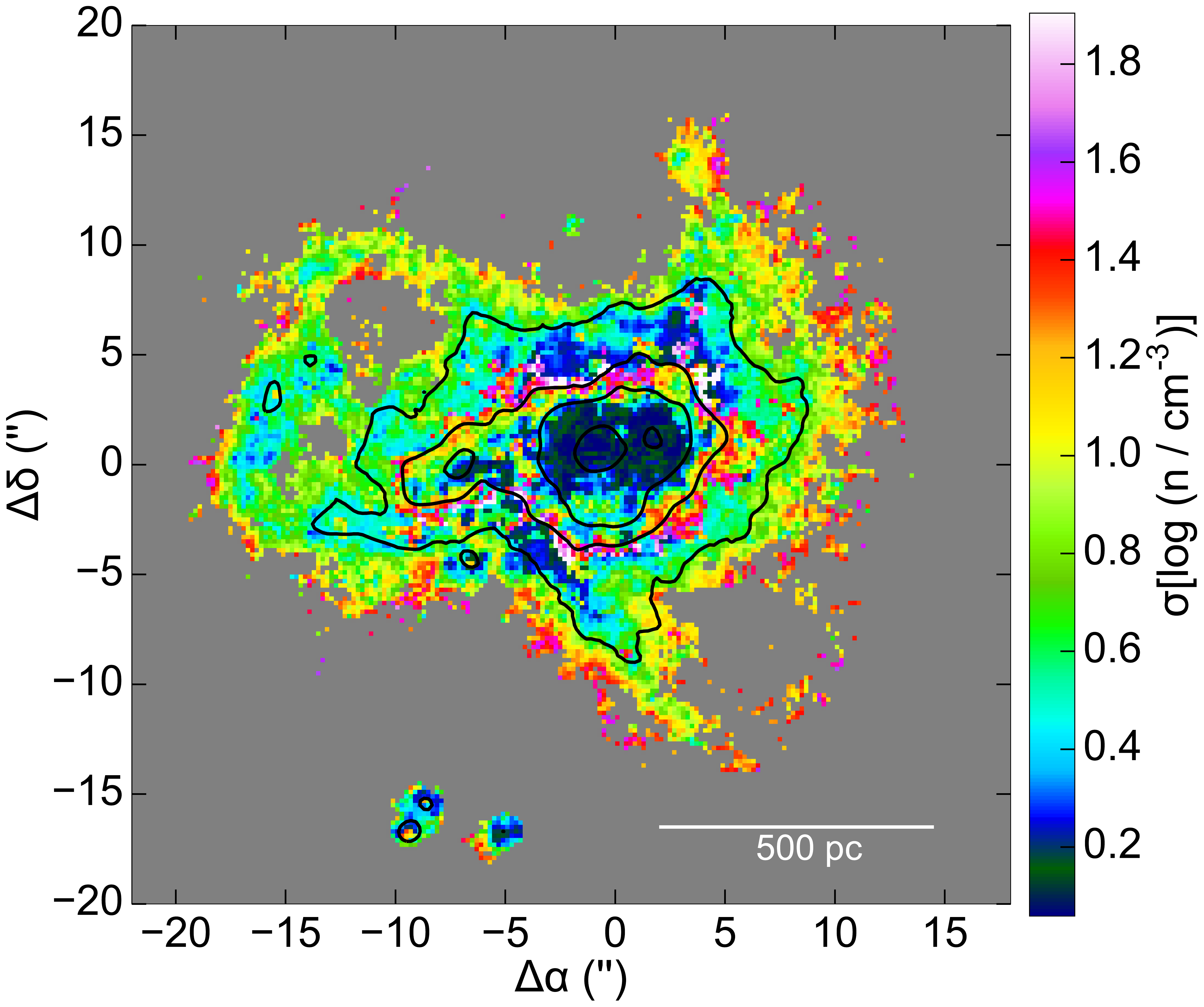}
    \raggedleft
    \includegraphics[width=0.482\textwidth]{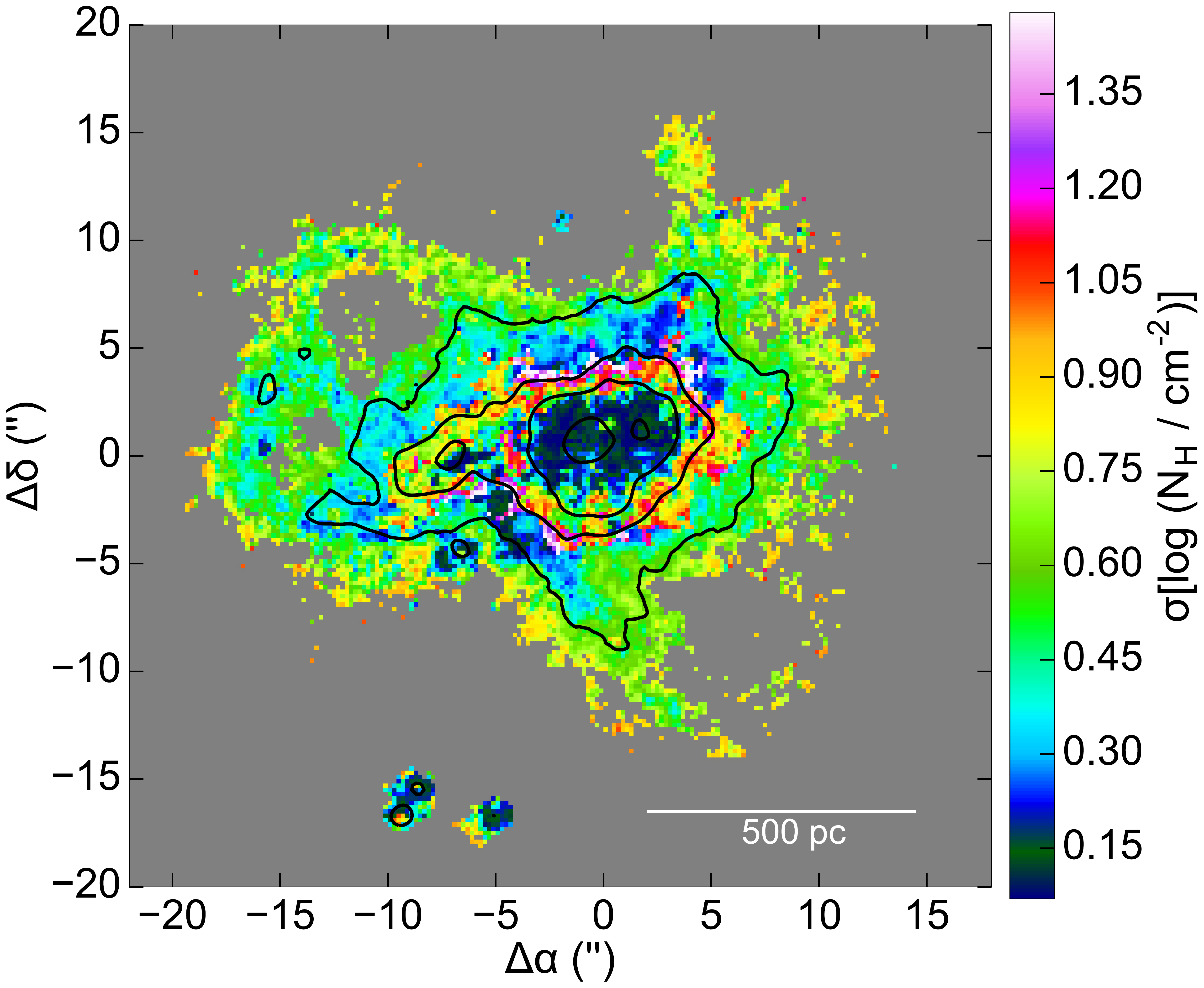}
    \raggedleft
    \includegraphics[width=0.482\textwidth]{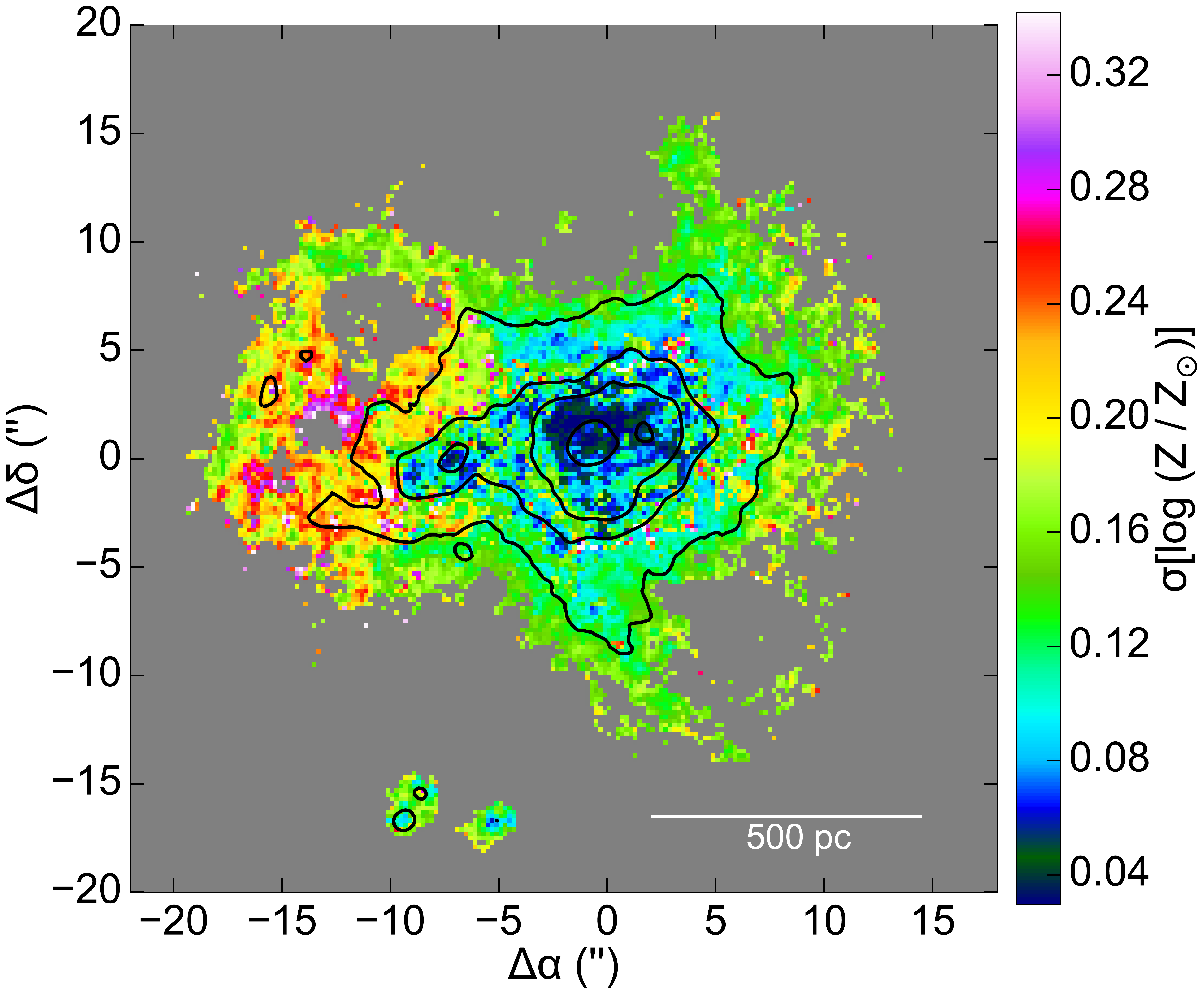}
	\raggedleft
	\includegraphics[width=0.478\textwidth]{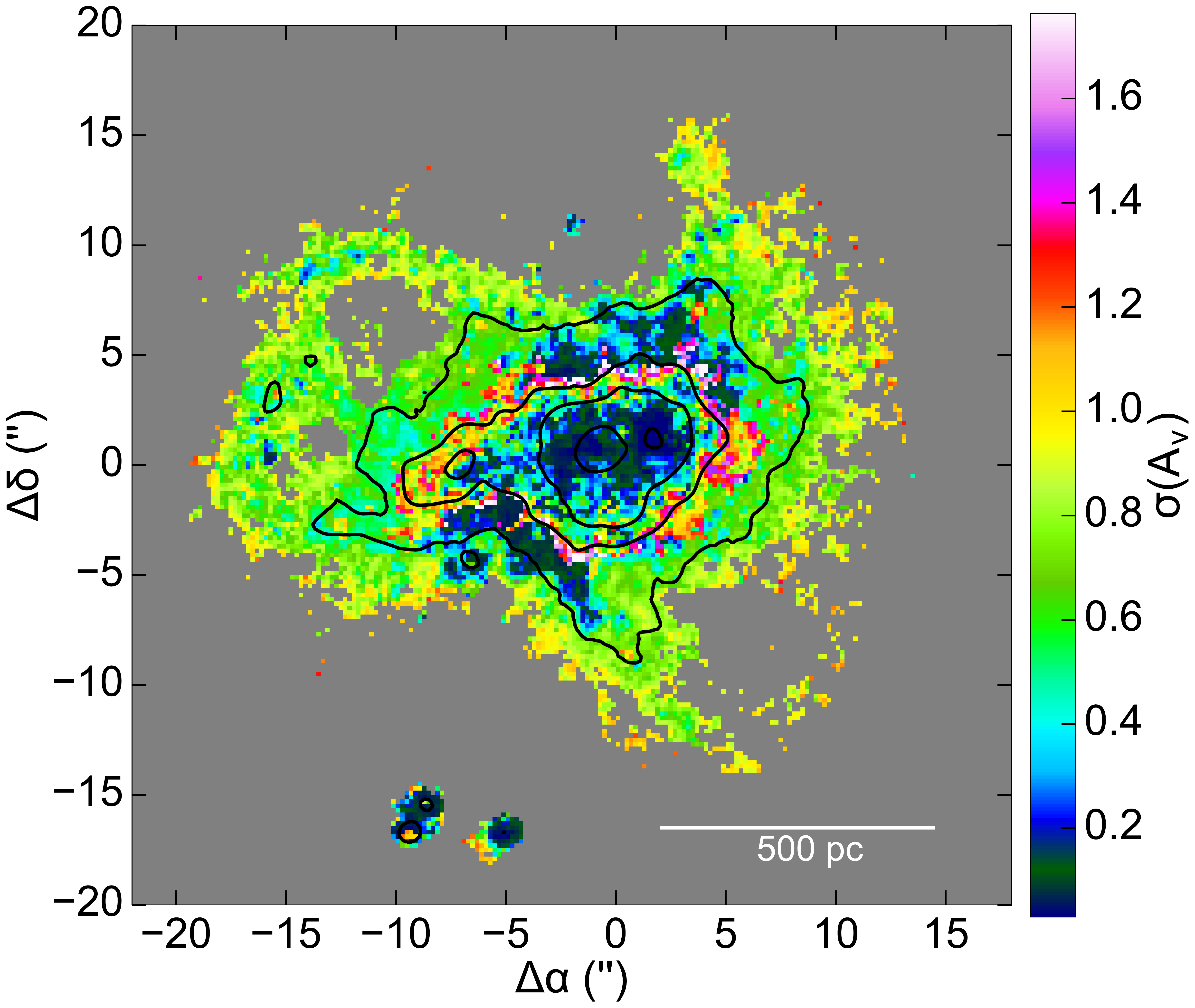}
    \raggedright
    \includegraphics[width=0.485\textwidth]{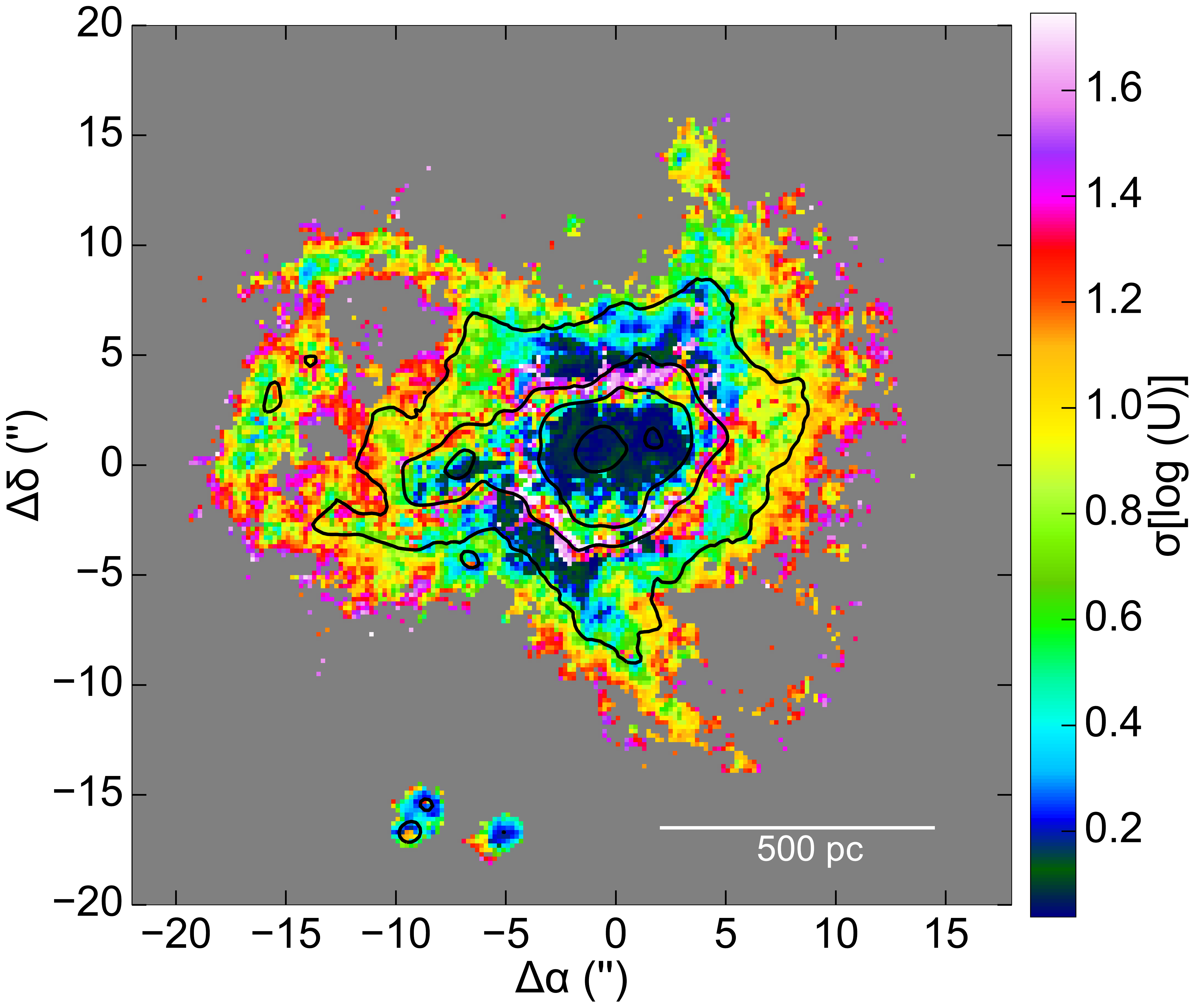}
    \includegraphics[width=0.485\textwidth]{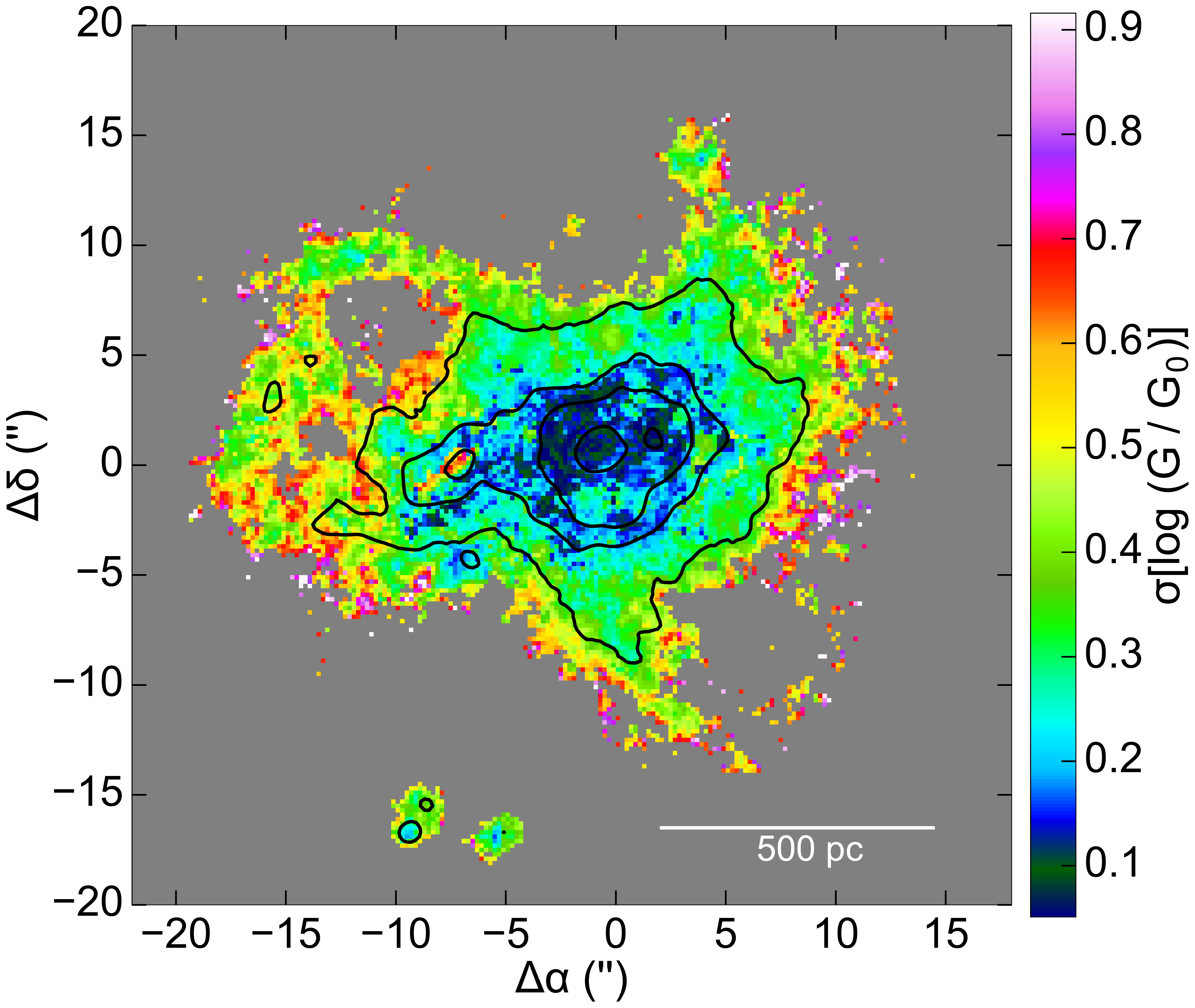}
    \caption{Relative uncertainties on the He 2-10 ISM physical properties reported in Fig. \ref{fig:He_2-10_physical}: density ($n$), column density ($N_H$), metallicity ($Z$), visual extinction ($A_V$), ionization parameter ($U$) and the FUV flux in the Habing band ($G/G_{0}$). As in Fig. \ref{fig:he_halpha}, black lines show the contours of the H$\alpha$ emission (80, 10, 4, and 1 in units of 10$^{-16}$ ergs s$^{-1}$ cm$^{-2}$). In the upper left corner we report the average relative uncertainty on the inferred values.}
	\label{fig:He_2-10_sigma}
\end{figure*}

Fig. \ref{fig:He_2-10_sigma} shows the relative uncertainties on the He 2-10 physical properties derived directly by game and reported in Fig. \ref{fig:He_2-10_physical}. The typical uncertainties in the central region of the galaxy (within $\sim 10\arcsec$ from the center) are smaller than 30\%, while in the case of the metallicity the vast majority of spaxels have lower uncertainties even at larger distances. The highest uncertainties remains associated to the ionization parameter $U$, although in the central part of the galaxy, $\sigma$[log($U$)]/log($U$) reduces to values $\lesssim$ 10\%.

\section{Gas-to-extinction ratio}\label{sec:av_nh}
In this section, as an additional check of consistency for \textlcsc{GAME}, we also study the recovered gas-to-extinction ratio inside the two galaxies.

\begin{figure*}
	\centering
\includegraphics[width=0.475\linewidth]{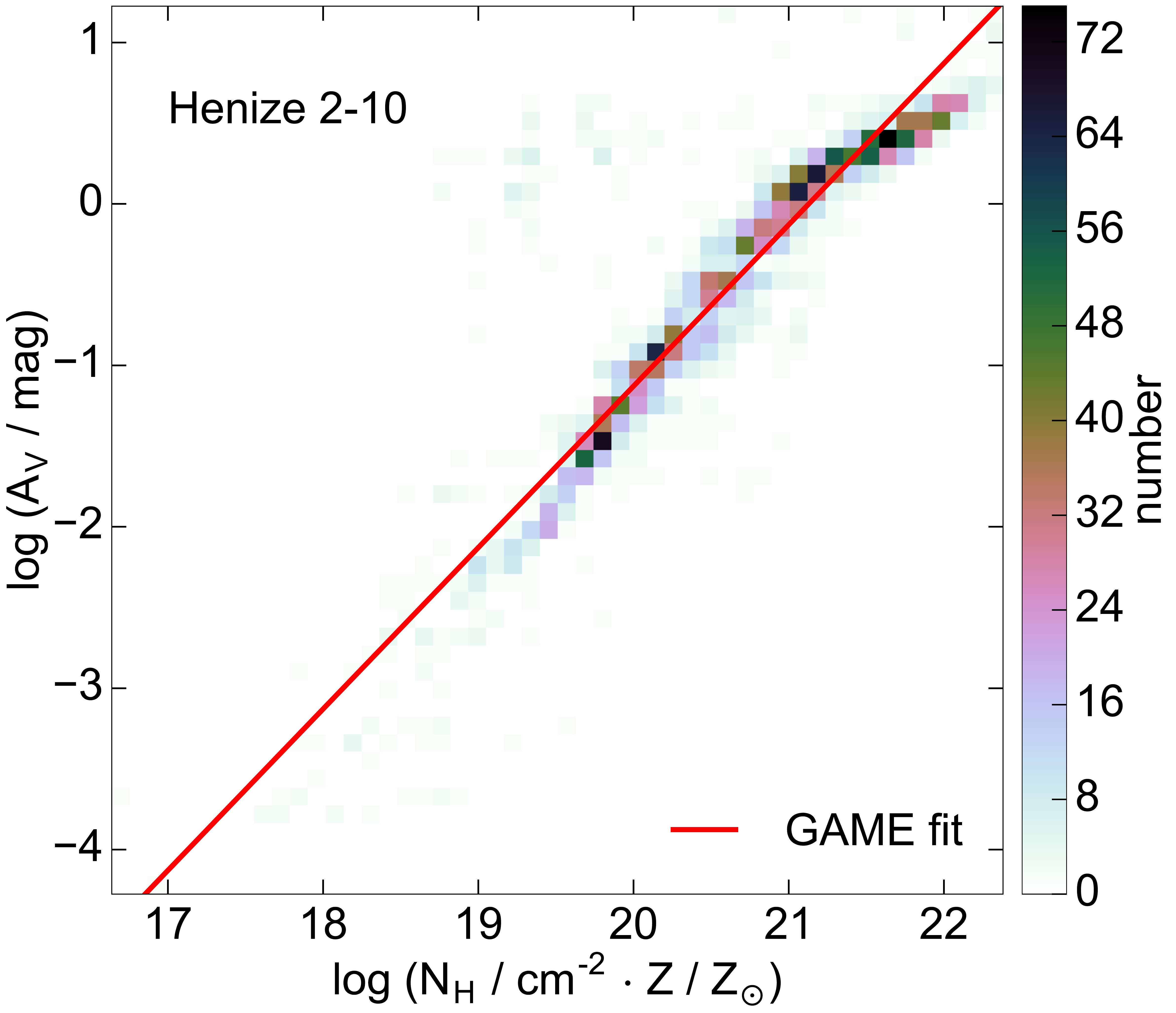}
    \includegraphics[width=0.44\linewidth]{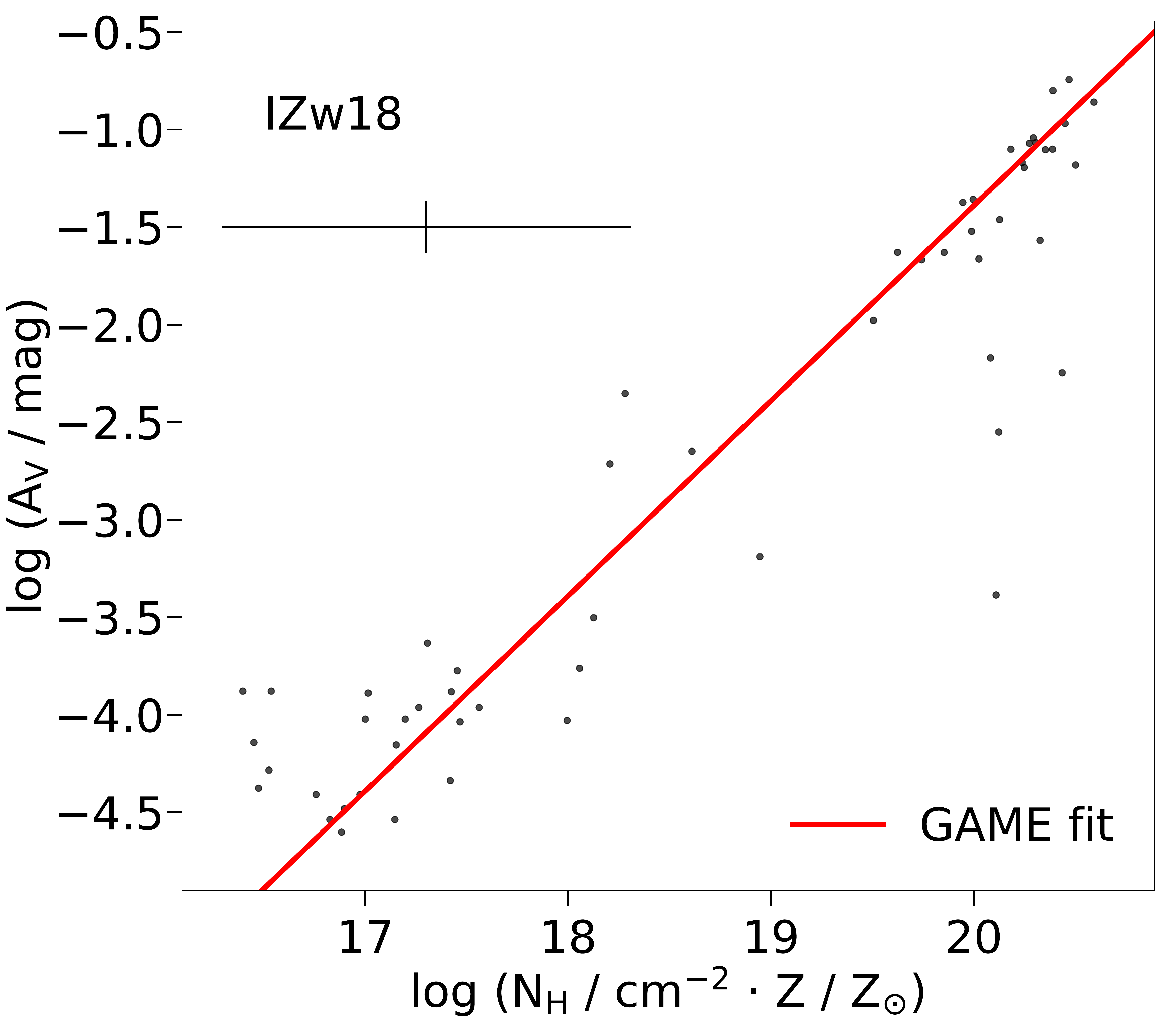}
	\caption{\textit{Left panel:} 2d histogram for the visual extinction $A_V$ as a function of the product $N_H Z$ for all the analyzed spaxels in He 2-10. The vast majority of points are located around the line. \textit{Right panel:} visual extinction $A_V$ as a function of the product $N_H Z$ for the analyzed spaxels in IZw18. Red lines denote the fit to the data using eq. \ref{eq:av_generic}. The upper-left cross denotes the typical error bar on the points.}
	\label{fig:nh_av}
\end{figure*}

In the Galaxy, the extinction in the optical band ($A_V$) and the hydrogen column density ($N_H$) are related by the following expression \citep{Reina1973,Gorenstein1975,Predehl1995,Guver2009,Watson2011,Zhu2017}:

\begin{equation}
A_V = \frac{N_H}{\alpha_{MW}} \approx \frac{N_H}{2.21 \times 10^{21} \text{cm}^{-2}} \text{mag}.
\label{eq:av_MW}
\end{equation}
where $\alpha_{MW}$ is the Milky Way hydrogen-to-extinction ratio. In the \textlcsc{GAME} library we assumed instead the following functional form:

\begin{equation}
A_V = \frac{1}{\alpha} N_H\, Z\,.
\label{eq:av_generic}
\end{equation}
where $\alpha = 2.52 \times 10^{21}$ cm$^{-2}$ mag$^{-1}$. To asses the \textlcsc{GAME} performances, we use our results on $A_V$, $N_H$, and $Z$ inferred by \textlcsc{GAME} (discussed in Sec. \ref{sec:henize} and \ref{sec:izw18}). The red lines in Fig. \ref{fig:nh_av}, represent the best fit of eq. \ref{eq:av_generic} to the \textlcsc{GAME} results. For He 2-10 and IZw18, we find $\alpha_{He 2-10}$ = (2.40 $\pm$ 0.02) $\times$ 10$^{21}$ cm$^{-2}$ mag$^{-1}$ and $\alpha_{IZw18}$ = 2.46$^{+0.43}_{-0.37}$ $\times$ 10$^{21}$ cm$^{-2}$ mag$^{-1}$, respectively. The $\alpha$ coefficients are compatible, implying that \textlcsc{GAME} is able to recover in a consistent way internal relations within the library in spite of the fact that the physical properties are recovered independently.

\bsp	
\label{lastpage}
\end{document}